\documentclass[12pt]{article}
\usepackage[latin9]{inputenc}
\usepackage{geometry}
\usepackage[american,english]{babel}
\usepackage{array}
\usepackage{float}
\usepackage{url}
\usepackage{bm}
\usepackage{multirow}
\usepackage{amsmath}
\usepackage{amsthm}
\usepackage{amssymb}
\usepackage{setspace}
\usepackage[authoryear]{natbib}
\usepackage[unicode=true,pdfusetitle,
 bookmarks=true,bookmarksnumbered=false,bookmarksopen=false,
 breaklinks=false,pdfborder={0 0 0},pdfborderstyle={},backref=false,colorlinks=false]
 {hyperref}

\makeatletter

\let\SF@@footnote\footnote
\def\footnote{\ifx\protect\@typeset@protect
    \expandafter\SF@@footnote
  \else
    \expandafter\SF@gobble@opt
  \fi
}
\expandafter\def\csname SF@gobble@opt \endcsname{\@ifnextchar[
  \SF@gobble@twobracket
  \@gobble
}
\edef\SF@gobble@opt{\noexpand\protect
  \expandafter\noexpand\csname SF@gobble@opt \endcsname}
\def\SF@gobble@twobracket[#1]#2{}
\providecommand{\tabularnewline}{\\}
\floatstyle{ruled}
\newfloat{algorithm}{tbp}{loa}
\providecommand{\algorithmname}{Algorithm}
\floatname{algorithm}{\protect\algorithmname}

\theoremstyle{plain}
\newtheorem{prop}{\protect\propositionname}
\theoremstyle{plain}
\newtheorem{cor}{\protect\corollaryname}
\theoremstyle{plain}
\newtheorem{thm}{\protect\theoremname}

\bibpunct{(}{)}{,}{a}{,}{,}
\usepackage{url}
\theoremstyle{plain}

\usepackage{lscape}
\usepackage[bottom]{footmisc}

\newtheorem{ass}{Assumption}

\usepackage{listings}

\usepackage{matlab-prettifier}

\makeatother

\addto\captionsamerican{\renewcommand{\algorithmname}{Algorithm}}
\addto\captionsamerican{\renewcommand{\corollaryname}{Corollary}}
\addto\captionsamerican{\renewcommand{\propositionname}{Proposition}}
\addto\captionsamerican{\renewcommand{\theoremname}{Theorem}}
\addto\captionsenglish{\renewcommand{\corollaryname}{Corollary}}
\addto\captionsenglish{\renewcommand{\propositionname}{Proposition}}
\addto\captionsenglish{\renewcommand{\theoremname}{Theorem}}
\providecommand{\corollaryname}{Corollary}
\providecommand{\propositionname}{Proposition}
\providecommand{\theoremname}{Theorem}

\begin{document}
\title{Computationally Efficient Methods for Solving Discrete-time Dynamic models with Continuous Actions}
\author{Takeshi Fukasawa\thanks{Graduate School of Public Policy, The University of Tokyo. 7-3-1 Hongo, Tokyo, Japan. E-mail: fukasawa3431@gmail.com.\protect \\
The current paper is a largely revised version of the paper entitled \textquotedblleft Simple method for efficiently solving dynamic models with continuous actions using policy gradient\textquotedblright . I thank participants of the Spring Meeting of The Operations Research Society of Japan, and the European Winter Meeting of the Econometric Society 2024 for their helpful comments.\protect \\
The replication code of the numerical experiments in this article is available at \protect\url{https://github.com/takeshi-fukasawa/dynamic_opt_conti_action}.\protect \\
\foreignlanguage{american}{This work was supported by JSPS KAKENHI Grant Number JP24K22629. }}}
\maketitle
\begin{abstract}
This study investigates computationally efficient algorithms for solving discrete-time infinite-horizon single-agent/multi-agent dynamic models with continuous actions. It shows that we can easily reduce the computational costs by slightly changing basic algorithms using value functions, such as the Value Function Iteration (VFI) and the Policy Iteration (PI).

The PI method with a Krylov iterative method (GMRES), which can be easily implemented using built-in packages, works much better than VFI-based algorithms even when considering continuous state models. Concerning the VFI algorithm, we can largely speed up the convergence by introducing acceleration methods of fixed-point iterations. The current study also proposes the VF-PGI-Spectral (Value Function-Policy Gradient Iteration Spectral) algorithm, which is a slight modification of the VFI. It shows numerical results where the VF-PGI-Spectral performs much better than the VFI- and PI-based algorithms especially in multi-agent dynamic games. Finally, it shows that using relative value functions further reduces the computational cost of these methods.

{\flushleft{{\bf Keywords:}  Dynamic optimization with continuous actions; Krylov iterative method; Policy Iteration; VF-PGI-Spectral algorithm; Acceleration methods of fixed-point iterations}}

\pagebreak{}
\end{abstract}

\section{Introduction}

Dynamic optimization problems involving continuous actions are widespread in economics and other fields. Examples include consumption, investment, R\&D, advertising, and dynamic pricing decisions, which are extensively studied in the literature of macroeconomics, industrial organization, labor economics, and others. Although such problems are common and important, one challenge to conducting quantitative analyses is the computational burden, especially when the size of the state space is large. In addition, researchers sometimes estimate the structural parameters of the model based on the nested fixed-point algorithm, where the dynamic optimization problem is solved given each candidate parameter values. In this case, they need to solve the dynamic optimization problem tens or hundreds of times, and the computational burden can be very huge even when the computational cost of solving the problem once is relatively small.

This study investigates computationally efficient algorithms for solving discrete-time infinite-horizon single-agent and multi-agent dynamic models with continuous actions. Concerning single-agent models, Policy Iteration (PI) method with Krylov-based methods (such as the GMRES) for the policy evaluation step performs much better than the Value Function Iteration (VFI)-based methods. Concerning multi-agent dynamic games\footnote{We restrict our attention to Markov perfect equilibrium.}, VF-PGI-Spectral (Value-Function-Policy Gradient-Iteration-Spectral) algorithm, which this study proposes, outperforms other methods. 

With regard to single-agent models, many algorithms have been proposed so far. VFI would be the most famous one, but it is known that the convergence speed of VFI is slow, especially when the discount factor is close to 1. PI would be also famous, but numerical experiments in previous studies (\citealp{santos2004convergence}, \citealp{coleman2021matlab}) have shown that PI is not necessarily very fast, though theoretical fast convergence rate is guaranteed. Finally, studies in economics literature, especially in macroeconomics, have developed some methods for solving single-agent models. Examples include the Euler Equation method (EE; cf. \citealp{judd1998numerical}), the Endogenous Gridpoint Method (EGM; \citealp{Carroll2006}), and the Envelope Condition Method (ECM; \citealp{maliar2013envelope} and \citealp{arellano2016envelope}) . Though useful in some models, their performance largely relies on whether we can avoid nonlinear root-finding procedure in each iteration.\footnote{Discretizing the possible actions may help avoid the nonlinear optimization problem, but coarse discretized action can reduce the precision of the approximation. In contrast, when the number of possible actions is large, evaluating long-term profits for all the possible actions can be time-consuming. Hence, this approach has limitations. See also the discussion in Section 12 of \citet{judd1998numerical}.}

The current study points out that we can easily attain fast convergence of the PI by additionally introducing Krylov-based methods (PI-Krylov) in the policy evaluation step, even when we consider a continuous state model. By introducing the procedure, we can vastly reduce the computational cost per iteration, and it contributes to smaller computation time until convergence. Though \citet{mrkaic2002policy} first showed that using Krylov-based methods largely reduces the computational burden of the PI method, he considered a model with discretized states. The current study shows that we can easily apply the Krylov method even in continuous state models without discretizations. Because PI is theoretically known to attain fast (quadratic or superlinear) convergence in single-agent models (cf., \citealp{puterman1979convergence}, \citealp{santos2004convergence}), and Krylov-based algorithms (e.g., GMRES) are easy to implement using built-in packages in many programming languages, it is worth considering when practitioners make much of computation time.\footnote{\citet{chen2025model} recently proposed a method to solve linear equations in the policy evaluation step of dynamic discrete choice models with large state spaces using a technique based on the conjugate gradient algorithm. However, the current study found by numerical experiments that the GMRES performs much better than the method. For more discussions on the method, see Appendix \ref{subsec:Model-Adaptive-PI}.} \footnote{\citet{rendahl2022continuous}, comparing solution methods for solving continuous-time and discrete-time models, pointed out that a key reason for the larger computational costs in the PI compared to the VFI in the discrete-time model observed by \citet{santos2004convergence} is that they did not make use of sparse matrix computations. \citet{mrkaic2002policy} also mentions that Krylov methods are very efficient at solving large sparse systems. Though exploiting sparsity would further computational costs, the current study shows results where the PI-Krylov works much better than the VFI.}

The current study also shows that we can largely accelerate the convergence of VFI-based algorithms by introducing appropriate acceleration methods of fixed-point iterations (e.g., spectral, Anderson acceleration).\footnote{The idea of introducing acceleration methods can be also found in the recent economics literature (e.g., \citet{aguirregabiria2021imposing} for estimation of dynamic discrete games, and \citet{conlon2020best,fukasawa2024fast} for estimations of demand (BLP) models.} It also shows further introducing the idea of relative value functions (\citealp{morton1977discounting,bray2019strong}) or endogenous value functions (\citealp{bray2019markov}) leads to faster convergence.

The current study also proposes a new, simple, computationally efficient ``Value Function-Policy Gradient Iteration-Spectral (VF-PGI-Spectral)'' algorithm for quantitatively solving the models. Though the proposed algorithm inherits the idea of the VFI, we jointly update value functions $V$ and action variables $a$ (policy functions) in each iteration. Value functions $V$ are updated as in the value function iteration with candidate values of actions $a$. Continuous actions $a$ are updated so that they step in the direction of the gradient of the action value functions $Q$, given candidate value functions $V$.\footnote{In standard continuous optimization algorithms, gradient-type methods use gradients as the updating direction. The concept of using gradients as an updating direction can be also found in the reinforcement learning literature (Deterministic policy gradient method; \citealp{silver2014deterministic}). However, the settings may differ, as discussed in Appendix \ref{sec:other_algorithms}.} Action value functions are the dynamic version of ``profit.'' The computational cost per iteration is low because these updating steps do not require nonlinear root-finding or optimization procedures. Note that we can easily introduce box constraints on the domain of actions, such as nonnegativity constraints.

VF-PGI-Spectral utilizes the spectral algorithm, known for its effectiveness in reducing the number of iterations until convergence to solve nonlinear equations, mainly in numerical analysis (cf. (\citealp{barzilai1988two}; \citealp{la2006spectral}). The spectral algorithm shares similarities with the Newton's method, well known for its rapid local convergence. Unlike the Newton's method, we do not have to specify the first derivative to be solved, which is the advantage of the spectral algorithm. Though we need to introduce a tuning parameter which can be interpreted as the learning rate in VF-PGI-Spectral, the current study analytically and numerically shows that the convergence is not so sensitive to the choice of the parameter by introducing the ideas of the spectral algorithm. This study shows the results of numerical experiments, where the VF-PGI-Spectral algorithm is faster than the PI- and VFI- based algorithms, especially in dynamic games. 

Concerning multi-agent dynamic games, there are only a few studies on computationally efficient algorithms, to my knowledge.\footnote{\citet{farias2012approximate} proposed an approximate dynamic programming approach based on linear programming for solving high-dimensional dynamic games. Though promising, we should discretize the possible actions and relax the stopping criteria in the method when we consider a model with continuous actions. Such discreteness is not attractive for estimating structural parameters, because the objective function for estimation can be discontinuous. 

Besides, the algorithms discussed in the current paper may not find all the equilibria, unlike the one proposed by \citet{yeltekin2017computing}. Note that slow convergence of iterations sometimes discourages practitioners from the attempt to find multiple equilibria. In that sense, considering faster algorithms for finding one solution would encourage practitioners to find more equilibria.} In dynamic games, counterparts of VFI and PI in single-agent models exist. However, there is no guarantee that the convergence speed of the PI-based method is fast (quadratic or superlinear) unlike the case of single-agent models, and the relative advantage of the PI-based algorithm is small. By conducting numerical experiments, the current study found that the VF-PGI-Spectral is much faster than PI-based algorithms in multi-agent settings. 

The contributions of the current study are threefold. First, the current study shows that we can easily reduce the computation time of the PI using Krylov-based methods, even when considering continuous-state models. Second, concerning single-agent models, the current study comprehensively compares many algorithms for solving dynamic models with continuous actions and continuous states following the recent development in the fields of economics and operations research, using the canonical single-agent optimal-growth model with elastic labor supply.\footnote{Though the model might be simple, comprehensive comparisons of algorithms using simple models are essential for further considerations of algorithms in more complicated models.} Though computationally efficient algorithms for solving single-agent models with continuous actions have been extensively studied in the literature (cf., \citealp{maliar2014numerical}), there is further room for speed-up. The current study points out that we can largely reduce the computational burden by combining existing ideas (Krylov method, relative value function, acceleration methods etc.), which are slight modifications of value function-based algorithms.\footnote{The current study makes much of not only the computational speed, but also easy implementations of algorithms.}\footnote{The current study also contributes to the literature on operations research studying computationally efficient methods for solving dynamic optimization problems. Though they have extensively studied single-agent finite-state finite-action models (see \citealp{grand2021convex} for literature review), studies focusing on computationally efficient methods for solving continuous actions models and multi-agent dynamic games are scarce, to my knowledge. In continuous action models and multi-agent dynamic games, insights on single-agent finite-state finite-action models are not directly applicable. The current study contributes to the literature by showing that we can largely speed-up the algorithms by slightly changing algorithms using value functions even in the case of continuous state models.} Third, the current study proposes the VF-PGI-Spectral algorithm, which is useful especially in dynamic games. The algorithm is also a slight modification of the traditional VFI algorithm, and it is easy to implement. The methods are worth considering when practitioners face difficulties associated with computational burden. 

\medskip{}

The remainder of this paper is presented as follows: Section \ref{sec:Single-agent-dynamic-models} discusses algorithms for single-agent models, and Section \ref{sec:Multi-agent-dynamic-games} discusses algorithms for multi-agent dynamic games. Section \ref{sec:Numerical-experiments} shows the results of numerical experiments, and Section \ref{sec:Conclusions} concludes. Appendix \ref{sec:Full-results-growth-model} shows the detailed results of the numerical experiments, and Appendix \ref{sec:spectral-krylov} concisely describes the spectral algorithm and the Krylov iterative method. 

Appendix \ref{sec:other_algorithms} further discusses algorithms not described in the main part of the current paper. Appendix \ref{sec:Further-discussions-on-spectral} further discusses the spectral algorithm. In Appendix \ref{sec:Extension:-Box-constraint}, we examine how to incorporate box constraints related to continuous actions, such as nonnegative constraints, into the VF-PGI-Spectral algorithm. Appendix \ref{sec:Details-experiments} describes the details of the numerical experiments, and Appendix \ref{sec:Additional-results} shows additional numerical results. Appendix \ref{sec:Proof} shows all proofs of mathematical statements.

\section{Single-agent dynamic models\label{sec:Single-agent-dynamic-models}}

In this section, we focus on a single-agent dynamic model. Section \ref{subsec:Model-single-agent} describes the model, and Section \ref{subsec:Algorithms-single-agent} discusses algorithms for solving the model, including VFI, PI, and VF-PGI-Spectral. In the following, the primes on variables denote the next-period variables, and $\epsilon$ denotes a tolerance level of iterations.

\subsection{Model\label{subsec:Model-single-agent}}

Consider a single-agent infinite-horizon dynamic optimization problem with continuous action variables $a(s)$ characterized by the following Bellman equation:

\begin{eqnarray*}
V(s) & = & \max_{a(s)\in\mathcal{A}(s)}Q(a(s),s;V)\equiv r(a^{*}(s),s)+\beta\int V\left(s^{\prime}\right)p\left(s^{\prime}|s,a^{*}(s)\right)ds^{\prime},
\end{eqnarray*}
where $a(s)\equiv\left(a^{d}(s)\right)_{d=1,\cdots,D}$ denotes the action of the agent at state $s$. Note that the action can be multi-dimensional, and let $D$ be the dimension of the actions. Let $a^{*}(s)$ be the solution to the maximization problem. $\mathcal{S}$ denotes the state space, which can be either discrete or continuous. $\mathcal{A}(s)$ denotes the convex closed set of possible actions at state $s$. For now, we assume $\mathcal{A}(s)=\mathbb{R}^{D}$.\footnote{We can easily introduce box constraints, as discussed in Appendix \ref{sec:Extension:-Box-constraint}.} $p\left(s^{\prime}|s,a(s)\right)$ denotes the state transition probability. $Q(a(s),s;V)$ denotes the action value function, and $V(s)$ denotes the value function. Note that $Q$ depends on the value function $V$, and $Q\left(a^{*}(s),s;V\right)=V(s)$ holds.

Assuming that $Q$ is differentiable with respect to $a$, the following equations hold:

\begin{eqnarray}
0 & = & \frac{\partial Q(a^{*}(s),s;V)}{\partial a^{*}(s)}\nonumber \\
 & = & \frac{\partial r\left(a^{*}(s),s\right)}{\partial a^{*}(s)}+\beta\frac{\partial\left[\int V\left(s^{\prime}\right)p\left(s^{\prime}|s,a^{*}(s)\right)ds^{\prime}\right]}{\partial a^{*}(s)},\label{eq:FOC_single_agent}\\
V(s) & = & r(a^{*}(s),s)+\beta\int V\left(s^{\prime}\right)p\left(s^{\prime}|s,a^{*}(s)\right)ds^{\prime}.\label{eq:V_eq_single_agent}
\end{eqnarray}

Then, the problem reduces to solving the nonlinear equations (\ref{eq:FOC_single_agent}) and (\ref{eq:V_eq_single_agent}) for $\left\{ a^{*}(s)\right\} _{s\in\mathcal{S}}$ and $\left\{ V(s)\right\} _{s\in\mathcal{S}}$. In general, the state space of $s$ is not finite. In this case, we should take suitable finite set of grid points $\widehat{\mathcal{S}}\subset\mathcal{S}$ to approximate the solution well. If the state space $\mathcal{S}$ is finite, let $\widehat{\mathcal{S}}=\mathcal{S}$. The alternative equations that we should solve are as follows:\footnote{We implicitly assume that we choose appropriate $\overline{V}$ so that $\int\overline{V}\left(s^{\prime}\right)p\left(s^{\prime}|\widehat{s},a\left(\widehat{s}\right)\right)ds^{\prime}$ is differentiable concerning $a(s)$. If we apply piecewise linear interpolations as in \citet{santos2004convergence}, the assumption may not hold.}

\begin{eqnarray}
0 & = & \frac{\partial Q(a^{*}\left(\widehat{s}\right),\widehat{s};V)}{\partial a^{*}\left(\widehat{s}\right)}\nonumber \\
 & = & \frac{\partial r\left(a^{*}\left(\widehat{s}\right),\widehat{s}\right)}{\partial a^{*}\left(\widehat{s}\right)}+\beta\frac{\partial\left[\int\overline{V}\left(s^{\prime}\right)p\left(s^{\prime}|\widehat{s},a^{*}\left(\widehat{s}\right)\right)ds^{\prime}\right]}{\partial a^{*}\left(\widehat{s}\right)},\label{eq:FOC_single_agent-grid}\\
V\left(\widehat{s}\right) & = & r\left(a^{*}\left(\widehat{s}\right),\widehat{s}\right)+\beta\int\overline{V}\left(s^{\prime}\right)p\left(s^{\prime}|\widehat{s},a^{*}\left(\widehat{s}\right)\right)ds^{\prime}.\label{eq:V_eq_single_agent-grid}
\end{eqnarray}

Here, $\overline{V}$ denotes the interpolated values of $V$.\footnote{For example, when using Chebyshev polynomials as basis functions, $\overline{V}$ can be constructed by $\overline{V}(s)=\sum_{m=0}^{M-1}\theta_{m}T_{m}(s)$, where $T_{m}(s)$ denotes the order-$m$ Chebyshev basis function. $\theta_{m=0,1,\cdots,M}$ are unknown coefficients, and these values are computed based on the values of $\left\{ V(s_{t}^{(grid)})\right\} _{s_{t}^{(grid)}\in\mathcal{S}^{(grid)}}$. Note that $T_{m}$ satisfies $T_{0}(s)=1,T_{1}(s)=s,$ and $T_{m}(s)=2sT_{m-1}(s)-T_{m-2}(s)\ (m\geq2)$.} If $\widehat{\mathcal{S}}\subsetneq\mathcal{S}$, we cannot directly evaluate the values of $V(s)\ s\notin\widehat{\mathcal{S}}$, and we interpolate these values based on the values of $\left\{ V(\widehat{s})\right\} _{\widehat{s}\in\text{\ensuremath{\widehat{\mathcal{S}}}}}$.\footnote{See \citet{maliar2014numerical} for detailed discussions on the procedure and other issues.} We solve these equations for $\left\{ a^{*}(\widehat{s})\right\} _{\widehat{s}\in\widehat{\mathcal{S}}}$ and $\left\{ V(\widehat{s})\right\} _{\widehat{s}\in\widehat{\mathcal{S}}}$.

\subsection{Algorithms\label{subsec:Algorithms-single-agent}}

In this subsection, we discuss some algorithms, including Value Function Iteration (VFI), Policy Iteration (PI), and Value Function-Policy Gradient Iteration-Spectral (VF-PGI-Spectral), for solving the dynamic model. We then discuss additional tricks to reduce the computational costs, including relative value function and acceleration methods of fixed-point iterations. In Appendix \ref{subsec:Other-methods-sigle-agent}, we discuss other methods not described in the current section. 

\subsubsection{VFI}

Value Function Iteration (VFI) would be the most popular method for solving the dynamic optimization problem. Algorithm \ref{alg:VFI_single_agent} shows the steps of the VFI. 

\begin{algorithm}[H]
\begin{enumerate}
\item Take grid points $\widehat{s}\in\widehat{\mathcal{S}}$. Set initial values $V^{(0)}\equiv\left\{ V^{(0)}\left(\widehat{s}\right)\right\} _{\widehat{s}\in\widehat{\mathcal{S}}}$
\item Iterate the following $(n=0,1,\cdots)$:
\begin{enumerate}
\item Solve for $a^{(n+1)}\left(\widehat{s}\right)$ satisfying the following nonlinear equation for each $\widehat{s}\in\widehat{\mathcal{S}}$:

\begin{eqnarray*}
0 & = & \frac{\partial Q\left(a^{(n+1)}\left(\widehat{s}\right),\widehat{s};V^{(n)}\right)}{\partial a^{(n+1)}\left(\widehat{s}\right)}\\
 & = & \frac{\partial r\left(a^{(n+1)}\left(\widehat{s}\right),\widehat{s}\right)}{\partial a^{(n+1)}\left(\widehat{s}\right)}+\beta\frac{\partial\left[\int\overline{V}^{(n)}\left(\text{\ensuremath{s^{\prime}}}\right)p\left(s^{\prime}|\widehat{s},a^{(n+1)}\left(\widehat{s}\right)\right)ds^{\prime}\right]}{\partial a^{(n+1)}\left(\widehat{s}\right)}
\end{eqnarray*}

\item Compute
\begin{eqnarray*}
V^{(n+1)}\left(\widehat{s}\right) & = & r\left(a^{(n+1)}\left(\widehat{s}\right),\widehat{s}\right)+\beta\int\overline{V}^{(n)}\left(s^{\prime}\right)p\left(s^{\prime}|\widehat{s},a^{(n+1)}\left(\widehat{s}\right)\right)ds^{\prime}
\end{eqnarray*}
 for $\widehat{s}\in\widehat{\mathcal{S}}$
\item Exit the iteration if $\left\Vert V^{(n+1)}-V^{(n)}\right\Vert <\epsilon_{V}$

\medskip{}

\end{enumerate}
\caption{VFI (Single-agent dynamic model)\label{alg:VFI_single_agent}}
\end{enumerate}
\end{algorithm}

Though the procedure is simple and the iteration is convergent under regularity conditions, one problem of the VFI is the need to solve nonlinear equations at all the grid points and in each iteration (Step 2(a)). Generally, finding a root of a nonlinear equation takes much time, and it is the undesirable feature of the VFI with continuous actions. 

\subsubsection{PI\label{subsec:PI-single-agent}}

Policy Iteration (PI) algorithm would be also well-known for solving dynamic optimization problems.\footnote{PI is also known as Howard Policy Iteration (HPI) as in \citet{sargent2024dynamic}.} Algorithm \ref{alg:PI_single_agent} shows the steps. 

\begin{algorithm}[H]
\begin{enumerate}
\item Take grid points $\widehat{s}\in\widehat{\mathcal{S}}$. Set initial values $V^{(0)}\equiv\left\{ V^{(0)}\left(\widehat{s}\right)\right\} _{\widehat{s}\in\widehat{\mathcal{S}}}$.
\item Iterate the following $(n=0,1,\cdots)$:
\begin{enumerate}
\item Policy Improvement Step:

Solve for $a^{(n+1)}\left(\widehat{s}\right)$ satisfying the following nonlinear equation for each $\widehat{s}\in\widehat{\mathcal{S}}$:

\begin{eqnarray*}
0 & = & \frac{\partial Q\left(a^{(n+1)}\left(\widehat{s}\right),\widehat{s};V^{(n)}\right)}{\partial a^{(n+1)}\left(\widehat{s}\right)}\\
 & = & \frac{\partial r\left(a^{(n+1)}\left(\widehat{s}\right),\widehat{s}\right)}{\partial a^{(n+1)}\left(\widehat{s}\right)}+\beta\frac{\partial\left[\int\overline{V}^{(n)}\left(\text{\ensuremath{s^{\prime}}}\right)p\left(s^{\prime}|\widehat{s},a^{(n+1)}\left(\widehat{s}\right)\right)ds^{\prime}\right]}{\partial a^{(n+1)}\left(\widehat{s}\right)}
\end{eqnarray*}

\item Policy Evaluation Step: Solve $V^{(n+1)}=r+\beta\int\overline{V}^{(n+1)}\left(s^{\prime}\right)p\left(s^{\prime}|\widehat{s},a^{(n+1)}\left(\widehat{s}\right)\right)ds^{\prime}$ for $V^{(n+1)}$
\begin{itemize}
\item PI: Iterate applying the mapping $\Phi_{V}\left(V,a\right)\left(\widehat{s}\right)\equiv r\left(a\left(\widehat{s}\right),\widehat{s}\right)+\beta\int\overline{V}\left(s^{\prime}\right)p\left(s^{\prime}|\widehat{s},a\left(\widehat{s}\right)\right)ds^{\prime}$ to $V^{(n)}$ given $a^{(n+1)}$ until convergence
\item PI-Krylov: Solve the linear equation by Krylov-based methods (cf. Algorithm \ref{alg:PI-Policy-evaluation-Krylov})
\end{itemize}
\item Exit the iteration if $\left\Vert V^{(n+1)}-V^{(n)}\right\Vert <\epsilon_{V}$
\end{enumerate}
\medskip{}

\caption{PI (Single-agent dynamic model)\label{alg:PI_single_agent}}
\end{enumerate}
\end{algorithm}

As discussed in \citet{santos2004convergence}, PI is convergent and the convergence speed is superlinear or quadratic under regularity conditions. The problem of the PI is the computational cost per iteration. In the policy evaluation step (Step 2(b)), we need to solve an equation, and the step can be computationally burdensome. Hence, we discuss an efficient implementation of the policy evaluation step below. Note that we need to solve a nonlinear equation concerning $a$ in the policy improvement step (Step 2(a)) as in the VFI. However, the computational burden of the policy evaluation step is generally also large, and the computational burden of the nonlinear root finding concerning $a$ is less critical than the case of the VFI. 

\paragraph*{Policy Evaluation Step}

In principle, we can repeatedly apply the mapping $\Phi_{V}$ until the convergence of $V$ to solve $V^{(n+1)}=r+\beta\int\overline{V}^{(n+1)}\left(s^{\prime}\right)p\left(s^{\prime}|\widehat{s},a^{(n+1)}\left(\widehat{s}\right)\right)ds^{\prime}$ for $V^{(n+1)}$. The strategy has been used in previous studies (e.g., \citealp{arellano2016envelope}; \citealp{coleman2021matlab}). However, the convergence speed of the iteration is typically slow, and we discuss more sophisticated and faster method building on Krylov-based methods that is easy to implement in many programming languages. Note that we can directly solve the linear equation $V^{(n)}=r+\beta\int V^{(n)}\left(s^{\prime}\right)p\left(s^{\prime}|\widehat{s},a^{(n)}\left(\widehat{s}\right)\right)ds^{\prime}$ for $V^{(n)}$ by directly applying the Krylov-based methods when the state space $\mathcal{S}$ is finite.\footnote{When the state space $\mathcal{S}$ is finite, we can also solve the linear equation by any non-iterative algorithms.} Hence, the following discussion focuses on the case where the state space $\mathcal{S}$ is continuous and we need some approximation techniques.\footnote{The discussion also holds even when $\mathcal{S}$ is finite but practitioners introduce approximation methods to mitigate problems associated with large $|\mathcal{S}|$. }

First, the integral $\int\overline{V}\left(s^{\prime}\right)p\left(s^{\prime}|\widehat{s},a\left(\widehat{s}\right)\right)ds^{\prime}$ is typically approximated by weighted sum of terms $\sum_{m=1}^{M}w_{m}\overline{V}\left(s_{(m)}^{\prime}\right)$, where $\sum_{m=1}^{M}w_{m}=1$, and we assume that we use the approximation method. We further assume that we use linear basis functions\footnote{The linear interpolation method can be regarded as one specific form of the method using linear basis functions.} to interpolate the values of $\overline{V}\left(s^{\prime}\right)$,\footnote{See Sections 6 and 12.9 of \citet{judd1998numerical} for the discussions on the choice of linear basis functions.} and let $\overline{V}\left(s^{\prime}\right)=\Psi\left(s^{\prime}\right)\theta$, where $\Psi(s^{\prime})$ denotes a $|\widehat{\mathcal{S}}|\times n_{\theta}$-dimensional matrix and $\theta$ denotes a $n_{\theta}\times1$-dimensional parameter.\footnote{In general, $|\widehat{\mathcal{S}}|\geq n_{\theta}$ should hold.} This assumption is typically satisfied in most applications, unless we use nonlinear basis functions (e.g., neural networks). Concerning $\theta$, $\theta$ is typically computed so that $\theta=\arg\min_{\theta}\left\Vert X\theta-V\right\Vert _{2}=\left(X^{\prime}X\right)^{-1}X^{\prime}V$, where $X\equiv\left(\Psi(s)\right)_{s\in\widehat{\mathcal{S}}}$ denotes a $|\widehat{\mathcal{S}}|\times n_{\theta}$-dimensional matrix and $V\equiv\left(V(s)\right)_{s\in\widehat{\mathcal{S}}}$ denotes a $|\widehat{\mathcal{S}}|$-dimensional vector. 

Here, let $V\equiv\left(V\left(\widehat{s}\right)\right)_{\widehat{s}\in\widehat{\mathcal{S}}}$, $r\equiv\left(r\left(\widehat{s}\right)\right)_{\widehat{s}\in\widehat{\mathcal{S}}}$, $\widetilde{P}\equiv\left(\begin{array}{ccc}
w_{m=1}I_{|\widehat{\mathcal{S}}|}, & \cdots & ,w_{m=M}I_{|\widehat{\mathcal{S}}|}\end{array}\right)$ $\left(|\widehat{\mathcal{S}}|\times\left(|\widehat{\mathcal{S}}|\times M\right)\text{-dimensional}\right)$, $\widetilde{X_{m}}\equiv\left(\Psi\left(s_{(m)}^{\prime}\left(\widehat{s}\right)\right)\right)_{\widehat{s}\in\widehat{\mathcal{S}}}$ $\left(|\widehat{\mathcal{S}}|\times n_{\theta}\text{-dimensional}\right)$, and $\widetilde{X}\equiv\left(\widetilde{X_{m}}\right)_{m=1,\cdots M}$ $\left(\left(|\widehat{\mathcal{S}}|\times M\right)\times n_{\theta}\text{-dimensional}\right)$ where $I_{|\widehat{\mathcal{S}}|}$ is the $|\widehat{\mathcal{S}}|\times|\widehat{\mathcal{S}}|$-dimensional identity matrix. Then, the policy evaluation step reduces to solving the following linear equation concerning $V\equiv\left(V\left(\widehat{s}\right)\right)_{\widehat{s}\in\widehat{\mathcal{S}}}$:

\begin{eqnarray*}
V & = & r+\beta\int\overline{V}\left(s^{\prime}\right)p\left(s^{\prime}|\widehat{s},a\left(\widehat{s}\right)\right)ds^{\prime}\\
 & = & r+\beta\left(\sum_{m=1}^{M}w_{m}p\left(s_{(m)}^{\prime}\left(\widehat{s}\right)|\widehat{s},a\left(\widehat{s}\right)\right)\overline{V}\left(s_{(m)}^{\prime}\left(\widehat{s}\right)\right)\right)_{\widehat{s}\in\widehat{\mathcal{S}}}\\
 & = & r+\beta\sum_{m=1}^{M}\left[\left(w_{m}\right)_{\widehat{s}\in\widehat{\mathcal{S}}}\cdot\left(\overline{V}\left(s_{(m)}^{\prime}\left(\widehat{s}\right)\right)\right)_{\widehat{s}\in\widehat{\mathcal{S}}}\right]\\
 & = & r+\beta\left(\begin{array}{ccc}
w_{m=1}I_{|\widehat{\mathcal{S}}|}, & \cdots & ,w_{m=M}I_{|\widehat{\mathcal{S}}|}\end{array}\right)\left(\begin{array}{c}
\left(\overline{V}\left(s_{(m=1)}^{\prime}\left(\widehat{s}\right)\right)\right)_{\widehat{s}\in\widehat{\mathcal{S}}}\\
\vdots\\
\left(\overline{V}\left(s_{(m=M)}^{\prime}\left(\widehat{s}\right)\right)\right)_{\widehat{s}\in\widehat{\mathcal{S}}}
\end{array}\right)\\
 & = & r+\beta\widetilde{P}\widetilde{X}\theta\\
 & = & r+\beta\widetilde{P}\widetilde{X}\left(X^{\prime}X\right)^{-1}X^{\prime}V
\end{eqnarray*}
Namely, $AV=r$ holds, where $A\equiv I_{|\widehat{\mathcal{S}}|}-\beta\widetilde{P}\widetilde{X}\left(X^{\prime}X\right)^{-1}X^{\prime}$.

In principle, we can solve the linear equation by using any linear equation solvers by explicitly computing $A\equiv I_{|\widehat{\mathcal{S}}|}-\beta\widetilde{P}\widetilde{X}\left(X^{\prime}X\right)^{-1}X^{\prime}$. However, the explicit computation of $A$ is sometimes cumbersome, if not possible. For instance, when dealing with the problems associated with ill-conditioned $X^{\prime}X$.\footnote{The problem can be severe when we use non-orthogonal basis functions.} When the matrix is ill-conditioned, the direct computation of $\left(X^{\prime}X\right)^{-1}$ can be numerically unreliable, and linear least squares solvers utilize the QR factorization or singular-value decomposition (SVD) to compute $\left(X^{\prime}X\right)^{-1}X^{\prime}$.\footnote{See Chapter 10 of \citet{nocedal1999numerical} and Section 4.2 of \citet{judd2011numerically} for details.} Delving into these details enables practitioners to directly compute $A$, but the computational code would be more involved. Hence, we focus on the matrix-free Krylov iterative method, which does not necessarily require the direct computation of $A$.\footnote{Matrix-free Krylov methods have been applied for solving nonlinear equations using Newton's method (cf. \citealp{kelley2003solving}). In the economics literature, \citet{fukasawa2025jacobian} also applied the idea to develop the Jacobian-free version of the EPL (Efficient Pseudo Likelihood) algorithm proposed by \citet{dearing2024efficient} for structurally estimating dynamic discrete game models without unobserved heterogeneity.}

As discussed in Appendix \ref{subsec:Krylov-iterative-method} concisely, in the Krylov-based algorithms, to solve a linear equation $Ax=b$, it is sufficient to specify the vector $b$, initial values of $x$, and a function $g$ such that $g(x)=Ax$. In the algorithms, explicit computation of $A$ is not necessary. It implies we can solve $V=r+\beta\int\overline{V}\left(s^{\prime}\right)p\left(s^{\prime}|\widehat{s},a\left(\widehat{s}\right)\right)ds^{\prime}$ for $V$ by the procedure shown in Algorithm \ref{alg:PI-Policy-evaluation-Krylov}.

\begin{algorithm}[H]
\begin{enumerate}
\item Construct a function $g:\mathbb{R}^{|\mathcal{S}|}\rightarrow\mathbb{R}^{|\mathcal{S}|}$ such that $g\left(V\right)=V-\beta\int\overline{V}\left(s^{\prime}\right)p\left(s^{\prime}|\widehat{s},a\left(\widehat{s}\right)\right)ds^{\prime}$.
\item Apply the matrix-free Krylov-based algorithm (for solving $Ax=b$; e.g., GMRES,BiCG, BiCGSTAB, CGS) by using the following inputs:
\begin{itemize}
\item vector $V_{0}$ $\leftarrow$ Counterpart of initial $x$
\item vector $r$ $\leftarrow$ Counterpart of $b$
\item function $g$ $\leftarrow$Counterpart of the function $g$ such that $g(x)=Ax$ 
\end{itemize}
\end{enumerate}
\caption{Computationally efficient method for the policy evaluation step of the PI algorithm (Single-agent dynamic model)\label{alg:PI-Policy-evaluation-Krylov}}

{\footnotesize{}Note. $\int\overline{V}\left(s^{\prime}\right)p\left(s^{\prime}|\widehat{s},a\left(\widehat{s}\right)\right)ds^{\prime}=\widetilde{X}\theta$, and $\widetilde{X}$ can be precomputed before applying the Krylov-based algorithm. The precomputation leads to smaller computational burden. We can compute $\theta\equiv\arg\min_{\theta}\left\Vert X\theta-V\right\Vert _{2}$ by any linear least squares solvers.}{\footnotesize\par}
\end{algorithm}

Krylov-based methods (e.g., GMRES, BiCG, BiCGSTAB, CGS)\footnote{Concerning these methods, see \citet{saad2003iterative} for details, and see \citet{mrkaic2002policy} for brief description.} are available in many programming languages (e.g., MATLAB, Scipy package in Python, IterativeSolvers package in Julia), and we can easily implement these methods. In the numerical experiments, the current study uses the GMRES (Generalized minimal residual) method, because the theoretical properties are investigated well, and widely used to solve high-dimensional linear equations.

For instance, in MATLAB, we can solve for $V$ satisfying $V=r+\beta\int\overline{V}\left(s^{\prime}\right)p\left(s^{\prime}|\widehat{s},a\left(\widehat{s}\right)\right)ds^{\prime}$ by applying the following code using the GMRES algorithm with initial values $V_{0}$:\\
\begin{lstlisting}[frame=single,style=Matlab-bw]
func_for_krylov_anonymous=@(V)func_for_krylov(V,a);
V_sol =...
gmres(func_for_krylov_anonymous, r,[],[],[],[],[],V0);
\end{lstlisting}

Here, \url{func_for_krylov(V,a)} denotes a function computing the counterpart of $V-\beta\int\overline{V}\left(s^{\prime}\right)p\left(s^{\prime}|\widehat{s},a\left(\widehat{s}\right)\right)ds^{\prime}$ given $V$ and $a$, and $V$ is a vector. \url{V_sol} is the solution of the equation based on the GMRES algorithm.

\subsubsection{VF-PGI-Spectral}

As mentioned earlier, one problem of the VFI is the computational cost associated with the nonlinear root finding concerning $a$. Hence, the current study proposes a novel algorithm, which we term as Value Function-Policy Gradient Iteration-Spectral (VF-PGI-Spectral) algorithm, which alleviates the problem. Algorithm \ref{alg:VF-PGI_single_agent} shows the procedure.

\begin{algorithm}[H]
\begin{enumerate}
\item Take grid points $\widehat{s}\in\widehat{\mathcal{S}}$. Set initial values $V^{(0)}\equiv\left\{ V^{(0)}\left(\widehat{s}\right)\right\} _{\widehat{s}\in\widehat{\mathcal{S}}}$ and\textsl{ $a^{(0)}\equiv\left\{ a^{(0)}\left(\widehat{s}\right)\right\} _{\widehat{s}\in\widehat{\mathcal{S}}}$. }Set tuning parameters $\alpha^{(0)}$ and $\lambda$.
\item Iterate the following $(n=0,1,\cdots)$:
\begin{enumerate}
\item Compute $a^{(n+1*)}\left(\widehat{s}\right)$ by:

\[
a^{(n+1*)}\left(\widehat{s}\right)=\Phi_{a}\left(V^{(n)},a^{(n)}\right)\left(\widehat{s}\right)\equiv a^{(n)}\left(\widehat{s}\right)+\lambda\cdot\frac{\partial Q\left(a^{(n)}\left(\widehat{s}\right),\widehat{s};V^{(n)}\right)}{\partial a^{(n)}\left(\widehat{s}\right)}
\]
where

\[
\frac{\partial Q\left(a^{(n)}\left(\widehat{s}\right),\widehat{s};V^{(n)}\right)}{\partial a^{(n)}\left(\widehat{s}\right)}=\frac{\partial r\left(a^{(n)}\left(\widehat{s}\right),\widehat{s}\right)}{\partial a^{(n)}\left(\widehat{s}\right)}+\beta\frac{\partial\left[\int\overline{V}^{(n)}\left(s^{\prime}\right)p\left(s^{\prime}|\widehat{s},a^{(n)}\left(\widehat{s}\right)\right)ds^{\prime}\right]}{\partial a^{(n)}\left(\widehat{s}\right)}
\]

\item Compute $V^{(n+1*)}\left(\widehat{s}\right)$ by:
\begin{eqnarray*}
V^{(n+1*)}\left(\widehat{s}\right) & = & \Phi_{V}\left(V^{(n)},a^{(n)}\right)\left(\widehat{s}\right)\\
 & \equiv & r\left(a^{(n)}\left(\widehat{s}\right),\widehat{s}\right)+\beta\int\overline{V}^{(n)}\left(s^{\prime}\right)p\left(s^{\prime}|\widehat{s},a^{(n)}\left(\widehat{s}\right)\right)ds^{\prime}
\end{eqnarray*}
 for $\widehat{s}\in\widehat{\mathcal{S}}$
\item Update $V$ and $a$ (spectral algorithm): 

For $z\in\left\{ V,a^{d=1,\cdots,D}\right\} ,$$z^{(n+1)}\left(\widehat{s}\right)\leftarrow\alpha_{z}^{(n)}z^{(n+1*)}\left(\widehat{s}\right)+(1-\alpha_{z}^{(n)})z^{(n)}\left(\widehat{s}\right)$, where $\alpha_{z}^{(n\geq1)}\equiv\frac{\left\Vert z^{(n)}-z^{(n-1)}\right\Vert _{2}}{\left\Vert F_{z}\left(V^{(n)},a^{(n)}\right)-F_{z}\left(V^{(n-1)},a^{(n-1)}\right)\right\Vert _{2}}$, $F_{z}(V,a)\equiv\Phi_{z}\left(V,a\right)-z.$
\item Exit the iteration if $\left\Vert V^{(n+1)}-V^{(n)}\right\Vert <\epsilon_{V}$ \textsl{and $\left\Vert a^{(n+1)}-a^{(n)}\right\Vert <\epsilon_{a}$}

\medskip{}

\end{enumerate}
\caption{VF-PGI-Spectral algorithm (Single-agent dynamic model)\label{alg:VF-PGI_single_agent}}
\end{enumerate}
\end{algorithm}

Regarding the VF-PGI-Spectral, we can intuitively explain the steps and treat the iterations as a learning process. We first set the initial values of value function $V$ and actions $a$. If the gradient of ``long-run profit'' $Q$ with respect to the action given the initial value functions and the action in a state is positive, we increase the value of the action, expecting higher long-term profit. If negative, we reduce the value of the action. When the gradient is close to zero, the action is likely to be optimal given value functions, and we rarely change the value of the action. As for value function $V$, we update the values so that they are consistent with Bellman equations given initial value functions and actions. Note that the initial actions in this step might not be optimal. We repeat these steps until the initial and updated values are sufficiently close.

In the algorithm, we introduce the idea of the spectral algorithm, which is motivated by the Newton's method and is effective at accelerating the convergence of iterations for solving continuous optimization problems and nonlinear equations. For details of the spectral algorithm, see Appendix \ref{subsec:Spectral-algorithm}. As Algorithm \ref{alg:VF-PGI_single_agent} shows, the values of $V,a$ are updated by $z^{(n+1)}\left(\widehat{s}\right)\leftarrow\alpha_{z}^{(n)}z^{(n*)}\left(\widehat{s}\right)+(1-\alpha_{z}^{(n)})z^{(n)}\left(\widehat{s}\right)\ z\in\left\{ V,a^{d=1,\cdots,D}\right\} $, where $\alpha_{z}^{(n)}\equiv\frac{\left\Vert z^{(n)}-z^{(n-1)}\right\Vert _{2}}{\left\Vert F_{z}\left(V^{(n)},a^{(n)}\right)-F_{z}\left(V^{(n-1)},a^{(n-1)}\right)\right\Vert _{2}}$. Here, $\alpha_{z}^{(n)}$ denote step sizes. If we do not introduce the spectral algorithm, $\alpha_{z}^{(n)}$ should be 1. The values of step sizes $\alpha_{z}^{(n)}$ work as learning rates, and the spectral algorithm accelerates the learning by automatically choosing the values of step sizes according to a predetermined formula. For details of the spectral algorithm, see Appendix \ref{subsec:Spectral-algorithm}. Note that without using the spectral algorithm, the VF-PGI-Spectral algorithm works poorly due to slow convergence, and introducing the spectral algorithm is essential. 

Note that there is no guarantee that VF-PGI-Spectral is a contraction, unlike the VFI under regularity conditions.\footnote{We can show the local convergence of the algorithm by introducing a line search procedure in the algorithm. For details, see Appendix \ref{sec:Further-discussions-on-spectral}.} However, in the numerical experiments, VF-PGI-Spectral converges for appropriate $\lambda$.

\paragraph*{Choice of $\lambda$ in VF-PGI-Spectral}

As shown in the numerical results in Appendix \ref{subsec:Further-results-VF-PGI-Spectral}, the convergence speed of the VF-PGI-Spectral is not so sensitive to the value of $\lambda$ as long as $\lambda$ is relatively small. We can understand the reason analytically.

For $n\geq1$, the following holds:

\begin{eqnarray*}
a^{(n+1)} & = & \alpha_{a}^{(n)}\Phi_{a}\left(V^{(n)},a^{(n)}\right)+\left(1-\alpha_{a}^{(n)}\right)a^{(n)}\\
 & = & a^{(n)}+\alpha^{(n)}F_{a}\left(V^{(n)},a^{(n)}\right)\\
 & = & a^{(n)}+\frac{\left\Vert a^{(n)}-a^{(n-1)}\right\Vert _{2}}{\left\Vert F_{a}\left(V^{(n)},a^{(n)}\right)-F_{a}\left(V^{(n-1)},a^{(n-1)}\right)\right\Vert _{2}}F_{a}\left(V^{(n)},a^{(n)}\right)\\
 & = & a^{(n)}+\frac{\left\Vert a^{(n)}-a^{(n-1)}\right\Vert _{2}}{\left\Vert \lambda\frac{\partial Q(a^{(n)};V^{(n)})}{\partial a^{(n)}}-\lambda\frac{\partial Q(a^{(n-1)};V^{(n-1)})}{\partial a^{(n-1)}}\right\Vert _{2}}\lambda\frac{\partial Q(a^{(n)};V^{(n)})}{\partial a^{(n)}}\\
 & = & a^{(n)}+\frac{\left\Vert a^{(n)}-a^{(n-1)}\right\Vert _{2}}{\left\Vert \frac{\partial Q(a^{(n)};V^{(n)})}{\partial a^{(n)}}-\frac{\partial Q(a^{(n-1)};V^{(n-1)})}{\partial a^{(n-1)}}\right\Vert _{2}}\frac{\partial Q(a^{(n)};V^{(n)})}{\partial a^{(n)}}
\end{eqnarray*}

The equation implies that updated $a^{(n+1)}$ does not directly depend on $\lambda$, implying that the convergence of VF-PGI-Spectral is not largely affected by the choice of $\lambda$. Note that $a^{(1)}$ is computed by $a^{(1)}=\alpha_{a}^{(0)}\Phi_{a}\left(V^{(0)},a^{(0)}\right)+\left(1-\alpha_{a}^{(0)}\right)a^{(0)}=a^{(0)}+\alpha_{a}^{(0)}\lambda\frac{\partial Q(a^{(0)};V^{(0)})}{\partial a^{(0)}}$, where $\alpha_{a}^{(0)}$ is a pre-specified tuning parameter. If we choose too large $\alpha_{a}^{(0)}$ or too large $\lambda$, $a^{(1)}$ can be too far from $a^{(0)}$. Because $a^{(0)}$ is typically chosen so that $a^{(0)}$ is close to the solution, $a^{(1)}$ can be too far from the solution, and it might lead to the divergence of the iteration. Hence, choosing sufficiently small $\alpha^{(0)}$ and $\lambda$ is necessary for the VF-PGI-Spectral to work well.

\paragraph*{Comparison of computational costs}

Generally, when applying value function-based methods, costly part for solving the dynamic optimization problem is the evaluation of $\int\overline{V}^{(n)}\left(s^{\prime}\right)p\left(s^{\prime}|\widehat{s},a^{(n)}\left(\widehat{s}\right)\right)ds^{\prime}$ (corresponding to the number of evaluating $V$) and $\frac{\partial\left[\int\overline{V}^{(n)}\left(s^{\prime}\right)p\left(s^{\prime}|\widehat{s},a^{(n)}\left(\widehat{s}\right)\right)ds^{\prime}\right]}{\partial a^{(n)}\left(\widehat{s}\right)}$ (corresponding to the number of evaluating $\frac{\partial Q}{\partial a}\left(\widehat{s},a;V\right)$). Hence, we compare the numbers of the evaluations of these terms.

In the VF-PGI-Spectral algorithm, we need to evaluate $\frac{\partial\left[\int\overline{V}^{(n)}\left(s^{\prime}\right)p\left(s^{\prime}|\widehat{s},a^{(n)}\left(\widehat{s}\right)\right)ds^{\prime}\right]}{\partial a^{(n)}\left(\widehat{s}\right)}$ once and $\int\overline{V}^{(n)}\left(s^{\prime}\right)p\left(s^{\prime}|\widehat{s},a^{(n)}\left(\widehat{s}\right)\right)ds^{\prime}$ once in each iteration. In contrast, in the VFI algorithm, when we use gradient-based optimization algorithms to solve maximization problems, we need to evaluate $\frac{\partial\left[\int\overline{V}^{(n)}\left(\text{\ensuremath{s^{\prime}}}\right)p\left(s^{\prime}|\widehat{s},a^{(n*)}\left(\widehat{s}\right)\right)ds^{\prime}\right]}{\partial a^{(n*)}\left(\widehat{s}\right)}$ many times in the gradient-based optimization process, and $\int\overline{V}^{(n)}\left(s^{\prime}\right)p\left(s^{\prime}|\widehat{s},a^{(n*)}\left(\widehat{s}\right)\right)ds^{\prime}$ once in each iteration. Hence, the computational cost per iteration is smaller in the case of the VF-PGI-Spectral.

Concerning the PI, computational cost per iteration is much larger, because in each iteration we need to evaluate $\frac{\partial\left[\int\overline{V}^{(n)}\left(\text{\ensuremath{s^{\prime}}}\right)p\left(s^{\prime}|\widehat{s},a^{(n*)}\left(\widehat{s}\right)\right)ds^{\prime}\right]}{\partial a^{(n*)}\left(\widehat{s}\right)}$ many times in the gradient-based optimization process, and $\int\overline{V}^{(n)}\left(s^{\prime}\right)p\left(s^{\prime}|\widehat{s},a^{(n*)}\left(\widehat{s}\right)\right)ds^{\prime}$ many times to iteratively solve the equation. In contrast, the number of iterations would be generally small, as theoretically guaranteed.

\subsubsection{Relative Value function\label{subsec:Relative-VFI-single-agent}}

So far, we have considered algorithms solving for $V\equiv\left(V(\widehat{s})\right)_{\widehat{s}\in\widehat{\mathcal{S}}}$ and $a\equiv\left(a(\widehat{s})\right)_{\widehat{s}\in\widehat{\mathcal{S}}}$ satisfying $a^{*}(\widehat{s})=\arg\max_{a(\widehat{s})}r\left(a\left(\widehat{s}\right),\widehat{s}\right)+\beta\int\overline{V}\left(s^{\prime}\right)p\left(s^{\prime}|\widehat{s},a\left(\widehat{s}\right)\right)ds^{\prime}$ and $V(\widehat{s})=r(a^{*}(\widehat{s}),\widehat{s})+\beta\int\overline{V}\left(s^{\prime}\right)p\left(s^{\prime}|\widehat{s},a^{*}(\widehat{s})\right)ds^{\prime}$. Here, let $T_{a}$ be an operator such that $T_{a}V(\widehat{s})=r(a(\widehat{s}),\widehat{s})+\beta\int\overline{V}\left(s^{\prime}\right)p\left(s^{\prime}|\widehat{s},a(\widehat{s})\right)ds^{\prime}$, and let $T\equiv T_{a^{*}}$. In addition, let $\Delta$ be a operator such that $\Delta V(\widehat{s})\equiv V(\widehat{s})-V(\widehat{s_{0}})$, where $\widehat{s_{0}}\in\widehat{\mathcal{S}}$ denotes an arbitrary state in $\widehat{\mathcal{S}}$. \citet{morton1977discounting} and \citet{bray2019strong} proposed to iterate the following until convergence: 
\begin{enumerate}
\item Compute $a^{*(n+1)}(\widehat{s})=\arg\max_{a(\widehat{s})}Q\left(a(\widehat{s}),\widehat{s};\widetilde{V}^{(n)}\right)$
\item Update $\widetilde{V}$ by 
\begin{eqnarray*}
\widetilde{V}^{(n+1)}(\widehat{s}) & = & \Delta T_{a^{*(n+1)}}\widetilde{V}^{(n)}(\widehat{s})\\
 & = & Q\left(a^{*(n+1)}(\widehat{s}),\widehat{s};\widetilde{V}^{(n)}\right)-Q\left(a^{*(n+1)}(\widehat{s_{0}}),\widehat{s_{0}};\widetilde{V}^{(n)}\right)
\end{eqnarray*}
\end{enumerate}
The method is known as Relative Value Function Iteration (we term the method as RVFI). Note that $\widetilde{V}\equiv\left(V(\widehat{s})-C\right)_{s\in\widehat{\mathcal{S}}}$, where $C=\frac{1}{1-\beta}\left[T\widetilde{V}\left(\widehat{s_{0}}\right)\right]$, holds.

We can easily see why it is sufficient to solve for $a^{*}$ and $\widetilde{V}$ by the iteration specified above. Because
\begin{eqnarray*}
a^{*}(s) & = & \arg\max_{a(\widehat{s})}\left[r\left(a\left(\widehat{s}\right),\widehat{s}\right)+\beta\int\overline{V}\left(s^{\prime}\right)p\left(s^{\prime}|\widehat{s},a\left(\widehat{s}\right)\right)ds^{\prime}\right]\\
 & = & \arg\max_{a(\widehat{s})}\left[r\left(a\left(\widehat{s}\right),\widehat{s}\right)+\beta\int\overline{V}\left(s^{\prime}\right)p\left(s^{\prime}|\widehat{s},a\left(\widehat{s}\right)\right)ds^{\prime}-\beta C\right]\\
 & = & \arg\max_{a(\widehat{s})}\left[r\left(a\left(\widehat{s}\right),\widehat{s}\right)+\beta\int\left(\overline{V}\left(s^{\prime}\right)-C\right)p\left(s^{\prime}|\widehat{s},a\left(\widehat{s}\right)\right)ds^{\prime}\right]\ \left(\because\int p\left(s^{\prime}|\widehat{s},a\left(\widehat{s}\right)\right)ds^{\prime}=1\right)
\end{eqnarray*}
 and $\left(V\left(\widehat{s}\right)-C\right)=r\left(a^{*}\left(\widehat{s}\right),\widehat{s}\right)+\beta\int\left(\overline{V}\left(s^{\prime}\right)-C\right)p\left(s^{\prime}|\widehat{s},a^{*}\left(\widehat{s}\right)\right)ds^{\prime}-(1-\beta)C$ holds for any $C\in\mathbb{R}$, we can think of solving for $\widetilde{V}\equiv\left(V(\widehat{s})-C\right)_{s\in\widehat{\mathcal{S}}}$ and $a^{*}\equiv\left(a^{*}(\widehat{s})\right)_{s\in\widehat{\mathcal{S}}}$ satisfying $C=\frac{1}{1-\beta}\left[r\left(a^{*}\left(\widehat{s_{0}}\right),\widehat{s_{0}}\right)+\beta\int\overline{\widetilde{V}}\left(s^{\prime}\right)p\left(s^{\prime}|\widehat{s_{0}},a^{*}\left(\widehat{s_{0}}\right)\right)ds^{\prime}\right]=\frac{1}{1-\beta}\left[T_{a^{*}}\widetilde{V}\left(\widehat{s_{0}}\right)\right]$ and the following equations:

\begin{eqnarray*}
a^{*}(s) & = & \arg\max_{a(s)}r\left(a^{*}\left(\widehat{s}\right),\widehat{s}\right)+\beta\int\overline{\widetilde{V}}\left(s^{\prime}\right)p\left(s^{\prime}|\widehat{s},a^{*}\left(\widehat{s}\right)\right)ds^{\prime}\\
\widetilde{V}\left(\widehat{s}\right) & = & \left[r\left(a^{*}\left(\widehat{s}\right),\widehat{s}\right)+\beta\int\overline{\widetilde{V}}\left(s^{\prime}\right)p\left(s^{\prime}|\widehat{s},a^{*}\left(\widehat{s}\right)\right)ds^{\prime}\right]\\
 &  & -\left[r\left(a^{*}\left(\widehat{s_{0}}\right),\widehat{s_{0}}\right)+\beta\int\overline{\widetilde{V}}\left(s^{\prime}\right)p\left(s^{\prime}|\widehat{s_{0}},a^{*}\left(\widehat{s_{0}}\right)\right)ds^{\prime}\right]\\
 & = & T_{a^{*}}\widetilde{V}\left(\widehat{s}\right)-T_{a^{*}}\widetilde{V}\left(\widehat{s_{0}}\right)\\
 & = & \Delta T_{a^{*}}\widetilde{V}\left(\widehat{s}\right)
\end{eqnarray*}

As discussed in \citet{morton1977discounting} and \citet{bray2019strong}, the RVFI converges faster than the VFI when the underlying stochastic process is ergodic.\footnote{More specifically, \citet{bray2019strong} showed in a finite-state finite-action setting that $\left\Vert T^{n}V-V^{*}\right\Vert $ is $O(\beta^{n})$ for VFI, and $\left\Vert \left(\Delta T_{a^{*}}\right)^{n}\widetilde{V}-\widetilde{V}^{*}\right\Vert $is $O(\beta^{n}\gamma^{n})$ for all $\gamma>\sigma\left(Q(a^{*})\right)$ for RVFI, where $\sigma(\cdot)$ denotes the second largest eigenvalue modulus, and $Q(a^{*})$ denotes the state transition probability matrix given the optimal policy function $a^{*}$. If the Markov chain is ergodic, $\sigma\left(Q(a^{*})\right)<1$ holds, and RVFI is faster than VFI.\\
Note that the performance of RVFI largely relies on the time persistence of the state variables characterized by $\sigma\left(Q(a^{*})\right)$, as discussed in \citet{aguirregabiria2023solving}. If the states are not so persistent over time, RVFI converges much faster than VFI. In contrast, if the states are time persistent, the convergence speed of RVFI is mostly the same as VFI, and the benefit of using RVFI is small. } Because the RVFI is a slight modification of the VFI algorithm, it is straightforward to implement.

As discussed in \citet{bray2019strong}, we can also introduce the idea of relative value function to the PI.\footnote{As discussed in Section 4.4 of \citet{bray2019strong}, the condition number of the linear equation to be solved in the policy evaluation step of the PI takes a large value when the discount factor is close to 1. In contrast, that of the RPI remains well-conditioned even when the discount factor is close to 1. Generally, linear equations with large condition numbers are harder to solve precisely due to numerical precision problems. In addition, as shown in Section 6.11.4 of \citet{saad2003iterative}, GMRES iterations converge faster when the condition number of the matrix of the linear equation to be solved is smaller, at least when the matrix is diagonalizable. Hence, we can expect that RPI requires fewer number of iterations in the policy evaluation step than the PI, even when we introduce the Krylov-based methods to solve linear equations.} The discussions above imply that we can obtain the solution $a^{*}$ by the following iteration:
\begin{enumerate}
\item Compute $a^{*(n+1)}(\widehat{s})=\arg\max_{a(\widehat{s})}Q\left(a(\widehat{s}),\widehat{s};\widetilde{V}^{(n)}\right)$
\item Update $\widetilde{V}$ by solving the following equation:
\begin{eqnarray*}
\widetilde{V}^{(n+1)}(\widehat{s}) & = & T_{a^{*(n+1)}}\widetilde{V}^{(n+1)}(\widehat{s})-T_{a^{*(n+1)}}\widetilde{V}^{(n+1)}(\widehat{s_{0}})\\
 & = & \beta\left[\int\overline{\widetilde{V}^{(n+1)}}\left(s^{\prime}\right)p\left(s^{\prime}|\widehat{s},a^{*}\left(\widehat{s}\right)\right)ds^{\prime}-\int\overline{\widetilde{V}^{(n+1)}}\left(s^{\prime}\right)p\left(s^{\prime}|\widehat{s_{0}},a^{*}\left(\widehat{s_{0}}\right)\right)ds^{\prime}\right]+\\
 &  & \left[r\left(a^{*(n+1)}\left(\widehat{s}\right),\widehat{s}\right)-r\left(a^{*(n+1)}\left(\widehat{s_{0}}\right),\widehat{s_{0}}\right)\right]
\end{eqnarray*}
\end{enumerate}
The derivation of the equations above also implies that we can also introduce the idea of the relative value function to other value function-based algorithms, such as the VF-PGI-Spectral.

\subsubsection{Acceleration methods of fixed-point iterations}

The algorithms discussed so far (except for VF-PGI-Spectral) can be regarded as fixed-point iterations. Hence, we can introduce acceleration methods of fixed-point iterations (e.g., Spectral, SQUAREM, Anderson acceleration).\footnote{For details of the SQUAREM and Anderson acceleration method, see the description in \citet{fukasawa2024fast}.}

Even though the acceleration methods would accelerate the convergence of fixed-point iterations, they generally still inherit the properties of the original fixed-point mappings. Hence, intuitively, we can expect that using fixed-point mappings with better convergence properties leads to faster convergence even when introducing acceleration methods. Hence, the effort to find appropriate fixed-point mappings with good convergence properties, such as the ones using relative value functions, is important to reduce computational costs.

\section{Multi-agent dynamic games\label{sec:Multi-agent-dynamic-games}}

In this section, we consider a multi-agent dynamic game, in which multiple agents interact over time. Section \ref{subsec:Model-dynamic-game} describes the model, and Section \ref{subsec:Algorithms-dynamic-game} discusses the algorithms for solving the model, including VFI{*}, PI{*}, and VF-PGI-Spectral. Note that it seems that it is not straightforward to directly apply some methods designed to solve single-agent models, including EE, ECM, EGM, to dynamic games. For details, see Appendix \ref{subsec:ECM,-EE,-EGM-dynamic-game}.

\subsection{Model\label{subsec:Model-dynamic-game}}

Let $\pi_{j}\left(a_{t+\tau},s_{t+\tau}\right)$ be agent $j$'s profit when they choose actions $a_{t+\tau}\equiv\left(a_{jt+\tau}^{*}\right)_{j\in\mathcal{J}}$ at state $s_{t+\tau}$. Here, $\mathcal{J}$ denotes the set of agents. Given a current state $s_{t}$, agent $j$'s expected future profit is $E\left[\sum_{\tau=0}^{\infty}\beta^{\tau}\pi_{j}\left(a_{t+\tau},s_{t+\tau}\right)|s_{t}\right]$. Actions can be multidimensional, and let $a_{j}\equiv\left(a_{j}^{d}\right)_{d=1,\cdots,D}$.

Here, we focus on pure strategy Markov perfect equilibria. We assume the existence of such an equilibrium, and let $a_{j}^{*}:\mathcal{S}\rightarrow\mathcal{A}_{j}$ be the Markov strategy of agent $j$. $a^{*}\equiv\left(a_{j}^{*}\right)_{j\in\mathcal{J}}$ denotes the profile of the vector, where $a^{*}:\mathcal{S}\rightarrow\mathcal{A}$.

Then, the value function of firm $j$ given a state $s$ can be written recursively:

\begin{eqnarray*}
V_{j}(s) & = & \max_{a_{j}(s)\in\mathcal{A}_{j}(s)}Q_{j}\left(a_{j}(s),a_{-j}^{*}(s),s;V_{j}\right)\\
 & \equiv & r_{j}(a_{j}^{*}(s),a_{-j}^{*}(s),s)+\beta\int V_{j}\left(s^{\prime}\right)p(s^{\prime}|s,a_{j}^{*}(s),a_{-j}^{*}(s))ds^{\prime},
\end{eqnarray*}
where $a_{j}(s)$ is the action of agent $j$ at state $s$. $Q_{j}\left(a_{j}(s),a_{-j}(s),s;V_{j}\right)$ is the action value function of firm $j$, and $V_{j}(s)$ is the value function. $p\left(s^{\prime}|s,a(s)\right)$ is the state transition probability. Note that $Q_{j}\left(a_{j}^{*}(s),a_{-j}^{*}(s),s;V_{j}\right)=V_{j}(s)$ holds.

If the action $a_{j}(s)$ is a continuous variable and $Q_{j}$ is differentiable with respect to $a_{j}(s)$, the following equations hold:

\begin{eqnarray}
0 & = & \frac{\partial Q_{j}(a_{j}^{*}(s),a_{-j}^{*}(s),s;V_{j})}{\partial a_{j}^{*}(s)}\nonumber \\
 & = & \frac{\partial r_{j}\left(a_{j}^{*}(s),a_{-j}^{*}(s),s\right)}{\partial a_{j}^{*}(s)}+\beta\frac{\partial\left[\int V_{j}\left(s^{\prime}\right)p(s^{\prime}|s,a_{j}^{*}(s),a_{-j}^{*}(s))ds^{\prime}\right]}{\partial a_{j}^{*}(s)},\label{eq:FOC-dynamic_game}\\
V_{j}(s) & = & r_{j}(a^{*}(s),s)+\beta\int V_{j}\left(s^{\prime}\right)p\left(s^{\prime}|s,a^{*}(s)\right)ds^{\prime}.\label{eq:Bellman-dynamic_game}
\end{eqnarray}

Then, the problem reduces to solving nonlinear equations (\ref{eq:FOC-dynamic_game}) and (\ref{eq:Bellman-dynamic_game}) concerning $\left\{ a_{j}^{*}(s)\right\} _{j\in\mathcal{J},s\in\mathcal{S}}$ and $\left\{ V_{j}(s)\right\} _{j\in\mathcal{J},s\in\mathcal{S}}$.

In general, the state space of $s$ is not finite. In this case, we should take suitable finite sets of grid points $\widehat{s}\in\widehat{\mathcal{S}}\subset\mathcal{S}$ to well-approximate the solution. Then, we should alternatively solve the following system of equations:

\begin{eqnarray}
0 & = & \frac{\partial Q_{j}\left(a_{j}^{*}\left(\widehat{s}\right),a_{-j}^{*}\left(\widehat{s}\right),\widehat{s};V_{j}\right)}{\partial a_{j}^{*}\left(\widehat{s}\right)}\nonumber \\
 &  & \frac{\partial r_{j}\left(a_{j}^{*}\left(\widehat{s}\right),a_{-j}^{*}\left(\widehat{s}\right),s\right)}{\partial a_{j}^{*}\left(\widehat{s}\right)}+\beta\frac{\partial\left[\int\overline{V_{j}}\left(s^{\prime}\right)p\left(s^{\prime}|\widehat{s},a_{j}^{*}\left(\widehat{s}\right),a_{-j}^{*}\left(\widehat{s}\right)\right)ds^{\prime}\right]}{\partial a_{j}^{*}\left(\widehat{s}\right)},\label{eq:FOC_dynamic_game_grid}\\
V_{j}\left(\widehat{s}\right) & = & r_{j}\left(a^{*}\left(\widehat{s}\right),\widehat{s}\right)+\beta\int\overline{V_{j}}\left(s^{\prime}\right)p\left(s^{\prime}|\widehat{s},a^{*}\left(\widehat{s}\right)\right)ds^{\prime}.\label{eq:Bellman_dynamic_game_grid}
\end{eqnarray}

\subsection{Algorithms\label{subsec:Algorithms-dynamic-game}}

\subsubsection{VFI{*}}

In previous studies, an algorithm, termed VFI{*} algorithm, because it is a slight extension of the single-agent VFI algorithm, has been applied.\footnote{Corresponding algorithms have been applied in \citet{Pakes_McGuire1994}, \citet{miranda2004applied}, and subsequent studies. } I show the algorithm in Algorithm \ref{alg:VFI_dynamic_game}. 

\begin{algorithm}[H]
\begin{enumerate}
\item Take grid points $\widehat{s}\in\widehat{\mathcal{S}}$. Set initial values $V^{(0)}\equiv\left\{ V_{j}^{(0)}\left(\widehat{s}\right)\right\} _{j\in\mathcal{J},\widehat{s}\in\widehat{\mathcal{S}}}$ and $a^{(0)}\equiv\left\{ a_{j}^{(0)}\left(\widehat{s}\right)\right\} _{j\in\mathcal{J},\widehat{s}\in\widehat{\mathcal{S}}}$
\item Iterate the following $(n=0,1,\cdots)$:
\begin{enumerate}
\item Solve for $a_{j}^{(n+1)}\left(\widehat{s}\right)$ satisfying the following nonlinear equation for each $j\in\mathcal{J},\widehat{s}\in\widehat{\mathcal{S}}$:

\begin{eqnarray*}
0 & = & \frac{\partial Q(a_{j}^{(n+1)}\left(\widehat{s}\right),a_{-j}^{(n)}\left(\widehat{s}\right),\widehat{s};V_{j}^{(n)})}{\partial a_{j}^{(n+1)}\left(\widehat{s}\right)}\\
 & = & \frac{\partial r_{j}\left(a_{j}^{(n+1)}\left(\widehat{s}\right),a_{-j}^{(n)}\left(\widehat{s}\right),\widehat{s}\right)}{\partial a_{j}^{(n+1)}\left(\widehat{s}\right)}+\beta\frac{\partial\left[\int\overline{V_{j}}^{(n)}\left(s^{\prime}\right)p\left(s^{\prime}|\widehat{s},a_{j}^{(n+1)}\left(\widehat{s}\right),a_{-j}^{(n)}\left(\widehat{s}\right)\right)ds^{\prime}\right]}{\partial a_{j}^{(n+1)}\left(\widehat{s}\right)}
\end{eqnarray*}

\item Compute 
\begin{eqnarray*}
V_{j}^{(n+1)}\left(\widehat{s}\right) & = & r_{j}\left(a^{(n+1)}\left(\widehat{s}\right),\widehat{s}\right)+\beta\int\overline{V_{j}}^{(n)}\left(s^{\prime}\right)p\left(s^{\prime}|\widehat{s},a^{(n+1)}\left(\widehat{s}\right)\right)ds^{\prime}
\end{eqnarray*}
 for $j\in\mathcal{J},\widehat{s}\in\widehat{\mathcal{S}}$
\item Exit the iteration if $\left\Vert V^{(n+1)}-V^{(n)}\right\Vert <\epsilon_{V}$ \textsl{and $\left\Vert a^{(n+1)}-a^{(n)}\right\Vert <\epsilon_{a}$}
\end{enumerate}
\end{enumerate}
\caption{VFI{*} (Multi-agent dynamic game)\label{alg:VFI_dynamic_game}}
\end{algorithm}

Note that there is no guarantee that the VFI{*} is a contraction in the multi-agent setting. Furthermore, given the existence of multiple equilibria, there is no guarantee that we can find all equilibria even when starting from several initial values. 

\subsubsection{PI{*}}

We can think of the counterpart of the PI algorithm in the single-agent models. We term it as PI{*}. Algorithm \ref{alg:PI_dynamic_game} shows the steps.

\begin{algorithm}[H]
\begin{enumerate}
\item Take grid points $\widehat{s}\in\widehat{\mathcal{S}}$. Set initial values $a^{(0)}\equiv\left\{ a_{j}^{(0)}\left(\widehat{s}\right)\right\} _{j\in\mathcal{J},\widehat{s}\in\widehat{\mathcal{S}}}$ and $V^{(0)}\equiv\left\{ V_{j}^{(0)}\left(\widehat{s}\right)\right\} _{j\in\mathcal{J},\widehat{s}\in\widehat{\mathcal{S}}}$.
\item Iterate the following $(n=0,1,\cdots)$:
\begin{enumerate}
\item Policy Improvement Step:

Solve for $a_{j}^{(n+1)}\left(\widehat{s}\right)$ satisfying the following nonlinear equation for each $\widehat{s}\in\widehat{\mathcal{S}}$:

\begin{eqnarray*}
0 & = & \frac{\partial Q_{j}\left(a^{(n+1)}\left(\widehat{s}\right),\widehat{s};V_{j}^{(n*)}\right)}{\partial a_{j}^{(n+1)}\left(\widehat{s}\right)}\\
 & = & \frac{\partial r_{j}\left(a^{(n+1)}\left(\widehat{s}\right),\widehat{s}\right)}{\partial a_{j}^{(n+1)}\left(\widehat{s}\right)}+\beta\frac{\partial\left[\int\overline{V_{j}}^{(n+1)}\left(\text{\ensuremath{s^{\prime}}}\right)p\left(s^{\prime}|\widehat{s},a^{(n*)}\left(\widehat{s}\right)\right)ds^{\prime}\right]}{\partial a_{j}^{(n+1)}\left(\widehat{s}\right)}
\end{eqnarray*}

\item Policy Evaluation Step: For each $j\in\mathcal{J}$, solve $V_{j}^{(n+1)}\left(\widehat{s}\right)=r_{j}\left(a^{(n+1)}\left(\widehat{s}\right),\widehat{s}\right)+\beta\int\overline{V}_{j}^{(n+1)}\left(s^{\prime}\right)p\left(s^{\prime}|\widehat{s},a^{(n+1)}\left(\widehat{s}\right)\right)ds^{\prime}$ for $V_{j}^{(n+1)}$:
\end{enumerate}
\begin{itemize}
\item PI: Iterate applying the mapping $\Phi_{V,j}\left(V,a\right)\left(\widehat{s}\right)\equiv r_{j}\left(a^{(n+1)}\left(\widehat{s}\right),\widehat{s}\right)+\beta\int\overline{V_{j}}\left(s^{\prime}\right)p\left(s^{\prime}|\widehat{s},a^{(n+1)}\left(\widehat{s}\right)\right)ds^{\prime}$ to $V^{(n)}$ given $a^{(n+1)}$ until convergence
\item PI-Krylov: Solve the linear equation by Krylov-based methods 
\end{itemize}
\begin{enumerate}
\item Exit the iteration if $\left\Vert V^{(n+1)}-V^{(n)}\right\Vert <\epsilon_{V}$ \textsl{and $\left\Vert a^{(n+1)}-a^{(n)}\right\Vert <\epsilon_{a}$}
\end{enumerate}
\medskip{}

\caption{PI{*} (Multi-agent dynamic game)\label{alg:PI_dynamic_game}}
\end{enumerate}
\end{algorithm}

Note that there is no guarantee that the PI is convergent. In addition, the superlinear or quadratic convergence may not hold, unlike the single-agent setting. 

\subsubsection{VF-PGI-Spectral}

Algorithm \ref{alg:VF-PGI_dynamic_game} shows the proposed VF-PGI-Spectral algorithm for the dynamic game.

\begin{algorithm}[H]
\begin{enumerate}
\item Take grid points $\widehat{s}\in\widehat{\mathcal{S}}$. Set initial values $V^{(0)}\equiv\left\{ V_{j}^{(0)}\left(\widehat{s}\right)\right\} _{j\in\mathcal{J},\widehat{s}\in\widehat{\mathcal{S}}}$ and $a^{(0)}\equiv\left\{ a_{j}^{(0)}\left(\widehat{s}\right)\right\} _{j\in\mathcal{J},\widehat{s}\in\widehat{\mathcal{S}}}$. Set tuning parameters $\alpha^{(0)}$ and $\lambda$.
\item Iterate the following $(n=0,1,\cdots)$:
\begin{enumerate}
\item Solve for $\left\{ a_{j}^{(n*)}\left(\widehat{s}\right)\right\} _{j\in\mathcal{J},\widehat{s}\in\widehat{\mathcal{S}}}$:

\[
a_{j}^{(n*)}\left(\widehat{s}\right)=\Phi_{a,j}\left(V^{(n)},a^{(n)}\right)\left(\widehat{s}\right)\equiv a_{j}^{(n)}\left(\widehat{s}\right)+\lambda\cdot\frac{\partial Q_{j}\left(a_{j}^{(n)}\left(\widehat{s}\right),a_{-j}^{(n)}\left(\widehat{s}\right),\widehat{s};V_{j}^{(n)}\right)}{\partial a_{j}^{(n)}\left(\widehat{s}\right)}
\]
where

\begin{eqnarray*}
 &  & \frac{\partial Q_{j}(a_{j}^{(n)}\left(\widehat{s}\right),a_{-j}^{(n)}\left(\widehat{s}\right),\widehat{s};V_{j}^{(n)})}{\partial a_{j}^{(n)}\left(\widehat{s}\right)}\\
 & \equiv & \frac{\partial r_{j}\left(a_{j}^{(n)}\left(\widehat{s}\right),a_{-j}^{(n)}\left(\widehat{s}\right),s\right)}{\partial a_{j}^{(n)}\left(\widehat{s}\right)}+\beta\frac{\partial\left[\int\overline{V_{j}}^{(n)}\left(s^{\prime}\right)p\left(s^{\prime}|\widehat{s},a_{j}^{(n)}\left(\widehat{s}\right),a_{-j}^{(n)}\left(\widehat{s}\right)\right)ds^{\prime}\right]}{\partial a_{j}^{(n)}\left(\widehat{s}\right)}
\end{eqnarray*}

\item Find 
\begin{eqnarray*}
V_{j}^{(n*)}\left(\widehat{s}\right) & \equiv & \Phi_{V,j}\left(V^{(n)},a^{(n)}\right)\left(\widehat{s}\right)\\
 & = & r_{j}\left(a^{(n)}\left(\widehat{s}\right),\widehat{s}\right)+\beta\int\overline{V_{j}}^{(n)}\left(s^{\prime}\right)p\left(s^{\prime}|\widehat{s},a^{(n)}\left(\widehat{s}\right)\right)ds^{\prime}
\end{eqnarray*}
 for $j\in\mathcal{J},\widehat{s}\in\widehat{\mathcal{S}}$
\item Update $a$ and $V$ (spectral algorithm):

For $z_{j}\in\left\{ V_{j},a_{j}^{d=1,\cdots,D}\right\} $, $z_{j}^{(n+1)}\left(\widehat{s}\right)\leftarrow\alpha_{z}^{(n)}z_{j}^{(n*)}\left(\widehat{s}\right)+(1-\alpha_{z}^{(n)})z_{j}^{(n)}\left(\widehat{s}\right)$, where

$\alpha_{z}^{(n\geq1)}\equiv\frac{\left\Vert z^{(n)}-z^{(n-1)}\right\Vert _{2}}{\left\Vert F_{z}\left(V^{(n)},a^{(n)}\right)-F_{z}\left(V^{(n-1)},a^{(n-1)}\right)\right\Vert _{2}}$, $F_{z}(V,a)\equiv\Phi_{z}\left(V,a\right)\left(\widehat{s}\right)-z$
\item Exit the iteration if $\left\Vert V^{(n+1)}-V^{(n)}\right\Vert <\epsilon_{V}$ \textsl{and $\left\Vert a^{(n+1)}-a^{(n)}\right\Vert <\epsilon_{a}$}
\end{enumerate}
\end{enumerate}
\caption{VF-PGI-Spectral (Multi-agent dynamic game)\label{alg:VF-PGI_dynamic_game}}
\end{algorithm}

Intuitively, in the VF-PGI-Spectral algorithm, we first set initial values of all agents' value functions $V$ and actions $a$. Then, if the gradient of agent $j$'s ``long-run profit'' $\frac{\partial Q}{\partial a_{j}}$ given their initial value functions and all agents' actions is positive, we increase the value of agent $j$'s action expecting agent $j$'s larger long-run profit. If it is negative, we decrease the value. When the gradient is close to zero, the action is mostly optimal given value functions and all agents' actions, and we rarely change the value of agent $j$'s action. 

Regarding value functions, we update the values of agent $j$'s value functions so that they align with agent $j$'s Bellman equations given their own initial value functions and all agents' actions. Note that the initial actions in this step might not be optimal. We repeat the procedure for all agents, and iterate these steps until the initial and updated values are sufficiently close.

As in the single-agent version, we need not solve nonlinear problems in each iteration, and it contributes to smaller computational burden.

\paragraph*{Comparison of computational costs}

In the VF-PGI-Spectral, we need to evaluate $\frac{\partial\left[\int\overline{V}_{j}^{(n)}\left(s^{\prime}\right)p\left(s^{\prime}|\widehat{s},a^{(n)}\left(\widehat{s}\right)\right)ds^{\prime}\right]}{\partial a_{j}^{(n)}\left(\widehat{s}\right)}$ once and $\int\overline{V}_{j}^{(n)}\left(s^{\prime}\right)p\left(s^{\prime}|\widehat{s},a^{(n)}\left(\widehat{s}\right)\right)ds^{\prime}$ once in each iteration, as in the single-agent model. In contrast, in the VFI, when we use gradient-based optimization algorithms to solve maximization problems, we need to evaluate $\frac{\partial\left[\int\overline{V}_{j}^{(n)}\left(\text{\ensuremath{s^{\prime}}}\right)p\left(s^{\prime}|\widehat{s},a^{(n*)}\left(\widehat{s}\right)\right)ds^{\prime}\right]}{\partial a_{j}^{(n*)}\left(\widehat{s}\right)}$ many times in the gradient-based optimization process, and $\int\overline{V}_{j}^{(n)}\left(s^{\prime}\right)p\left(s^{\prime}|\widehat{s},a^{(n*)}\left(\widehat{s}\right)\right)ds^{\prime}$ once in each iteration. 

In some specific settings, the advantage of VF-PGI-Spectral gets larger. First, in dynamic models with continuous states, we typically use the collocation method. the values of agent $j$'s value function $V_{j}(s)$ is usually approximated by $\overline{V_{j}}(s)\equiv\Psi(s)\theta_{j}$, where $\Psi(s)$ denotes the basis function as the function of states $s$, and $\theta_{j}$ denotes a vector. In VF-PGI-Spectral, we should compute $\overline{V}_{j}^{(n)}\left(s^{\prime}\right)$ $\left(s^{\prime}\sim p\left(s^{\prime}|\widehat{s},a^{(n)}\left(\widehat{s}\right)\right)\right)$ for all $j\in\mathcal{J}$, and it is sufficient to compute $\Psi(s^{\prime})$ once regardless of the number of agents. In contrast, in VFI{*} and PI{*}, we should compute $\overline{V}_{j}^{(n)}\left(s^{\prime}\right)$ $\left(s^{\prime}\sim p\left(s^{\prime}|\widehat{s},a_{j}\left(\widehat{s}\right),a_{-j}^{(n)}\left(\widehat{s}\right)\right)\right)$ for all candidate $a_{j}\left(\widehat{s}\right)$ and $j\in\mathcal{J}$ in the optimization process. The values of $\Psi(s^{\prime})$ are not the same for each firm, and we should compute $\Psi(s^{\prime})$ for each firm.

Second, in models where the state transition of $s$ is stochastic, VF-PGI-Spectral is computationally attractive. In VF-PGI-Spectral, $p\left(s^{\prime}|s,a^{(n)}(s)\right)$ is common for all firms, and we have to compute the term once. In contrast, in VFI{*} and PI{*}, we need to compute $p\left(s^{\prime}|\widehat{s},a_{j}\left(\widehat{s}\right),a_{-j}^{(n)}\left(\widehat{s}\right)\right)$ for all candidate $a_{j}\left(\widehat{s}\right)$ and $j\in\mathcal{J}$ in the optimization process, and we cannot reduce the number of computation of the term.

Note that VF-PGI-Spectral may not be a contraction. That said, neither VFI{*} nor PI{*} may not be a contraction. In addition, the convergence speed of the PI{*} is not necessarily clear, to my knowledge. Hence, the relative advantage of the VF-PGI-Spectral would be large in the multi-agent settings. 

\subsubsection{Relative value function}

The idea of the relative value function discussed in the single-agent setting also applies to dynamic games. In fact, \citet{bray2019strong} applied the idea to solve a discrete-time dynamic game with finite states and continuous actions considered in \citet{doraszelski2012avoiding}, and showed much superior performance of RVFI{*} compared to VFI{*}.

\section{Numerical experiments\label{sec:Numerical-experiments}}

This section uses numerical experiments to compare the performance of algorithms under two settings. Section \ref{subsec:Single-agent-neoclassical} shows results under the single-agent neoclassical growth model with elastic labor supply, which is extensively considered in the literature on macroeconomics. Section \ref{subsec:investment-competition-conti-states} considers a dynamic investment competition model with continuous states. All experiments were conducted on a laptop computer with the CPU AMD Ryzen 5 6600H 3.30 GHz, 16.0 GB of RAM, Windows 11 64-bit, and MATLAB 2022b. When applying the spectral algorithm, we use $\alpha_{0}=1$ in this section. For details of each experiment, see Appendix \ref{sec:Details-experiments}.

In this section, we compare not only value function-based algorithms, but also relative value function-based algorithms. In the growth model, we also compare endogenous value function-based algorithms. We distinguish algorithms by adding ``R'' to the names for relative value function-based algorithms, and adding ``E'' for endogenous value function-based algorithms. 

\subsection{Single-agent neoclassical growth model with elastic labor supply\label{subsec:Single-agent-neoclassical}}

\subsubsection*{Settings}

As in \citet{santos2004convergence} and \citet{maliar2013envelope}, we consider a standard single-agent neoclassical growth model with elastic labor supply. The dynamic optimization problem is characterized by:

\begin{eqnarray}
V(k,z) & = & \max_{l,c}u(c,l)+\beta E\left[V(k^{\prime},z^{\prime})\right]\label{eq:growth_model_Bellman}\\
 & s.t. & k^{\prime}=(1-\delta)k+zf(k,l)-c\nonumber \\
 &  & \ln z^{\prime}=\rho\ln z+\epsilon^{\prime},\ \epsilon^{\prime}\sim N(0,\sigma^{2})\nonumber \\
 &  & u(c,l)=\frac{c^{1-\gamma}-1}{1-\gamma}+B\frac{(1-l)^{1-\mu}-1}{1-\mu}\nonumber \\
 &  & f(k,l)=Ak^{\alpha}l^{1-\alpha},\nonumber 
\end{eqnarray}
where $k,c,l,z$ are the capital, consumption, labor, and productivity level. 

The current study compares 134 methods for solving the model, which differ concerning the introduction of relative value functions, acceleration methods of fixed-point iterations, and others. The results are shown in Appendix \ref{sec:Full-results-growth-model}. There are some methods that are not discussed in Section \ref{subsec:Algorithms-single-agent}, and please see Appendix \ref{sec:other_algorithms} for details of each method. Because there are many methods, the current study picks up some important methods, and compares their performance in Table \ref{tab:results_growth_elastic_labor-summary}. Note that the current study further compares additional PI-based methods that explicitly compute the matrix $A\equiv I_{|\widehat{\mathcal{S}}|}-\beta\widetilde{P}\widetilde{X}\left(X^{\prime}X\right)^{-1}X^{\prime}$, as shown in Appendix \ref{subsec:Additional-comparison-of-PI}.

The experiments are conducted by modifying and extending the replication code of \citet{maliar2013envelope}. As in \citet{maliar2013envelope}, we choose steady-state capital-output ratio $\pi_{k}=10$, steady-state consumption-output ratio $\pi_{c}=3/4$, steady-state labor $\overline{l}=1/3$, and parameters $\delta=\frac{1-\pi_{c}}{\pi_{k}}=0.025,\beta=\frac{1}{1-\delta+\frac{\alpha}{\pi_{k}}}\approx0.9917,A=\frac{1/\beta-(1-\delta)}{\alpha},B=(1-\alpha)\pi_{k}^{(1-\gamma)\alpha/(1-\alpha)}\pi_{c}^{-\gamma}(1-\overline{l})^{\mu}\overline{l}^{-\mu}$ $\gamma=2,\mu=2,\alpha=1/3,\rho=0.95,\sigma=0.01$.

\subsubsection*{Results}

Table \ref{tab:results_growth_elastic_labor-summary} shows the computation time, numerical accuracy, and the number of iterations. Because the computation time may be affected by the computational environment, it also shows the number of value function evaluations (Eval($V$)) and the number of evaluating $\frac{\partial Q}{\partial a}$ (Eval$\left(\frac{\partial Q}{\partial a}\right)$).

The results in Table \ref{tab:results_growth_elastic_labor-summary} show that PI-Krylov converges much faster than the VFI and PI. Because the fast convergence of the PI is theoretically known and Krylov methods are easy to implement using packages in many programming languages, PI-Krylov is worth considering as the solution method. 

In addition, utilizing the idea of relative value functions further reduces the computation time and the function evaluations, regardless of whether we choose VFI-based or PI-based algorithms. Because it is simple to introduce the ideas of the relative value function, it is also helpful.

With regard to the VF-PGI-Spectral algorithm, the method shows slightly superior performance than the PI-Krylov. Though the convergence property of the VF-PGI-Spectral is less clear than the PI, it would be also useful when the computational cost of nonlinear root finding in the policy improvement step is the bottleneck.

Concerning the acceleration method of fixed-point iterations, the spectral algorithm vastly reduces the number of iterations and the computational time of the VFI-based algorithms.\footnote{As shown in the Appendix, the PI algorithms introducing acceleration methods of fixed-point iterations are slightly slower than the PI algorithms without the introduction. Because PI algorithms themselves are very fast as theoretically guaranteed, it seems there is little room to further accelerate the convergence.} It largely accelerates the convergence even when the speed-up due to the introduction of relative value functions itself is relatively small.

Finally, the table also shows the performance of the algorithms introducing the idea of endogenous value functions (\citealp{bray2019markov}), which is a slight extension of relative value functions and discussed in detail in Appendix \ref{sec:other_algorithms}. Though it seems that its direct applicability depends on how we construct the grid points of states\footnote{For instance, it seems not straightforward to introduce endogenous value functions when we construct the grid points of states by the Smolyak method.}, the use of endogenous value function further contributes to reducing the number of iterations and computation time. 
\begin{center}
{\footnotesize{}}
\begin{table}[H]
{\footnotesize{}\caption{Speed and accuracy of algorithms (Summary; Single-agent neoclassical growth model)\label{tab:results_growth_elastic_labor-summary}}
}{\footnotesize\par}
\begin{centering}
{\tiny{}}%
\begin{tabular}{ccccccccccc}
\hline 
{\tiny{}Method} & {\tiny{}Acceleration} & {\tiny{}Root finding} & {\tiny{}CPU (sec)} & {\tiny{}$L_{1}$} & {\tiny{}$L_{\infty}$} & {\tiny{}Conv.} & {\tiny{}Iter.} & {\tiny{}CPU / Iter.} & {\tiny{}Eval$\left(V\right)$} & {\tiny{}Eval$\left(\frac{\partial Q}{\partial a}\right)$}\tabularnewline
\hline 
\hline 
{\tiny{}VF-PGI} & {\tiny{}Spectral} & {\tiny{}No} & {\tiny{}0.102} & {\tiny{}-5.425} & {\tiny{}-3.983} & {\tiny{}1} & {\tiny{}102} & {\tiny{}0.001} & {\tiny{}10200} & {\tiny{}10200}\tabularnewline
{\tiny{}VFI} & {\tiny{}Spectral} & {\tiny{}Yes} & {\tiny{}0.625} & {\tiny{}-5.425} & {\tiny{}-3.983} & {\tiny{}1} & {\tiny{}75} & {\tiny{}0.008} & {\tiny{}7500} & {\tiny{}33458}\tabularnewline
{\tiny{}ECM} & {\tiny{}Spectral} & {\tiny{}Yes} & {\tiny{}0.397} & {\tiny{}-5.406} & {\tiny{}-3.96} & {\tiny{}1} & {\tiny{}63} & {\tiny{}0.006} & {\tiny{}6300} & {\tiny{}NaN}\tabularnewline
{\tiny{}EGM} & {\tiny{}Spectral} & {\tiny{}Yes} & {\tiny{}0.26} & {\tiny{}-5.442} & {\tiny{}-3.999} & {\tiny{}1} & {\tiny{}53} & {\tiny{}0.005} & {\tiny{}5300} & {\tiny{}NaN}\tabularnewline
{\tiny{}VFI} & {\tiny{}-} & {\tiny{}Yes} & {\tiny{}9.032} & {\tiny{}-5.425} & {\tiny{}-3.983} & {\tiny{}1} & {\tiny{}1399} & {\tiny{}0.006} & {\tiny{}139900} & {\tiny{}457500}\tabularnewline
{\tiny{}PI} & {\tiny{}-} & {\tiny{}Yes} & {\tiny{}0.484} & {\tiny{}-5.949} & {\tiny{}-4.836} & {\tiny{}1} & {\tiny{}5} & {\tiny{}0.097} & {\tiny{}567700} & {\tiny{}3322}\tabularnewline
{\tiny{}PI-Krylov} & {\tiny{}-} & {\tiny{}Yes} & {\tiny{}0.142} & {\tiny{}-5.949} & {\tiny{}-4.836} & {\tiny{}1} & {\tiny{}5} & {\tiny{}0.028} & {\tiny{}5000} & {\tiny{}3322}\tabularnewline
\hline 
{\tiny{}RVF-PGI} & {\tiny{}Spectral} & {\tiny{}No} & {\tiny{}0.057} & {\tiny{}-5.425} & {\tiny{}-3.983} & {\tiny{}1} & {\tiny{}60} & {\tiny{}0.001} & {\tiny{}6000} & {\tiny{}6000}\tabularnewline
{\tiny{}RVFI} & {\tiny{}Spectral} & {\tiny{}Yes} & {\tiny{}0.495} & {\tiny{}-5.425} & {\tiny{}-3.983} & {\tiny{}1} & {\tiny{}67} & {\tiny{}0.007} & {\tiny{}6700} & {\tiny{}26470}\tabularnewline
{\tiny{}RVFI} & {\tiny{}-} & {\tiny{}Yes} & {\tiny{}3.084} & {\tiny{}-5.425} & {\tiny{}-3.983} & {\tiny{}1} & {\tiny{}410} & {\tiny{}0.008} & {\tiny{}41000} & {\tiny{}158720}\tabularnewline
{\tiny{}RPI} & {\tiny{}-} & {\tiny{}Yes} & {\tiny{}0.206} & {\tiny{}-5.949} & {\tiny{}-4.836} & {\tiny{}1} & {\tiny{}5} & {\tiny{}0.041} & {\tiny{}134400} & {\tiny{}3322}\tabularnewline
{\tiny{}RPI-Krylov} & {\tiny{}-} & {\tiny{}Yes} & {\tiny{}0.084} & {\tiny{}-5.949} & {\tiny{}-4.836} & {\tiny{}1} & {\tiny{}5} & {\tiny{}0.017} & {\tiny{}4600} & {\tiny{}3322}\tabularnewline
\hline 
{\tiny{}EVF-PGI} & {\tiny{}Spectral} & {\tiny{}No} & {\tiny{}0.056} & {\tiny{}-5.425} & {\tiny{}-3.983} & {\tiny{}1} & {\tiny{}54} & {\tiny{}0.001} & {\tiny{}5400} & {\tiny{}5400}\tabularnewline
{\tiny{}EVFI} & {\tiny{}Spectral} & {\tiny{}Yes} & {\tiny{}0.252} & {\tiny{}-5.425} & {\tiny{}-3.983} & {\tiny{}1} & {\tiny{}31} & {\tiny{}0.008} & {\tiny{}3100} & {\tiny{}13476}\tabularnewline
{\tiny{}EVFI} & {\tiny{}-} & {\tiny{}Yes} & {\tiny{}2.974} & {\tiny{}-5.425} & {\tiny{}-3.983} & {\tiny{}1} & {\tiny{}400} & {\tiny{}0.007} & {\tiny{}40000} & {\tiny{}155720}\tabularnewline
{\tiny{}EPI} & {\tiny{}-} & {\tiny{}Yes} & {\tiny{}0.144} & {\tiny{}-5.949} & {\tiny{}-4.836} & {\tiny{}1} & {\tiny{}5} & {\tiny{}0.029} & {\tiny{}112800} & {\tiny{}3322}\tabularnewline
{\tiny{}EPI-Krylov} & {\tiny{}-} & {\tiny{}Yes} & {\tiny{}0.063} & {\tiny{}-5.949} & {\tiny{}-4.836} & {\tiny{}1} & {\tiny{}5} & {\tiny{}0.013} & {\tiny{}3700} & {\tiny{}3322}\tabularnewline
\hline 
{\tiny{}EE} & {\tiny{}Spectral} & {\tiny{}Yes} & {\tiny{}0.183} & {\tiny{}-7.514} & {\tiny{}-5.809} & {\tiny{}1} & {\tiny{}38} & {\tiny{}0.005} & {\tiny{}NaN} & {\tiny{}NaN}\tabularnewline
{\tiny{}EE} & {\tiny{}-} & {\tiny{}Yes} & {\tiny{}1.283} & {\tiny{}-7.509} & {\tiny{}-5.809} & {\tiny{}1} & {\tiny{}280} & {\tiny{}0.005} & {\tiny{}NaN} & {\tiny{}NaN}\tabularnewline
\hline 
\end{tabular}{\tiny\par}
\par\end{centering}

{\scriptsize{}Notes. $\alpha_{0}=1,\lambda=10^{-7}$.}{\scriptsize\par}

{\scriptsize{}$L_{1}$ and $L_{\infty}$ denotes the average and the maximum absolute values of Euler residuals in log 10 units on a stochastic simulation of 10,000 observations.}{\scriptsize\par}

{\scriptsize{}``Root finding'' denotes whether a nonlinear root finding step exists in each algorithm.}{\scriptsize\par}

{\scriptsize{}Eval$\left(V\right)$ denotes the number of $\int\overline{V}^{(n)}\left(s^{\prime}\right)p\left(s^{\prime}|\widehat{s},a^{(n)}\left(\widehat{s}\right)\right)ds^{\prime}$ evaluations. Eval$\left(\frac{\partial Q}{\partial a}\right)$ denotes the number of $\frac{\partial\left[\int\overline{V}_{j}^{(n)}\left(\text{\ensuremath{s^{\prime}}}\right)p\left(s^{\prime}|\widehat{s},a^{(n*)}\left(\widehat{s}\right)\right)ds^{\prime}\right]}{\partial a^{(n*)}\left(\widehat{s}\right)}$ evaluations.}{\scriptsize\par}

\end{table}
{\footnotesize\par}
\par\end{center}

\subsection{Dynamic investment competition model with continuous states\label{subsec:investment-competition-conti-states}}

\subsubsection*{Settings}

We examine a simple stationary dynamic investment competition model.\footnote{A nonstationary version of the model was empirically applied in \citet{fukasawa2023long}, and an analogous model was also considered in \citet{miranda2004applied}.} Firms $j=1,\cdots,J$ produce and sell a homogeneous product in a single market in each period. Firms makes investment decisions to lower their marginal production costs. Firm $j$'s dynamic optimization problem related to its investment is given by:

\begin{eqnarray*}
\max_{i_{j}(s)\in\mathbb{R}} & \pi_{j}(s)-c\left(k_{j},i_{j}(s)\right)+\beta E_{z}V_{j}\left(k_{j}^{\prime},k_{-j}^{\prime},z^{\prime}\right)\\
 & s.t.\ k_{j}^{\prime}=(1-\delta)k_{j}+i_{j}
\end{eqnarray*}

Here, $\pi_{j}(s)$ denotes firm $j$'s per-period profit from producing and selling profit given states $s\equiv(k,z)$. $k\equiv\left(k_{j},k_{-j}\right)$ denotes all firms' capital stocks, and $z\equiv(\xi,\mu)$ denotes the exogenous state variables. $c\left(k_{j},i_{j}(s)\right)$ denotes firm $j$'s investment cost given its capital $k_{j}$ and investment $i_{j}(s)$, and assume $c(k,i)=\theta_{k}i+\theta_{a}\frac{i^{2}}{k}$. $k_{j}^{\prime}=(1-\delta)k_{j}+i_{j}$ represents the deterministic state transition of firm $j$'s capital stock. $\delta$ denotes the depreciation rate of the capital. 

Assuming constant marginal production costs, we define $mc_{j}\left(k_{j},\mu\right)=k_{j}^{-\gamma}\cdot\exp\left(\mu\right)$ as the marginal cost of firm $j$ given its capital stock $k_{j}$. $\mu$ denotes stochastic cost shock. Firms choose outputs in the market, and assume static Cournot competition with regard to the output choice. Let $Q(P)=P^{-\eta}\cdot\exp\left(\xi\right)$ be the demand function of the product, where $Q$ denotes the aggregate quantity, $P$ denotes the price, and $\xi$ denotes stochastic demand shock. $\eta$ denotes a parameter corresponding to the price elasticity of demand. Then, firm $j$'s profit function can be represented as $\pi_{j}(s)=\left(P(s)-mc_{j}(s)\right)q_{j}(s)$. Given the values of the state variables $s\equiv\left(k_{j},k_{-j};\xi,\mu\right)$ and parameters, we can compute the profits of the firms. 

Regarding the exogenous shocks, we assume stationary AR(1) transitions $\xi^{\prime}=\rho^{\xi}(\xi-\overline{\xi})+\overline{\xi}+u^{\xi}$ where $u^{\xi}\sim N(0,\sigma_{\xi}^{2})$, and $\mu^{\prime}=\rho^{\mu}(\mu-\overline{\mu})+\overline{\mu}+u^{\mu}$ where $u^{\mu}\sim N(0,\sigma_{\mu}^{2})$.

The parameters are specified as follows: $\alpha=1.5$, $\beta=0.9$, $\delta=0.08$, $\rho_{1}^{\xi}=\rho_{1}^{\mu}=0.9$, $\sigma_{\xi}=\sigma_{\mu}=0.01$, $\overline{\xi}=4$, $\overline{\mu}=2$, and $\theta_{k}=0.12$, $\theta_{a}=0.3$. Regarding VF-PGI-Spectral, we let $\alpha_{0}=1$ and $\lambda=0.001$.

\subsubsection*{Results}

Table \ref{tab:conti_state_dynamic_game} shows the results. When the number of firms is relatively small ($J=2$), the performance of VF-PGI-Spectral and PI{*}-Krylov is mostly the same. In contrast, as the number of firms increases, the relative advantage of VF-PGI-Spectral gets larger. Note that VFI{*} is much slower, and VFI{*}-Spectral is still much slower than the VF-PGI-Spectral and PI{*}-Krylov.

Concerning the relative value function, the introduction it further accelerates the convergence, as shown in the table. Hence, the idea is also worth considering even in multi-agent settings. 
\begin{center}
{\footnotesize{}}
\begin{table}[p]
{\footnotesize{}\caption{Speed and accuracy of algorithms (Dynamic investment competition model with continuous states)\label{tab:conti_state_dynamic_game}}
}{\footnotesize\par}
\begin{centering}
{\footnotesize{}}%
\begin{tabular}{cccccccccc}
\hline 
{\footnotesize{}$J$} & {\footnotesize{}Method} & {\footnotesize{}Root finding} & {\footnotesize{}CPU (sec)} & {\footnotesize{}$L_{1}$} & {\footnotesize{}$L_{\infty}$} & {\footnotesize{}Iter.} & {\footnotesize{}CPU / Iter.} & {\footnotesize{}Eval$\left(V\right)$} & {\footnotesize{}Eval$\left(\frac{\partial Q}{\partial a}\right)$}\tabularnewline
\hline 
\hline 
\multirow{10}{*}{{\footnotesize{}2}} & {\footnotesize{}VF-PGI-Spectral} & {\footnotesize{}No} & {\footnotesize{}0.088} & {\footnotesize{}-6.706} & {\footnotesize{}-5.934} & {\footnotesize{}84} & {\footnotesize{}0.001} & {\footnotesize{}23016} & {\footnotesize{}23016}\tabularnewline
 & {\footnotesize{}VFI{*}-Spectral} & {\footnotesize{}Yes} & {\footnotesize{}0.341} & {\footnotesize{}-6.706} & {\footnotesize{}-5.94} & {\footnotesize{}45} & {\footnotesize{}0.008} & {\footnotesize{}12330} & {\footnotesize{}62472}\tabularnewline
 & {\footnotesize{}VFI{*}} & {\footnotesize{}Yes} & {\footnotesize{}0.558} & {\footnotesize{}-6.703} & {\footnotesize{}-5.941} & {\footnotesize{}92} & {\footnotesize{}0.006} & {\footnotesize{}25208} & {\footnotesize{}97818}\tabularnewline
 & {\footnotesize{}PI{*}-Krylov-Spectral} & {\footnotesize{}Yes} & {\footnotesize{}0.154} & {\footnotesize{}-6.701} & {\footnotesize{}-5.943} & {\footnotesize{}19} & {\footnotesize{}0.008} & {\footnotesize{}37264} & {\footnotesize{}26852}\tabularnewline
 & {\footnotesize{}PI{*}-Krylov} & {\footnotesize{}Yes} & {\footnotesize{}0.136} & {\footnotesize{}-6.7} & {\footnotesize{}-5.944} & {\footnotesize{}12} & {\footnotesize{}0.011} & {\footnotesize{}27126} & {\footnotesize{}18084}\tabularnewline
\cline{2-10} \cline{3-10} \cline{4-10} \cline{5-10} \cline{6-10} \cline{7-10} \cline{8-10} \cline{9-10} \cline{10-10} 
 & {\footnotesize{}RVF-PGI-Spectral} & {\footnotesize{}No} & {\footnotesize{}0.073} & {\footnotesize{}-6.701} & {\footnotesize{}-5.94} & {\footnotesize{}78} & {\footnotesize{}0.001} & {\footnotesize{}21372} & {\footnotesize{}21372}\tabularnewline
 & {\footnotesize{}RVFI{*}-Spectral} & {\footnotesize{}Yes} & {\footnotesize{}0.226} & {\footnotesize{}-6.701} & {\footnotesize{}-5.946} & {\footnotesize{}38} & {\footnotesize{}0.006} & {\footnotesize{}10412} & {\footnotesize{}53704}\tabularnewline
 & {\footnotesize{}RVFI{*}} & {\footnotesize{}Yes} & {\footnotesize{}0.395} & {\footnotesize{}-6.703} & {\footnotesize{}-5.941} & {\footnotesize{}68} & {\footnotesize{}0.006} & {\footnotesize{}18632} & {\footnotesize{}83844}\tabularnewline
 & {\footnotesize{}RPI{*}-Krylov-Spectral} & {\footnotesize{}Yes} & {\footnotesize{}0.156} & {\footnotesize{}-6.704} & {\footnotesize{}-5.939} & {\footnotesize{}13} & {\footnotesize{}0.012} & {\footnotesize{}30688} & {\footnotesize{}19728}\tabularnewline
 & {\footnotesize{}RPI{*}-Krylov} & {\footnotesize{}Yes} & {\footnotesize{}0.139} & {\footnotesize{}-6.7} & {\footnotesize{}-5.944} & {\footnotesize{}12} & {\footnotesize{}0.012} & {\footnotesize{}27400} & {\footnotesize{}18084}\tabularnewline
\hline 
\multirow{10}{*}{{\footnotesize{}3}} & {\footnotesize{}VF-PGI-Spectral} & {\footnotesize{}No} & {\footnotesize{}0.438} & {\footnotesize{}-7.398} & {\footnotesize{}-6.488} & {\footnotesize{}59} & {\footnotesize{}0.007} & {\footnotesize{}42657} & {\footnotesize{}42657}\tabularnewline
 & {\footnotesize{}VFI{*}-Spectral} & {\footnotesize{}Yes} & {\footnotesize{}2.749} & {\footnotesize{}-7.454} & {\footnotesize{}-6.511} & {\footnotesize{}40} & {\footnotesize{}0.069} & {\footnotesize{}28920} & {\footnotesize{}138816}\tabularnewline
 & {\footnotesize{}VFI{*}} & {\footnotesize{}Yes} & {\footnotesize{}5.011} & {\footnotesize{}-7.454} & {\footnotesize{}-6.516} & {\footnotesize{}95} & {\footnotesize{}0.053} & {\footnotesize{}68685} & {\footnotesize{}255460}\tabularnewline
 & {\footnotesize{}PI{*}-Krylov-Spectral} & {\footnotesize{}Yes} & {\footnotesize{}1.675} & {\footnotesize{}-7.384} & {\footnotesize{}-6.53} & {\footnotesize{}18} & {\footnotesize{}0.093} & {\footnotesize{}97605} & {\footnotesize{}65070}\tabularnewline
 & {\footnotesize{}PI{*}-Krylov} & {\footnotesize{}Yes} & {\footnotesize{}1.329} & {\footnotesize{}-7.328} & {\footnotesize{}-6.542} & {\footnotesize{}12} & {\footnotesize{}0.111} & {\footnotesize{}75915} & {\footnotesize{}46272}\tabularnewline
\cline{2-10} \cline{3-10} \cline{4-10} \cline{5-10} \cline{6-10} \cline{7-10} \cline{8-10} \cline{9-10} \cline{10-10} 
 & {\footnotesize{}RVF-PGI-Spectral} & {\footnotesize{}No} & {\footnotesize{}0.499} & {\footnotesize{}-7.409} & {\footnotesize{}-6.525} & {\footnotesize{}50} & {\footnotesize{}0.01} & {\footnotesize{}36150} & {\footnotesize{}36150}\tabularnewline
 & {\footnotesize{}RVFI{*}-Spectral} & {\footnotesize{}Yes} & {\footnotesize{}3.36} & {\footnotesize{}-7.456} & {\footnotesize{}-6.512} & {\footnotesize{}34} & {\footnotesize{}0.099} & {\footnotesize{}24582} & {\footnotesize{}121464}\tabularnewline
 & {\footnotesize{}RVFI{*}} & {\footnotesize{}Yes} & {\footnotesize{}4.545} & {\footnotesize{}-7.454} & {\footnotesize{}-6.516} & {\footnotesize{}61} & {\footnotesize{}0.075} & {\footnotesize{}44103} & {\footnotesize{}202440}\tabularnewline
 & {\footnotesize{}RPI{*}-Krylov-Spectral} & {\footnotesize{}Yes} & {\footnotesize{}1.609} & {\footnotesize{}-7.363} & {\footnotesize{}-6.534} & {\footnotesize{}14} & {\footnotesize{}0.115} & {\footnotesize{}81699} & {\footnotesize{}54948}\tabularnewline
 & {\footnotesize{}RPI{*}-Krylov} & {\footnotesize{}Yes} & {\footnotesize{}1.568} & {\footnotesize{}-7.312} & {\footnotesize{}-6.546} & {\footnotesize{}12} & {\footnotesize{}0.131} & {\footnotesize{}71577} & {\footnotesize{}46272}\tabularnewline
\hline 
\multirow{3}{*}{{\footnotesize{}4}} & {\footnotesize{}RVF-PGI-Spectral} & {\footnotesize{}No} & {\footnotesize{}1.526} & {\footnotesize{}-6.027} & {\footnotesize{}-5.617} & {\footnotesize{}47} & {\footnotesize{}0.032} & {\footnotesize{}73132} & {\footnotesize{}73132}\tabularnewline
 & {\footnotesize{}RVFI{*}-Spectral} & {\footnotesize{}Yes} & {\footnotesize{}12.235} & {\footnotesize{}-6.028} & {\footnotesize{}-5.619} & {\footnotesize{}29} & {\footnotesize{}0.422} & {\footnotesize{}45124} & {\footnotesize{}230288}\tabularnewline
 & {\footnotesize{}RPI{*}-Krylov} & {\footnotesize{}Yes} & {\footnotesize{}8.956} & {\footnotesize{}-6.027} & {\footnotesize{}-5.617} & {\footnotesize{}13} & {\footnotesize{}0.689} & {\footnotesize{}163380} & {\footnotesize{}105808}\tabularnewline
\hline 
\multirow{3}{*}{{\footnotesize{}5}} & {\footnotesize{}RVF-PGI-Spectral} & {\footnotesize{}No} & {\footnotesize{}2.924} & {\footnotesize{}-4.853} & {\footnotesize{}-4.535} & {\footnotesize{}54} & {\footnotesize{}0.054} & {\footnotesize{}159030} & {\footnotesize{}159030}\tabularnewline
 & {\footnotesize{}RVFI{*}-Spectral} & {\footnotesize{}Yes} & {\footnotesize{}57.372} & {\footnotesize{}-4.853} & {\footnotesize{}-4.535} & {\footnotesize{}35} & {\footnotesize{}1.639} & {\footnotesize{}103075} & {\footnotesize{}524210}\tabularnewline
 & {\footnotesize{}RPI{*}-Krylov} & {\footnotesize{}Yes} & {\footnotesize{}23.714} & {\footnotesize{}-4.854} & {\footnotesize{}-4.535} & {\footnotesize{}16} & {\footnotesize{}1.482} & {\footnotesize{}424080} & {\footnotesize{}241490}\tabularnewline
\hline 
\end{tabular}{\footnotesize\par}
\par\end{centering}
{\footnotesize{}Notes. $L_{1}$ and $L_{\infty}$ denotes the average and the maximum of $\log_{10}\left(\left\Vert \frac{\partial Q_{j}(a_{j}\left(\widehat{s}\right),a_{-j}\left(\widehat{s}\right),\widehat{s};V_{j})}{\partial a_{j}\left(\widehat{s}\right)}\right\Vert _{\infty}\right)$ on a stochastic simulation of 100 observations.}{\footnotesize\par}

{\footnotesize{}Eval$\left(V\right)$ denotes the number of $\int\overline{V_{j}}^{(n)}\left(s^{\prime}\right)p\left(s^{\prime}|\widehat{s},a^{(n)}\left(\widehat{s}\right)\right)ds^{\prime}$ evaluations. Eval$\left(\frac{\partial Q}{\partial a}\right)$ denotes the number of $\frac{\partial\left[\int\overline{V}^{(n)}\left(\text{\ensuremath{s^{\prime}}}\right)p\left(s^{\prime}|\widehat{s},a^{(n*)}\left(\widehat{s}\right)\right)ds^{\prime}\right]}{\partial a_{j}^{(n*)}\left(\widehat{s}\right)}$ evaluations.}{\footnotesize\par}
\end{table}
{\footnotesize\par}
\par\end{center}

\section{Conclusions\label{sec:Conclusions}}

This study has investigated computationally efficient algorithms for solving discrete-time infinite-horizon single-agent/multi-agent dynamic models with continuous actions. It shows that we can easily reduce the computational costs by slightly changing basic algorithms using value functions, such as the Value Function Iteration (VFI) and the Policy Iteration (PI).

The PI method with a Krylov iterative method (GMRES), which can be easily implemented using built-in packages, works much better than VFI-based algorithms even when considering continuous state models. Concerning the VFI algorithm, we can largely speed up the convergence by introducing acceleration methods of fixed-point iterations. The current study also proposes the VF-PGI-Spectral (Value Function-Policy Gradient Iteration Spectral) algorithm, which is a slight modification of the VFI. It shows numerical results where the VF-PGI-Spectral performs much better than the VFI- and PI-based algorithms especially in multi-agent dynamic games. Finally, it shows that using relative value functions further reduces the computational costs of solving the model.

The current study has considered the algorithms where grid points of states are fixed throughout the iterations. However, simulation-based approaches, where the grid points are flexibly adjusted during the iterations, also exist and they are effective at reducing the number of grid points (cf. \citealp{maliar2014numerical}). In addition, recent studies (e.g., \citealp{achdou2022income}) apply continuous-time models. Concerning whether the insights in the current study are also informative in these settings, The current study leaves it for further research.

\appendix

\section{Detailed numerical results of the single-agent growth model\label{sec:Full-results-growth-model}}

Tables in this section show the full results of the comparison of algorithms. Other than the points discussed in Section \ref{subsec:Single-agent-neoclassical}, we can point out the followings:
\begin{itemize}
\item Using the idea of optimistic policy iterations sometimes reduces the computational time of PI and PI-Krylov, though appropriate choice of the tuning parameter $m$ is a priori unclear.
\item Using the idea of ECM slightly reduces the computation time of VFI and PI, though the convergence properties are theoretically less clear.
\item Algorithms with the Anderson acceleration are faster than the ones with the spectral algorithm in some algorithms, but slower in other algorithms. In the case of VF-PGI, using Anderson acceleration leads to divergence.
\item Algorithms with SQUAREM are mostly the same as or slower than the ones with the spectral. In the case of VF-PGI, using SQUAREM leads to divergence. 
\end{itemize}
\begin{center}
{\footnotesize{}}
\begin{table}[p]
{\footnotesize{}\caption{Speed and accuracy of algorithms (Single-agent neoclassical growth model with elastic labor supply)\label{tab:results_growth_elastic_labor-2-1-2-1}}
}{\footnotesize\par}
\begin{centering}
{\tiny{}}%
\begin{tabular}{ccccccccccc}
\hline 
{\tiny{}Method} & {\tiny{}Acceleration} & {\tiny{}Root finding} & {\tiny{}CPU (sec)} & {\tiny{}$L_{1}$} & {\tiny{}$L_{\infty}$} & {\tiny{}Conv.} & {\tiny{}Iter.} & {\tiny{}CPU / Iter.} & {\tiny{}Eval$\left(V\right)$} & {\tiny{}Eval$\left(\frac{\partial Q}{\partial a}\right)$}\tabularnewline
\hline 
\hline 
{\tiny{}VF-PGI} & {\tiny{}Spectral} & {\tiny{}No} & {\tiny{}0.102} & {\tiny{}-5.425} & {\tiny{}-3.983} & {\tiny{}1} & {\tiny{}102} & {\tiny{}0.001} & {\tiny{}10200} & {\tiny{}10200}\tabularnewline
{\tiny{}VF-PGI{*}} & {\tiny{}Spectral} & {\tiny{}No} & {\tiny{}0.066} & {\tiny{}-5.424} & {\tiny{}-3.983} & {\tiny{}1} & {\tiny{}75} & {\tiny{}0.001} & {\tiny{}7500} & {\tiny{}7500}\tabularnewline
{\tiny{}VFI} & {\tiny{}Spectral} & {\tiny{}Yes} & {\tiny{}0.625} & {\tiny{}-5.425} & {\tiny{}-3.983} & {\tiny{}1} & {\tiny{}75} & {\tiny{}0.008} & {\tiny{}7500} & {\tiny{}33458}\tabularnewline
{\tiny{}VFI-ECM} & {\tiny{}Spectral} & {\tiny{}Yes} & {\tiny{}0.397} & {\tiny{}-5.406} & {\tiny{}-3.96} & {\tiny{}1} & {\tiny{}63} & {\tiny{}0.006} & {\tiny{}6300} & {\tiny{}NaN}\tabularnewline
{\tiny{}VFI-EGM} & {\tiny{}Spectral} & {\tiny{}Yes} & {\tiny{}0.26} & {\tiny{}-5.442} & {\tiny{}-3.999} & {\tiny{}1} & {\tiny{}53} & {\tiny{}0.005} & {\tiny{}5300} & {\tiny{}NaN}\tabularnewline
{\tiny{}PI} & {\tiny{}Spectral} & {\tiny{}Yes} & {\tiny{}0.435} & {\tiny{}-5.949} & {\tiny{}-4.836} & {\tiny{}1} & {\tiny{}6} & {\tiny{}0.072} & {\tiny{}637900} & {\tiny{}3680}\tabularnewline
{\tiny{}PI-Krylov} & {\tiny{}Spectral} & {\tiny{}Yes} & {\tiny{}0.071} & {\tiny{}-5.949} & {\tiny{}-4.836} & {\tiny{}1} & {\tiny{}6} & {\tiny{}0.012} & {\tiny{}6100} & {\tiny{}3680}\tabularnewline
{\tiny{}OPI} & {\tiny{}Spectral} & {\tiny{}Yes} & {\tiny{}0.186} & {\tiny{}-5.949} & {\tiny{}-4.836} & {\tiny{}1} & {\tiny{}12} & {\tiny{}0.016} & {\tiny{}107100} & {\tiny{}6368}\tabularnewline
{\tiny{}OPI-Krylov} & {\tiny{}Spectral} & {\tiny{}Yes} & {\tiny{}0.134} & {\tiny{}-5.949} & {\tiny{}-4.836} & {\tiny{}1} & {\tiny{}14} & {\tiny{}0.01} & {\tiny{}6500} & {\tiny{}6984}\tabularnewline
{\tiny{}PI-ECM} & {\tiny{}Spectral} & {\tiny{}Yes} & {\tiny{}0.426} & {\tiny{}-5.926} & {\tiny{}-4.802} & {\tiny{}1} & {\tiny{}6} & {\tiny{}0.071} & {\tiny{}654200} & {\tiny{}NaN}\tabularnewline
{\tiny{}PI-Krylov-ECM} & {\tiny{}Spectral} & {\tiny{}Yes} & {\tiny{}0.053} & {\tiny{}-5.926} & {\tiny{}-4.802} & {\tiny{}1} & {\tiny{}6} & {\tiny{}0.009} & {\tiny{}6300} & {\tiny{}NaN}\tabularnewline
{\tiny{}OPI-ECM} & {\tiny{}Spectral} & {\tiny{}Yes} & {\tiny{}0.152} & {\tiny{}-5.926} & {\tiny{}-4.802} & {\tiny{}1} & {\tiny{}11} & {\tiny{}0.014} & {\tiny{}107800} & {\tiny{}NaN}\tabularnewline
{\tiny{}OPI-Krylov-ECM} & {\tiny{}Spectral} & {\tiny{}Yes} & {\tiny{}0.11} & {\tiny{}-5.926} & {\tiny{}-4.802} & {\tiny{}1} & {\tiny{}15} & {\tiny{}0.007} & {\tiny{}7300} & {\tiny{}NaN}\tabularnewline
{\tiny{}PI-Krylov-EGM} & {\tiny{}Spectral} & {\tiny{}Yes} & {\tiny{}0.117} & {\tiny{}-5.442} & {\tiny{}-3.999} & {\tiny{}1} & {\tiny{}20} & {\tiny{}0.006} & {\tiny{}15100} & {\tiny{}NaN}\tabularnewline
{\tiny{}DVFI-ECM} & {\tiny{}Spectral} & {\tiny{}Yes} & {\tiny{}0.165} & {\tiny{}-5.452} & {\tiny{}-4.005} & {\tiny{}1} & {\tiny{}25} & {\tiny{}0.007} & {\tiny{}2500} & {\tiny{}NaN}\tabularnewline
{\tiny{}DVFI-EGM} & {\tiny{}Spectral} & {\tiny{}Yes} & {\tiny{}0.183} & {\tiny{}-5.472} & {\tiny{}-4.047} & {\tiny{}1} & {\tiny{}36} & {\tiny{}0.005} & {\tiny{}3600} & {\tiny{}NaN}\tabularnewline
{\tiny{}EE} & {\tiny{}Spectral} & {\tiny{}Yes} & {\tiny{}0.183} & {\tiny{}-7.514} & {\tiny{}-5.809} & {\tiny{}1} & {\tiny{}38} & {\tiny{}0.005} & {\tiny{}NaN} & {\tiny{}NaN}\tabularnewline
{\tiny{}EE} & {\tiny{}Spectral} & {\tiny{}No} & {\tiny{}0.032} & {\tiny{}-7.514} & {\tiny{}-5.809} & {\tiny{}1} & {\tiny{}38} & {\tiny{}0.001} & {\tiny{}NaN} & {\tiny{}NaN}\tabularnewline
\hline 
{\tiny{}VF-PGI} & {\tiny{}-} & {\tiny{}No} & {\tiny{}0.169} & {\tiny{}NaN} & {\tiny{}NaN} & {\tiny{}0} & {\tiny{}8} & {\tiny{}0.021} & {\tiny{}800} & {\tiny{}800}\tabularnewline
{\tiny{}VF-PGI{*}} & {\tiny{}-} & {\tiny{}No} & {\tiny{}1.237} & {\tiny{}-5.425} & {\tiny{}-3.983} & {\tiny{}1} & {\tiny{}1405} & {\tiny{}0.001} & {\tiny{}140500} & {\tiny{}140500}\tabularnewline
{\tiny{}VFI} & {\tiny{}-} & {\tiny{}Yes} & {\tiny{}9.032} & {\tiny{}-5.425} & {\tiny{}-3.983} & {\tiny{}1} & {\tiny{}1399} & {\tiny{}0.006} & {\tiny{}139900} & {\tiny{}457500}\tabularnewline
{\tiny{}VFI-ECM} & {\tiny{}-} & {\tiny{}Yes} & {\tiny{}6.317} & {\tiny{}-5.406} & {\tiny{}-3.96} & {\tiny{}1} & {\tiny{}1399} & {\tiny{}0.005} & {\tiny{}139900} & {\tiny{}NaN}\tabularnewline
{\tiny{}VFI-EGM} & {\tiny{}-} & {\tiny{}Yes} & {\tiny{}4.253} & {\tiny{}-5.442} & {\tiny{}-3.999} & {\tiny{}1} & {\tiny{}970} & {\tiny{}0.004} & {\tiny{}97000} & {\tiny{}NaN}\tabularnewline
{\tiny{}PI} & {\tiny{}-} & {\tiny{}Yes} & {\tiny{}0.484} & {\tiny{}-5.949} & {\tiny{}-4.836} & {\tiny{}1} & {\tiny{}5} & {\tiny{}0.097} & {\tiny{}567700} & {\tiny{}3322}\tabularnewline
{\tiny{}PI-Krylov} & {\tiny{}-} & {\tiny{}Yes} & {\tiny{}0.142} & {\tiny{}-5.949} & {\tiny{}-4.836} & {\tiny{}1} & {\tiny{}5} & {\tiny{}0.028} & {\tiny{}5000} & {\tiny{}3322}\tabularnewline
{\tiny{}OPI} & {\tiny{}-} & {\tiny{}Yes} & {\tiny{}0.307} & {\tiny{}-5.949} & {\tiny{}-4.836} & {\tiny{}1} & {\tiny{}20} & {\tiny{}0.015} & {\tiny{}199600} & {\tiny{}7900}\tabularnewline
{\tiny{}OPI-Krylov} & {\tiny{}-} & {\tiny{}Yes} & {\tiny{}0.142} & {\tiny{}-5.949} & {\tiny{}-4.836} & {\tiny{}1} & {\tiny{}13} & {\tiny{}0.011} & {\tiny{}6200} & {\tiny{}6272}\tabularnewline
{\tiny{}PI-ECM} & {\tiny{}-} & {\tiny{}Yes} & {\tiny{}0.375} & {\tiny{}-5.926} & {\tiny{}-4.802} & {\tiny{}1} & {\tiny{}5} & {\tiny{}0.075} & {\tiny{}582800} & {\tiny{}NaN}\tabularnewline
{\tiny{}PI-Krylov-ECM} & {\tiny{}-} & {\tiny{}Yes} & {\tiny{}0.052} & {\tiny{}-5.926} & {\tiny{}-4.802} & {\tiny{}1} & {\tiny{}5} & {\tiny{}0.01} & {\tiny{}5200} & {\tiny{}NaN}\tabularnewline
{\tiny{}OPI-ECM} & {\tiny{}-} & {\tiny{}Yes} & {\tiny{}0.234} & {\tiny{}-5.926} & {\tiny{}-4.802} & {\tiny{}1} & {\tiny{}21} & {\tiny{}0.011} & {\tiny{}200000} & {\tiny{}NaN}\tabularnewline
{\tiny{}OPI-Krylov-ECM} & {\tiny{}-} & {\tiny{}Yes} & {\tiny{}0.078} & {\tiny{}-5.926} & {\tiny{}-4.802} & {\tiny{}1} & {\tiny{}10} & {\tiny{}0.008} & {\tiny{}4800} & {\tiny{}NaN}\tabularnewline
{\tiny{}PI-Krylov-EGM} & {\tiny{}-} & {\tiny{}Yes} & {\tiny{}0.432} & {\tiny{}-5.442} & {\tiny{}-3.999} & {\tiny{}1} & {\tiny{}85} & {\tiny{}0.005} & {\tiny{}24500} & {\tiny{}NaN}\tabularnewline
{\tiny{}AVFI} & {\tiny{}-} & {\tiny{}Yes} & {\tiny{}1.577} & {\tiny{}-5.425} & {\tiny{}-3.983} & {\tiny{}1} & {\tiny{}212} & {\tiny{}0.007} & {\tiny{}21200} & {\tiny{}85238}\tabularnewline
{\tiny{}AVFI-ECM} & {\tiny{}-} & {\tiny{}Yes} & {\tiny{}1.598} & {\tiny{}-5.425} & {\tiny{}-3.983} & {\tiny{}1} & {\tiny{}212} & {\tiny{}0.008} & {\tiny{}21200} & {\tiny{}NaN}\tabularnewline
{\tiny{}DVFI-ECM} & {\tiny{}-} & {\tiny{}Yes} & {\tiny{}2.138} & {\tiny{}-5.452} & {\tiny{}-4.005} & {\tiny{}1} & {\tiny{}394} & {\tiny{}0.005} & {\tiny{}39400} & {\tiny{}NaN}\tabularnewline
{\tiny{}DVFI-EGM} & {\tiny{}-} & {\tiny{}Yes} & {\tiny{}1.767} & {\tiny{}-5.472} & {\tiny{}-4.047} & {\tiny{}1} & {\tiny{}409} & {\tiny{}0.004} & {\tiny{}40900} & {\tiny{}NaN}\tabularnewline
{\tiny{}EE} & {\tiny{}-} & {\tiny{}Yes} & {\tiny{}1.283} & {\tiny{}-7.509} & {\tiny{}-5.809} & {\tiny{}1} & {\tiny{}280} & {\tiny{}0.005} & {\tiny{}NaN} & {\tiny{}NaN}\tabularnewline
{\tiny{}EE} & {\tiny{}-} & {\tiny{}Yes} & {\tiny{}0.221} & {\tiny{}-7.509} & {\tiny{}-5.809} & {\tiny{}1} & {\tiny{}280} & {\tiny{}0.001} & {\tiny{}NaN} & {\tiny{}NaN}\tabularnewline
\hline 
\end{tabular}{\tiny\par}
\par\end{centering}

{\scriptsize{}Notes. $\alpha_{0}=1,\lambda=10^{-7}$.}{\scriptsize\par}

{\scriptsize{}$L_{1}$ and $L_{\infty}$ denotes the average and the maximum absolute values of Euler residuals in log 10 units on a stochastic simulation of 10,000 observations.}{\scriptsize\par}

{\scriptsize{}``Root finding'' denotes whether a nonlinear root finding step exists in each algorithm.}{\scriptsize\par}

{\scriptsize{}Eval$\left(V\right)$ denotes the number of $\int\overline{V}^{(n)}\left(s^{\prime}\right)p\left(s^{\prime}|\widehat{s},a^{(n)}\left(\widehat{s}\right)\right)ds^{\prime}$ evaluations. Eval$\left(\frac{\partial Q}{\partial a}\right)$ denotes the number of $\frac{\partial\left[\int\overline{V}_{j}^{(n)}\left(\text{\ensuremath{s^{\prime}}}\right)p\left(s^{\prime}|\widehat{s},a^{(n*)}\left(\widehat{s}\right)\right)ds^{\prime}\right]}{\partial a^{(n*)}\left(\widehat{s}\right)}$ evaluations.}{\scriptsize\par}

\end{table}
{\footnotesize\par}
\par\end{center}

\begin{center}
{\footnotesize{}}
\begin{table}[H]
{\footnotesize{}\caption{Speed and accuracy of algorithms (Single-agent neoclassical growth model with elastic labor supply)\label{tab:results_growth_elastic_labor-2-1-2-1-1}}
}{\footnotesize\par}
\begin{centering}
{\tiny{}}%
\begin{tabular}{ccccccccccc}
\hline 
{\tiny{}Method} & {\tiny{}Acceleration} & {\tiny{}Root finding} & {\tiny{}CPU (sec)} & {\tiny{}$L_{1}$} & {\tiny{}$L_{\infty}$} & {\tiny{}Conv.} & {\tiny{}Iter.} & {\tiny{}CPU / Iter.} & {\tiny{}Eval$\left(V\right)$} & {\tiny{}Eval$\left(\frac{\partial Q}{\partial a}\right)$}\tabularnewline
\hline 
\hline 
{\tiny{}RVF-PGI} & {\tiny{}Spectral} & {\tiny{}No} & {\tiny{}0.057} & {\tiny{}-5.425} & {\tiny{}-3.983} & {\tiny{}1} & {\tiny{}60} & {\tiny{}0.001} & {\tiny{}6000} & {\tiny{}6000}\tabularnewline
{\tiny{}RVF-PGI{*}} & {\tiny{}Spectral} & {\tiny{}No} & {\tiny{}0.054} & {\tiny{}-5.425} & {\tiny{}-3.983} & {\tiny{}1} & {\tiny{}57} & {\tiny{}0.001} & {\tiny{}5700} & {\tiny{}5700}\tabularnewline
{\tiny{}RVFI} & {\tiny{}Spectral} & {\tiny{}Yes} & {\tiny{}0.495} & {\tiny{}-5.425} & {\tiny{}-3.983} & {\tiny{}1} & {\tiny{}67} & {\tiny{}0.007} & {\tiny{}6700} & {\tiny{}26470}\tabularnewline
{\tiny{}RVFI-ECM} & {\tiny{}Spectral} & {\tiny{}Yes} & {\tiny{}0.305} & {\tiny{}-5.406} & {\tiny{}-3.96} & {\tiny{}1} & {\tiny{}56} & {\tiny{}0.005} & {\tiny{}5600} & {\tiny{}NaN}\tabularnewline
{\tiny{}RVFI-EGM} & {\tiny{}Spectral} & {\tiny{}Yes} & {\tiny{}0.236} & {\tiny{}-5.442} & {\tiny{}-3.999} & {\tiny{}1} & {\tiny{}49} & {\tiny{}0.005} & {\tiny{}4900} & {\tiny{}NaN}\tabularnewline
{\tiny{}RPI} & {\tiny{}Spectral} & {\tiny{}Yes} & {\tiny{}0.188} & {\tiny{}-5.949} & {\tiny{}-4.836} & {\tiny{}1} & {\tiny{}7} & {\tiny{}0.027} & {\tiny{}163200} & {\tiny{}4130}\tabularnewline
{\tiny{}RPI-Krylov} & {\tiny{}Spectral} & {\tiny{}Yes} & {\tiny{}0.084} & {\tiny{}-5.949} & {\tiny{}-4.836} & {\tiny{}1} & {\tiny{}7} & {\tiny{}0.012} & {\tiny{}5900} & {\tiny{}4130}\tabularnewline
{\tiny{}ORPI} & {\tiny{}Spectral} & {\tiny{}Yes} & {\tiny{}0.119} & {\tiny{}-5.949} & {\tiny{}-4.836} & {\tiny{}1} & {\tiny{}7} & {\tiny{}0.017} & {\tiny{}59900} & {\tiny{}4116}\tabularnewline
{\tiny{}ORPI-Krylov} & {\tiny{}Spectral} & {\tiny{}Yes} & {\tiny{}0.086} & {\tiny{}-5.949} & {\tiny{}-4.836} & {\tiny{}1} & {\tiny{}8} & {\tiny{}0.011} & {\tiny{}3500} & {\tiny{}4462}\tabularnewline
{\tiny{}RPI-ECM} & {\tiny{}Spectral} & {\tiny{}Yes} & {\tiny{}0.164} & {\tiny{}-5.926} & {\tiny{}-4.802} & {\tiny{}1} & {\tiny{}7} & {\tiny{}0.023} & {\tiny{}168800} & {\tiny{}NaN}\tabularnewline
{\tiny{}RPI-Krylov-ECM} & {\tiny{}Spectral} & {\tiny{}Yes} & {\tiny{}0.067} & {\tiny{}-5.926} & {\tiny{}-4.802} & {\tiny{}1} & {\tiny{}7} & {\tiny{}0.01} & {\tiny{}6300} & {\tiny{}NaN}\tabularnewline
{\tiny{}ORPI-ECM} & {\tiny{}Spectral} & {\tiny{}Yes} & {\tiny{}0.098} & {\tiny{}-5.926} & {\tiny{}-4.802} & {\tiny{}1} & {\tiny{}7} & {\tiny{}0.014} & {\tiny{}60700} & {\tiny{}NaN}\tabularnewline
{\tiny{}ORPI-Krylov-ECM} & {\tiny{}Spectral} & {\tiny{}Yes} & {\tiny{}0.064} & {\tiny{}-5.926} & {\tiny{}-4.802} & {\tiny{}1} & {\tiny{}8} & {\tiny{}0.008} & {\tiny{}3400} & {\tiny{}NaN}\tabularnewline
{\tiny{}RPI-Krylov-EGM} & {\tiny{}Spectral} & {\tiny{}Yes} & {\tiny{}0.114} & {\tiny{}-5.442} & {\tiny{}-3.999} & {\tiny{}1} & {\tiny{}17} & {\tiny{}0.007} & {\tiny{}14400} & {\tiny{}NaN}\tabularnewline
\hline 
{\tiny{}RVF-PGI} & {\tiny{}-} & {\tiny{}No} & {\tiny{}0.041} & {\tiny{}NaN} & {\tiny{}NaN} & {\tiny{}0} & {\tiny{}8} & {\tiny{}0.005} & {\tiny{}800} & {\tiny{}800}\tabularnewline
{\tiny{}RVF-PGI{*}} & {\tiny{}-} & {\tiny{}No} & {\tiny{}0.37} & {\tiny{}-5.425} & {\tiny{}-3.983} & {\tiny{}1} & {\tiny{}417} & {\tiny{}0.001} & {\tiny{}41700} & {\tiny{}41700}\tabularnewline
{\tiny{}RVFI} & {\tiny{}-} & {\tiny{}Yes} & {\tiny{}3.084} & {\tiny{}-5.425} & {\tiny{}-3.983} & {\tiny{}1} & {\tiny{}410} & {\tiny{}0.008} & {\tiny{}41000} & {\tiny{}158720}\tabularnewline
{\tiny{}RVFI-ECM} & {\tiny{}-} & {\tiny{}Yes} & {\tiny{}2.234} & {\tiny{}-5.406} & {\tiny{}-3.96} & {\tiny{}1} & {\tiny{}410} & {\tiny{}0.005} & {\tiny{}41000} & {\tiny{}NaN}\tabularnewline
{\tiny{}RVFI-EGM} & {\tiny{}-} & {\tiny{}Yes} & {\tiny{}1.873} & {\tiny{}-5.442} & {\tiny{}-3.999} & {\tiny{}1} & {\tiny{}412} & {\tiny{}0.005} & {\tiny{}41200} & {\tiny{}NaN}\tabularnewline
{\tiny{}RPI} & {\tiny{}-} & {\tiny{}Yes} & {\tiny{}0.206} & {\tiny{}-5.949} & {\tiny{}-4.836} & {\tiny{}1} & {\tiny{}5} & {\tiny{}0.041} & {\tiny{}134400} & {\tiny{}3322}\tabularnewline
{\tiny{}RPI-Krylov} & {\tiny{}-} & {\tiny{}Yes} & {\tiny{}0.084} & {\tiny{}-5.949} & {\tiny{}-4.836} & {\tiny{}1} & {\tiny{}5} & {\tiny{}0.017} & {\tiny{}4600} & {\tiny{}3322}\tabularnewline
{\tiny{}ORPI} & {\tiny{}-} & {\tiny{}Yes} & {\tiny{}0.117} & {\tiny{}-5.949} & {\tiny{}-4.836} & {\tiny{}1} & {\tiny{}7} & {\tiny{}0.017} & {\tiny{}60200} & {\tiny{}3996}\tabularnewline
{\tiny{}ORPI-Krylov} & {\tiny{}-} & {\tiny{}Yes} & {\tiny{}0.087} & {\tiny{}-5.949} & {\tiny{}-4.836} & {\tiny{}1} & {\tiny{}8} & {\tiny{}0.011} & {\tiny{}3300} & {\tiny{}4284}\tabularnewline
{\tiny{}RPI-ECM} & {\tiny{}-} & {\tiny{}Yes} & {\tiny{}0.142} & {\tiny{}-5.926} & {\tiny{}-4.802} & {\tiny{}1} & {\tiny{}6} & {\tiny{}0.024} & {\tiny{}136500} & {\tiny{}NaN}\tabularnewline
{\tiny{}RPI-Krylov-ECM} & {\tiny{}-} & {\tiny{}Yes} & {\tiny{}0.051} & {\tiny{}-5.926} & {\tiny{}-4.802} & {\tiny{}1} & {\tiny{}6} & {\tiny{}0.009} & {\tiny{}5000} & {\tiny{}NaN}\tabularnewline
{\tiny{}ORPI-ECM} & {\tiny{}-} & {\tiny{}Yes} & {\tiny{}0.097} & {\tiny{}-5.926} & {\tiny{}-4.802} & {\tiny{}1} & {\tiny{}7} & {\tiny{}0.014} & {\tiny{}61300} & {\tiny{}NaN}\tabularnewline
{\tiny{}ORPI-Krylov-ECM} & {\tiny{}-} & {\tiny{}Yes} & {\tiny{}0.063} & {\tiny{}-5.926} & {\tiny{}-4.802} & {\tiny{}1} & {\tiny{}8} & {\tiny{}0.008} & {\tiny{}3300} & {\tiny{}NaN}\tabularnewline
{\tiny{}RPI-Krylov-EGM} & {\tiny{}-} & {\tiny{}Yes} & {\tiny{}0.347} & {\tiny{}-5.442} & {\tiny{}-3.999} & {\tiny{}1} & {\tiny{}66} & {\tiny{}0.005} & {\tiny{}23500} & {\tiny{}NaN}\tabularnewline
{\tiny{}ARVFI} & {\tiny{}-} & {\tiny{}Yes} & {\tiny{}1.163} & {\tiny{}-5.422} & {\tiny{}-3.983} & {\tiny{}1} & {\tiny{}138} & {\tiny{}0.008} & {\tiny{}13800} & {\tiny{}63040}\tabularnewline
{\tiny{}ARVFI-ECM} & {\tiny{}-} & {\tiny{}Yes} & {\tiny{}1.16} & {\tiny{}-5.422} & {\tiny{}-3.983} & {\tiny{}1} & {\tiny{}138} & {\tiny{}0.008} & {\tiny{}13800} & {\tiny{}NaN}\tabularnewline
\hline 
\end{tabular}{\tiny\par}
\par\end{centering}

{\scriptsize{}Notes. $\alpha_{0}=1,\lambda=10^{-7}$.}{\scriptsize\par}

{\scriptsize{}$L_{1}$ and $L_{\infty}$ denotes the average and the maximum absolute values of Euler residuals in log 10 units on a stochastic simulation of 10,000 observations.}{\scriptsize\par}

{\scriptsize{}``Root finding'' denotes whether a nonlinear root finding step exists in each algorithm.}{\scriptsize\par}

{\scriptsize{}Eval$\left(V\right)$ denotes the number of $\int\overline{V}^{(n)}\left(s^{\prime}\right)p\left(s^{\prime}|\widehat{s},a^{(n)}\left(\widehat{s}\right)\right)ds^{\prime}$ evaluations. Eval$\left(\frac{\partial Q}{\partial a}\right)$ denotes the number of $\frac{\partial\left[\int\overline{V}_{j}^{(n)}\left(\text{\ensuremath{s^{\prime}}}\right)p\left(s^{\prime}|\widehat{s},a^{(n*)}\left(\widehat{s}\right)\right)ds^{\prime}\right]}{\partial a^{(n*)}\left(\widehat{s}\right)}$ evaluations.}{\scriptsize\par}

\end{table}
{\footnotesize\par}
\par\end{center}

\begin{center}
{\footnotesize{}}
\begin{table}[H]
{\footnotesize{}\caption{Speed and accuracy of algorithms (Single-agent neoclassical growth model with elastic labor supply)\label{tab:results_growth_elastic_labor-2-1-2}}
}{\footnotesize\par}
\begin{centering}
{\tiny{}}%
\begin{tabular}{ccccccccccc}
\hline 
{\tiny{}Method} & {\tiny{}Acceleration} & {\tiny{}Root finding} & {\tiny{}CPU (sec)} & {\tiny{}$L_{1}$} & {\tiny{}$L_{\infty}$} & {\tiny{}Conv.} & {\tiny{}Iter.} & {\tiny{}CPU / Iter.} & {\tiny{}Eval$\left(V\right)$} & {\tiny{}Eval$\left(\frac{\partial Q}{\partial a}\right)$}\tabularnewline
\hline 
\hline 
{\tiny{}EVF-PGI} & {\tiny{}Spectral} & {\tiny{}No} & {\tiny{}0.056} & {\tiny{}-5.425} & {\tiny{}-3.983} & {\tiny{}1} & {\tiny{}54} & {\tiny{}0.001} & {\tiny{}5400} & {\tiny{}5400}\tabularnewline
{\tiny{}EVF-PGI{*}} & {\tiny{}Spectral} & {\tiny{}No} & {\tiny{}0.048} & {\tiny{}-5.425} & {\tiny{}-3.983} & {\tiny{}1} & {\tiny{}51} & {\tiny{}0.001} & {\tiny{}5100} & {\tiny{}5100}\tabularnewline
{\tiny{}EVFI} & {\tiny{}Spectral} & {\tiny{}Yes} & {\tiny{}0.252} & {\tiny{}-5.425} & {\tiny{}-3.983} & {\tiny{}1} & {\tiny{}31} & {\tiny{}0.008} & {\tiny{}3100} & {\tiny{}13476}\tabularnewline
{\tiny{}EVFI-ECM} & {\tiny{}Spectral} & {\tiny{}Yes} & {\tiny{}0.184} & {\tiny{}-5.406} & {\tiny{}-3.96} & {\tiny{}1} & {\tiny{}29} & {\tiny{}0.006} & {\tiny{}2900} & {\tiny{}NaN}\tabularnewline
{\tiny{}EVFI-EGM} & {\tiny{}Spectral} & {\tiny{}Yes} & {\tiny{}0.296} & {\tiny{}-5.442} & {\tiny{}-3.999} & {\tiny{}1} & {\tiny{}60} & {\tiny{}0.005} & {\tiny{}6000} & {\tiny{}NaN}\tabularnewline
{\tiny{}EPI} & {\tiny{}Spectral} & {\tiny{}Yes} & {\tiny{}0.173} & {\tiny{}-5.949} & {\tiny{}-4.836} & {\tiny{}1} & {\tiny{}6} & {\tiny{}0.029} & {\tiny{}137200} & {\tiny{}3856}\tabularnewline
{\tiny{}EPI-Krylov} & {\tiny{}Spectral} & {\tiny{}Yes} & {\tiny{}0.081} & {\tiny{}-5.949} & {\tiny{}-4.836} & {\tiny{}1} & {\tiny{}7} & {\tiny{}0.012} & {\tiny{}4800} & {\tiny{}4156}\tabularnewline
{\tiny{}OEPI} & {\tiny{}Spectral} & {\tiny{}Yes} & {\tiny{}0.121} & {\tiny{}-5.949} & {\tiny{}-4.836} & {\tiny{}1} & {\tiny{}7} & {\tiny{}0.017} & {\tiny{}54600} & {\tiny{}4156}\tabularnewline
{\tiny{}OEPI-Krylov} & {\tiny{}Spectral} & {\tiny{}Yes} & {\tiny{}0.082} & {\tiny{}-5.949} & {\tiny{}-4.836} & {\tiny{}1} & {\tiny{}7} & {\tiny{}0.012} & {\tiny{}3200} & {\tiny{}4274}\tabularnewline
{\tiny{}EPI-ECM} & {\tiny{}Spectral} & {\tiny{}Yes} & {\tiny{}0.177} & {\tiny{}-5.926} & {\tiny{}-4.802} & {\tiny{}1} & {\tiny{}7} & {\tiny{}0.025} & {\tiny{}142600} & {\tiny{}NaN}\tabularnewline
{\tiny{}EPI-Krylov-ECM} & {\tiny{}Spectral} & {\tiny{}Yes} & {\tiny{}0.062} & {\tiny{}-5.926} & {\tiny{}-4.802} & {\tiny{}1} & {\tiny{}7} & {\tiny{}0.009} & {\tiny{}5100} & {\tiny{}NaN}\tabularnewline
{\tiny{}OEPI-ECM} & {\tiny{}Spectral} & {\tiny{}Yes} & {\tiny{}0.096} & {\tiny{}-5.926} & {\tiny{}-4.802} & {\tiny{}1} & {\tiny{}7} & {\tiny{}0.014} & {\tiny{}57000} & {\tiny{}NaN}\tabularnewline
{\tiny{}OEPI-Krylov-ECM} & {\tiny{}Spectral} & {\tiny{}Yes} & {\tiny{}0.066} & {\tiny{}-5.926} & {\tiny{}-4.802} & {\tiny{}1} & {\tiny{}8} & {\tiny{}0.008} & {\tiny{}3400} & {\tiny{}NaN}\tabularnewline
{\tiny{}EPI-Krylov-EGM} & {\tiny{}Spectral} & {\tiny{}Yes} & {\tiny{}0.104} & {\tiny{}-5.442} & {\tiny{}-3.999} & {\tiny{}1} & {\tiny{}16} & {\tiny{}0.007} & {\tiny{}11200} & {\tiny{}NaN}\tabularnewline
\hline 
{\tiny{}EVF-PGI} & {\tiny{}-} & {\tiny{}No} & {\tiny{}0.021} & {\tiny{}NaN} & {\tiny{}NaN} & {\tiny{}0} & {\tiny{}8} & {\tiny{}0.003} & {\tiny{}800} & {\tiny{}800}\tabularnewline
{\tiny{}EVF-PGI{*}} & {\tiny{}-} & {\tiny{}No} & {\tiny{}0.378} & {\tiny{}-5.425} & {\tiny{}-3.983} & {\tiny{}1} & {\tiny{}447} & {\tiny{}0.001} & {\tiny{}44700} & {\tiny{}44700}\tabularnewline
{\tiny{}EVFI} & {\tiny{}-} & {\tiny{}Yes} & {\tiny{}2.974} & {\tiny{}-5.425} & {\tiny{}-3.983} & {\tiny{}1} & {\tiny{}400} & {\tiny{}0.007} & {\tiny{}40000} & {\tiny{}155720}\tabularnewline
{\tiny{}EVFI-ECM} & {\tiny{}-} & {\tiny{}Yes} & {\tiny{}2.148} & {\tiny{}-5.406} & {\tiny{}-3.96} & {\tiny{}1} & {\tiny{}400} & {\tiny{}0.005} & {\tiny{}40000} & {\tiny{}NaN}\tabularnewline
{\tiny{}EVFI-EGM} & {\tiny{}-} & {\tiny{}Yes} & {\tiny{}1.835} & {\tiny{}-5.442} & {\tiny{}-3.999} & {\tiny{}1} & {\tiny{}415} & {\tiny{}0.004} & {\tiny{}41500} & {\tiny{}NaN}\tabularnewline
{\tiny{}EPI} & {\tiny{}-} & {\tiny{}Yes} & {\tiny{}0.144} & {\tiny{}-5.949} & {\tiny{}-4.836} & {\tiny{}1} & {\tiny{}5} & {\tiny{}0.029} & {\tiny{}112800} & {\tiny{}3322}\tabularnewline
{\tiny{}EPI-Krylov} & {\tiny{}-} & {\tiny{}Yes} & {\tiny{}0.063} & {\tiny{}-5.949} & {\tiny{}-4.836} & {\tiny{}1} & {\tiny{}5} & {\tiny{}0.013} & {\tiny{}3700} & {\tiny{}3322}\tabularnewline
{\tiny{}OEPI} & {\tiny{}-} & {\tiny{}Yes} & {\tiny{}0.119} & {\tiny{}-5.949} & {\tiny{}-4.836} & {\tiny{}1} & {\tiny{}7} & {\tiny{}0.017} & {\tiny{}54000} & {\tiny{}3996}\tabularnewline
{\tiny{}OEPI-Krylov} & {\tiny{}-} & {\tiny{}Yes} & {\tiny{}0.079} & {\tiny{}-5.949} & {\tiny{}-4.836} & {\tiny{}1} & {\tiny{}7} & {\tiny{}0.011} & {\tiny{}3000} & {\tiny{}3948}\tabularnewline
{\tiny{}EPI-ECM} & {\tiny{}-} & {\tiny{}Yes} & {\tiny{}0.133} & {\tiny{}-5.926} & {\tiny{}-4.802} & {\tiny{}1} & {\tiny{}6} & {\tiny{}0.022} & {\tiny{}115000} & {\tiny{}NaN}\tabularnewline
{\tiny{}EPI-Krylov-ECM} & {\tiny{}-} & {\tiny{}Yes} & {\tiny{}0.057} & {\tiny{}-5.926} & {\tiny{}-4.802} & {\tiny{}1} & {\tiny{}5} & {\tiny{}0.011} & {\tiny{}4000} & {\tiny{}NaN}\tabularnewline
{\tiny{}OEPI-ECM} & {\tiny{}-} & {\tiny{}Yes} & {\tiny{}0.097} & {\tiny{}-5.926} & {\tiny{}-4.802} & {\tiny{}1} & {\tiny{}7} & {\tiny{}0.014} & {\tiny{}55000} & {\tiny{}NaN}\tabularnewline
{\tiny{}OEPI-Krylov-ECM} & {\tiny{}-} & {\tiny{}Yes} & {\tiny{}0.059} & {\tiny{}-5.926} & {\tiny{}-4.802} & {\tiny{}1} & {\tiny{}7} & {\tiny{}0.008} & {\tiny{}3000} & {\tiny{}NaN}\tabularnewline
{\tiny{}EPI-Krylov-EGM} & {\tiny{}-} & {\tiny{}Yes} & {\tiny{}0.311} & {\tiny{}-5.442} & {\tiny{}-3.999} & {\tiny{}1} & {\tiny{}62} & {\tiny{}0.005} & {\tiny{}19500} & {\tiny{}NaN}\tabularnewline
{\tiny{}AEVFI} & {\tiny{}-} & {\tiny{}Yes} & {\tiny{}1.148} & {\tiny{}-5.422} & {\tiny{}-3.983} & {\tiny{}1} & {\tiny{}137} & {\tiny{}0.008} & {\tiny{}13700} & {\tiny{}62742}\tabularnewline
{\tiny{}AEVFI-ECM} & {\tiny{}-} & {\tiny{}Yes} & {\tiny{}1.144} & {\tiny{}-5.422} & {\tiny{}-3.983} & {\tiny{}1} & {\tiny{}137} & {\tiny{}0.008} & {\tiny{}13700} & {\tiny{}NaN}\tabularnewline
\hline 
\end{tabular}{\tiny\par}
\par\end{centering}

{\scriptsize{}Notes. $\alpha_{0}=1,\lambda=10^{-7}$.}{\scriptsize\par}

{\scriptsize{}$L_{1}$ and $L_{\infty}$ denotes the average and the maximum absolute values of Euler residuals in log 10 units on a stochastic simulation of 10,000 observations.}{\scriptsize\par}

{\scriptsize{}``Root finding'' denotes whether a nonlinear root finding step exists in each algorithm.}{\scriptsize\par}

{\scriptsize{}Eval$\left(V\right)$ denotes the number of $\int\overline{V}^{(n)}\left(s^{\prime}\right)p\left(s^{\prime}|\widehat{s},a^{(n)}\left(\widehat{s}\right)\right)ds^{\prime}$ evaluations. Eval$\left(\frac{\partial Q}{\partial a}\right)$ denotes the number of $\frac{\partial\left[\int\overline{V}_{j}^{(n)}\left(\text{\ensuremath{s^{\prime}}}\right)p\left(s^{\prime}|\widehat{s},a^{(n*)}\left(\widehat{s}\right)\right)ds^{\prime}\right]}{\partial a^{(n*)}\left(\widehat{s}\right)}$ evaluations.}{\scriptsize\par}

\end{table}
{\footnotesize\par}
\par\end{center}

\begin{center}
{\footnotesize{}}
\begin{table}[H]
{\footnotesize{}\caption{Speed and accuracy of algorithms (Single-agent neoclassical growth model with elastic labor supply)\label{tab:results_growth_elastic_labor-2-1-2-1-2}}
}{\footnotesize\par}
\begin{centering}
{\tiny{}}%
\begin{tabular}{ccccccccccc}
\hline 
{\tiny{}Method} & {\tiny{}Acceleration} & {\tiny{}Root finding} & {\tiny{}CPU (sec)} & {\tiny{}$L_{1}$} & {\tiny{}$L_{\infty}$} & {\tiny{}Conv.} & {\tiny{}Iter.} & {\tiny{}CPU / Iter.} & {\tiny{}Eval$\left(V\right)$} & {\tiny{}Eval$\left(\frac{\partial Q}{\partial a}\right)$}\tabularnewline
\hline 
\hline 
{\tiny{}VF-PGI} & {\tiny{}SQUAREM} & {\tiny{}No} & {\tiny{}0.409} & {\tiny{}10.685} & {\tiny{}14.677} & {\tiny{}0} & {\tiny{}500} & {\tiny{}0.001} & {\tiny{}50000} & {\tiny{}50000}\tabularnewline
{\tiny{}VF-PGI{*}} & {\tiny{}SQUAREM} & {\tiny{}No} & {\tiny{}0.059} & {\tiny{}NaN} & {\tiny{}NaN} & {\tiny{}0} & {\tiny{}22} & {\tiny{}0.003} & {\tiny{}2200} & {\tiny{}2200}\tabularnewline
{\tiny{}VFI} & {\tiny{}SQUAREM} & {\tiny{}Yes} & {\tiny{}0.595} & {\tiny{}-5.425} & {\tiny{}-3.983} & {\tiny{}1} & {\tiny{}74} & {\tiny{}0.008} & {\tiny{}7400} & {\tiny{}31288}\tabularnewline
{\tiny{}VFI-ECM} & {\tiny{}SQUAREM} & {\tiny{}Yes} & {\tiny{}0.443} & {\tiny{}-5.406} & {\tiny{}-3.96} & {\tiny{}1} & {\tiny{}78} & {\tiny{}0.006} & {\tiny{}7800} & {\tiny{}NaN}\tabularnewline
{\tiny{}VFI-EGM} & {\tiny{}SQUAREM} & {\tiny{}Yes} & {\tiny{}0.3} & {\tiny{}-5.442} & {\tiny{}-3.999} & {\tiny{}1} & {\tiny{}63} & {\tiny{}0.005} & {\tiny{}6300} & {\tiny{}NaN}\tabularnewline
{\tiny{}PI} & {\tiny{}SQUAREM} & {\tiny{}Yes} & {\tiny{}0.449} & {\tiny{}-5.949} & {\tiny{}-4.836} & {\tiny{}1} & {\tiny{}6} & {\tiny{}0.075} & {\tiny{}623600} & {\tiny{}3688}\tabularnewline
{\tiny{}PI-Krylov} & {\tiny{}SQUAREM} & {\tiny{}Yes} & {\tiny{}0.074} & {\tiny{}-5.949} & {\tiny{}-4.836} & {\tiny{}1} & {\tiny{}6} & {\tiny{}0.012} & {\tiny{}5900} & {\tiny{}3688}\tabularnewline
{\tiny{}OPI} & {\tiny{}SQUAREM} & {\tiny{}Yes} & {\tiny{}0.167} & {\tiny{}-5.949} & {\tiny{}-4.836} & {\tiny{}1} & {\tiny{}10} & {\tiny{}0.017} & {\tiny{}91400} & {\tiny{}5472}\tabularnewline
{\tiny{}OPI-Krylov} & {\tiny{}SQUAREM} & {\tiny{}Yes} & {\tiny{}0.123} & {\tiny{}-5.949} & {\tiny{}-4.836} & {\tiny{}1} & {\tiny{}12} & {\tiny{}0.01} & {\tiny{}5900} & {\tiny{}6112}\tabularnewline
{\tiny{}PI-ECM} & {\tiny{}SQUAREM} & {\tiny{}Yes} & {\tiny{}0.435} & {\tiny{}-5.926} & {\tiny{}-4.802} & {\tiny{}1} & {\tiny{}6} & {\tiny{}0.073} & {\tiny{}644000} & {\tiny{}NaN}\tabularnewline
{\tiny{}PI-Krylov-ECM} & {\tiny{}SQUAREM} & {\tiny{}Yes} & {\tiny{}0.056} & {\tiny{}-5.926} & {\tiny{}-4.802} & {\tiny{}1} & {\tiny{}6} & {\tiny{}0.009} & {\tiny{}6200} & {\tiny{}NaN}\tabularnewline
{\tiny{}OPI-ECM} & {\tiny{}SQUAREM} & {\tiny{}Yes} & {\tiny{}0.132} & {\tiny{}-5.926} & {\tiny{}-4.802} & {\tiny{}1} & {\tiny{}11} & {\tiny{}0.012} & {\tiny{}103600} & {\tiny{}NaN}\tabularnewline
{\tiny{}OPI-Krylov-ECM} & {\tiny{}SQUAREM} & {\tiny{}Yes} & {\tiny{}0.097} & {\tiny{}-5.926} & {\tiny{}-4.802} & {\tiny{}1} & {\tiny{}14} & {\tiny{}0.007} & {\tiny{}6500} & {\tiny{}NaN}\tabularnewline
{\tiny{}PI-Krylov-EGM} & {\tiny{}SQUAREM} & {\tiny{}Yes} & {\tiny{}0.116} & {\tiny{}-5.442} & {\tiny{}-3.999} & {\tiny{}1} & {\tiny{}18} & {\tiny{}0.006} & {\tiny{}14400} & {\tiny{}NaN}\tabularnewline
{\tiny{}DVFI-ECM} & {\tiny{}SQUAREM} & {\tiny{}Yes} & {\tiny{}0.255} & {\tiny{}-5.452} & {\tiny{}-4.005} & {\tiny{}1} & {\tiny{}42} & {\tiny{}0.006} & {\tiny{}4200} & {\tiny{}NaN}\tabularnewline
{\tiny{}DVFI-EGM} & {\tiny{}SQUAREM} & {\tiny{}Yes} & {\tiny{}0.312} & {\tiny{}-5.472} & {\tiny{}-4.047} & {\tiny{}1} & {\tiny{}60} & {\tiny{}0.005} & {\tiny{}6000} & {\tiny{}NaN}\tabularnewline
{\tiny{}EE} & {\tiny{}SQUAREM} & {\tiny{}Yes} & {\tiny{}0.316} & {\tiny{}-7.515} & {\tiny{}-5.809} & {\tiny{}1} & {\tiny{}53} & {\tiny{}0.006} & {\tiny{}NaN} & {\tiny{}NaN}\tabularnewline
{\tiny{}EE} & {\tiny{}SQUAREM} & {\tiny{}No} & {\tiny{}0.045} & {\tiny{}-7.515} & {\tiny{}-5.809} & {\tiny{}1} & {\tiny{}53} & {\tiny{}0.001} & {\tiny{}NaN} & {\tiny{}NaN}\tabularnewline
\hline 
{\tiny{}VF-PGI} & {\tiny{}Anderson} & {\tiny{}No} & {\tiny{}1.184} & {\tiny{}NaN} & {\tiny{}NaN} & {\tiny{}0} & {\tiny{}500} & {\tiny{}0.002} & {\tiny{}50000} & {\tiny{}50000}\tabularnewline
{\tiny{}VF-PGI{*}} & {\tiny{}Anderson} & {\tiny{}No} & {\tiny{}0.131} & {\tiny{}-5.425} & {\tiny{}-3.983} & {\tiny{}1} & {\tiny{}116} & {\tiny{}0.001} & {\tiny{}11600} & {\tiny{}11600}\tabularnewline
{\tiny{}VFI} & {\tiny{}Anderson} & {\tiny{}Yes} & {\tiny{}0.55} & {\tiny{}-5.423} & {\tiny{}-3.983} & {\tiny{}1} & {\tiny{}60} & {\tiny{}0.009} & {\tiny{}6000} & {\tiny{}29342}\tabularnewline
{\tiny{}VFI-ECM} & {\tiny{}Anderson} & {\tiny{}Yes} & {\tiny{}0.365} & {\tiny{}-5.406} & {\tiny{}-3.959} & {\tiny{}1} & {\tiny{}48} & {\tiny{}0.008} & {\tiny{}4800} & {\tiny{}NaN}\tabularnewline
{\tiny{}VFI-EGM} & {\tiny{}Anderson} & {\tiny{}Yes} & {\tiny{}2.557} & {\tiny{}-5.234} & {\tiny{}-3.943} & {\tiny{}0} & {\tiny{}500} & {\tiny{}0.005} & {\tiny{}50100} & {\tiny{}NaN}\tabularnewline
{\tiny{}PI} & {\tiny{}Anderson} & {\tiny{}Yes} & {\tiny{}0.541} & {\tiny{}-5.949} & {\tiny{}-4.836} & {\tiny{}1} & {\tiny{}7} & {\tiny{}0.077} & {\tiny{}786700} & {\tiny{}4380}\tabularnewline
{\tiny{}PI-Krylov} & {\tiny{}Anderson} & {\tiny{}Yes} & {\tiny{}0.099} & {\tiny{}-5.949} & {\tiny{}-4.836} & {\tiny{}1} & {\tiny{}8} & {\tiny{}0.012} & {\tiny{}8700} & {\tiny{}4684}\tabularnewline
{\tiny{}OPI} & {\tiny{}Anderson} & {\tiny{}Yes} & {\tiny{}0.123} & {\tiny{}-5.949} & {\tiny{}-4.836} & {\tiny{}1} & {\tiny{}6} & {\tiny{}0.021} & {\tiny{}65800} & {\tiny{}4372}\tabularnewline
{\tiny{}OPI-Krylov} & {\tiny{}Anderson} & {\tiny{}Yes} & {\tiny{}0.12} & {\tiny{}-5.949} & {\tiny{}-4.836} & {\tiny{}1} & {\tiny{}11} & {\tiny{}0.011} & {\tiny{}5600} & {\tiny{}6166}\tabularnewline
{\tiny{}PI-ECM} & {\tiny{}Anderson} & {\tiny{}Yes} & {\tiny{}0.533} & {\tiny{}-5.926} & {\tiny{}-4.802} & {\tiny{}1} & {\tiny{}7} & {\tiny{}0.076} & {\tiny{}821800} & {\tiny{}NaN}\tabularnewline
{\tiny{}PI-Krylov-ECM} & {\tiny{}Anderson} & {\tiny{}Yes} & {\tiny{}0.073} & {\tiny{}-5.926} & {\tiny{}-4.802} & {\tiny{}1} & {\tiny{}8} & {\tiny{}0.009} & {\tiny{}8900} & {\tiny{}NaN}\tabularnewline
{\tiny{}OPI-ECM} & {\tiny{}Anderson} & {\tiny{}Yes} & {\tiny{}0.096} & {\tiny{}-5.926} & {\tiny{}-4.802} & {\tiny{}1} & {\tiny{}6} & {\tiny{}0.016} & {\tiny{}63400} & {\tiny{}NaN}\tabularnewline
{\tiny{}OPI-Krylov-ECM} & {\tiny{}Anderson} & {\tiny{}Yes} & {\tiny{}0.12} & {\tiny{}-5.926} & {\tiny{}-4.802} & {\tiny{}1} & {\tiny{}17} & {\tiny{}0.007} & {\tiny{}8700} & {\tiny{}NaN}\tabularnewline
{\tiny{}PI-Krylov-EGM} & {\tiny{}Anderson} & {\tiny{}Yes} & {\tiny{}0.133} & {\tiny{}-5.442} & {\tiny{}-3.999} & {\tiny{}1} & {\tiny{}21} & {\tiny{}0.006} & {\tiny{}15700} & {\tiny{}NaN}\tabularnewline
{\tiny{}DVFI-ECM} & {\tiny{}Anderson} & {\tiny{}Yes} & {\tiny{}0.346} & {\tiny{}-5.452} & {\tiny{}-4.005} & {\tiny{}1} & {\tiny{}57} & {\tiny{}0.006} & {\tiny{}5800} & {\tiny{}NaN}\tabularnewline
{\tiny{}DVFI-EGM} & {\tiny{}Anderson} & {\tiny{}Yes} & {\tiny{}0.27} & {\tiny{}-5.472} & {\tiny{}-4.047} & {\tiny{}1} & {\tiny{}57} & {\tiny{}0.005} & {\tiny{}5800} & {\tiny{}NaN}\tabularnewline
{\tiny{}EE} & {\tiny{}Anderson} & {\tiny{}Yes} & {\tiny{}0.185} & {\tiny{}-7.514} & {\tiny{}-5.808} & {\tiny{}1} & {\tiny{}42} & {\tiny{}0.004} & {\tiny{}NaN} & {\tiny{}NaN}\tabularnewline
{\tiny{}EE} & {\tiny{}Anderson} & {\tiny{}No} & {\tiny{}0.037} & {\tiny{}-7.514} & {\tiny{}-5.808} & {\tiny{}1} & {\tiny{}42} & {\tiny{}0.001} & {\tiny{}NaN} & {\tiny{}NaN}\tabularnewline
\hline 
\end{tabular}{\tiny\par}
\par\end{centering}

{\scriptsize{}Notes. $\alpha_{0}=1,\lambda=10^{-7}$.}{\scriptsize\par}

{\scriptsize{}$L_{1}$ and $L_{\infty}$ denotes the average and the maximum absolute values of Euler residuals in log 10 units on a stochastic simulation of 10,000 observations.}{\scriptsize\par}

{\scriptsize{}``Root finding'' denotes whether a nonlinear root finding step exists in each algorithm.}{\scriptsize\par}

{\scriptsize{}Eval$\left(V\right)$ denotes the number of $\int\overline{V}^{(n)}\left(s^{\prime}\right)p\left(s^{\prime}|\widehat{s},a^{(n)}\left(\widehat{s}\right)\right)ds^{\prime}$ evaluations. Eval$\left(\frac{\partial Q}{\partial a}\right)$ denotes the number of $\frac{\partial\left[\int\overline{V}_{j}^{(n)}\left(\text{\ensuremath{s^{\prime}}}\right)p\left(s^{\prime}|\widehat{s},a^{(n*)}\left(\widehat{s}\right)\right)ds^{\prime}\right]}{\partial a^{(n*)}\left(\widehat{s}\right)}$ evaluations.}{\scriptsize\par}

{\footnotesize{}The maximum number of iterations is set to 500.}{\footnotesize\par}
\end{table}
{\footnotesize\par}
\par\end{center}

\section{Spectral algorithm and Krylov method\label{sec:spectral-krylov}}

This section briefly describes the spectral algorithm and the Krylov iterative method. In Appendix \ref{sec:Further-discussions-on-spectral}, we discuss the line search procedure to guarantee local convergence of the spectral algorithm. The discussions in Section \ref{sec:Further-discussions-on-spectral} basically follows \citet{la2006spectral}, except for the introduction of variable-type-specific step sizes mentioned in the current section.

\subsection{Spectral algorithm\label{subsec:Spectral-algorithm}}

The spectral algorithm is designed to solve nonlinear equations and nonlinear continuous optimization problems. When we want to solve a nonlinear equation $F(x)=0$, $x^{(n)}$ is iteratively updated as follows, given initial $x^{(0)}$:\footnote{Newton's method, which uses the updating equation $x^{(n+1)}=x^{(n)}-\left(\nabla F(x^{(n)})\right)^{-1}F\left(x^{(n)}\right)\ (n=0,1,2,\cdots)$, attains fast convergence around the solution. Although Newton's method can be applied to solve the equation, computing $\nabla F(x^{(n)})$ requires coding analytical first derivatives of $F$, which is not an easy task, especially when the function $F$ is complicated. Moreover, especially when $n_{x}$, the dimension of $x$, is large, computing the inverse of $n_{x}\times n_{x}$ matrix $\nabla F(x^{(n)})$ is computationally costly. Hence, the use of the spectral algorithm is attractive from the perspective of simplicity and computational cost.}

\[
x^{(n+1)}=x^{(n)}+\sigma^{(n)}F\left(x^{(n)}\right)\ (n=0,1,2,\cdots)
\]

Here, $\alpha^{(n)}$ is generally taken based on the values of $s^{(n)}\equiv x^{(n)}-x^{(n-1)}$ and $y^{(n)}\equiv F(x^{(n)})-F(x^{(n-1)})$, and it can vary across $n=0,1,2,\cdots$. 

The idea of the spectral algorithm can be utilized in the context of fixed point iterations. Suppose we want to solve a fixed point constraint $x=\Phi(x)$. Then, by letting $F(x)=\Phi(x)-x$, we can solve $x=\Phi(x)$ by iteratively updating the values of $x$ by:\footnote{Spectral algorithm can be thought of as a generalization of extrapolation method (cf. \citet{judd1998numerical}) with varying values of $\sigma$.}

\begin{eqnarray}
x^{(n+1)} & = & x^{(n)}+\sigma^{(n)}\left(\Phi\left(x^{(n)}\right)-x^{(n)}\right)\ (n=0,1,2,\cdots)\label{eq:spectral_update}\\
 & = & \sigma^{(n)}\Phi(x^{(n)})+(1-\sigma^{(n)})x^{(n)}\ (n=0,1,2,\cdots).\nonumber 
\end{eqnarray}

As discussed in detail in \citet{varadhan2008simple} and \citet{fukasawa2024fast},\footnote{Step size $\sigma_{S3}\equiv\frac{\left\Vert s^{(n)}\right\Vert _{2}}{\left\Vert y^{(n)}\right\Vert _{2}}$ can be derived from a simple optimization problem $\min_{\sigma^{(n)}}\frac{\left\Vert x^{(n+1)}-x^{(n)}\right\Vert _{2}^{2}}{\left|\sigma^{(n)}\right|}=\frac{\left\Vert s^{(n)}+\sigma^{(n)}y^{(n)}\right\Vert _{2}^{2}}{\left|\sigma^{(n)}\right|}.$ For details, see \citet{fukasawa2024fast}.} the step size $\sigma^{(n)}=\sigma_{S3}\equiv\frac{\left\Vert s^{(n)}\right\Vert _{2}}{\left\Vert y^{(n)}\right\Vert _{2}}>0$ works well. Algorithm \ref{alg:Spectral-algorithm} shows the detailed steps of the spectral algorithm under the step size $\alpha_{S3}$.

\begin{algorithm}[H]
\begin{enumerate}
\item Set initial values of $x^{(0)}$. Choose $\sigma^{(0)}$ and tolerance level $\epsilon$.
\item Iterate the following $(n=0,1,2,\cdots)$:
\begin{enumerate}
\item Compute $\Phi\left(x^{(n)}\right)$ and $F\left(x^{(n)}\right)=\Phi\left(x^{(n)}\right)-x^{(n)}$.
\item Compute $s^{(n)}\equiv x^{(n)}-x^{(n-1)},y^{(n)}\equiv F(x^{(n)})-F(x^{(n-1)}),\text{and }\sigma^{(n)}=\frac{\left\Vert s^{(n)}\right\Vert _{2}}{\left\Vert y^{(n)}\right\Vert _{2}}\text{\ if\ }n\geq1.$
\item Compute $x^{(n+1)}=x^{(n)}+\sigma^{(n)}F(x^{(n)})$
\item If $\left\Vert \Phi(x^{(n)})-x^{(n)}\right\Vert <\epsilon$, exit the iteration. Otherwise, go back to Step 2(a).
\end{enumerate}
\end{enumerate}
\caption{Spectral algorithm with fixed point mapping\label{alg:Spectral-algorithm}}
\end{algorithm}

\subsubsection*{Variable-type-specific step size}

Although the standard spectral algorithm uses a scalar $\sigma^{(n)}$, I introduce the idea of variable-type-specific step sizes to the spectral algorithm.

To introduce the idea, suppose we would like to solve a nonlinear equation $F(x)=\left(\begin{array}{c}
F_{1}(x)\\
F_{2}(x)
\end{array}\right)=0\in\mathbb{R}^{n_{1}+n_{2}}$, where $x=\left(\begin{array}{c}
x_{1}\\
x_{2}
\end{array}\right)\in\mathbb{R}^{n_{1}+n_{2}}$. When using the standard step size $\sigma_{S3}$, we set $\sigma^{(n)}=\frac{\left\Vert s^{(n)}\right\Vert _{2}}{\left\Vert y^{(n)}\right\Vert _{2}}$, where $s^{(n)}\equiv\left(\begin{array}{c}
x_{1}^{(n)}\\
x_{2}^{(n)}
\end{array}\right)-\left(\begin{array}{c}
x_{1}^{(n-1)}\\
x_{2}^{(n-1)}
\end{array}\right)$, $y^{(n)}\equiv\left(\begin{array}{c}
F_{1}(x^{(n)})\\
F_{2}(x^{(n)})
\end{array}\right)-\left(\begin{array}{c}
F_{1}(x^{(n-1)})\\
F_{2}(x^{(n-1)})
\end{array}\right)$, and update $x$ by $x^{(n+1)}=x^{(n)}+\sigma^{(n)}F\left(x^{(n)}\right)$. 

Here, we can alternatively update the values of $x=\left(\begin{array}{c}
x_{1}\\
x_{2}
\end{array}\right)$ by $x^{(n+1)}=\left(\begin{array}{c}
x_{1}^{(n+1)}\\
x_{2}^{(n+1)}
\end{array}\right)=\left(\begin{array}{c}
x_{1}^{(n)}\\
x_{2}^{(n)}
\end{array}\right)+\left(\begin{array}{c}
\sigma_{1}^{(n)}F_{1}(x^{(n)})\\
\sigma_{2}^{(n)}F_{2}(x^{(n)})
\end{array}\right),$ where $\sigma_{m}^{(n)}=\frac{\left\Vert s_{m}^{(n)}\right\Vert _{2}}{\left\Vert y_{m}^{(n)}\right\Vert _{2}}$, $s_{m}^{(n)}\equiv x_{m}^{(n)}-x_{m}^{(n-1)}$, $y_{m}^{(n)}\equiv F_{m}(x^{(n)})-F_{m}(x^{(n-1)})\ (m=1,2)$. The idea can be generalized to the case where $N\in\mathbb{N}$ types of variables exist. 

As discussed earlier, $\sigma^{(n)}$ is chosen to accelerate the convergence. In principle, any choices of $\sigma^{(n)}$ are allowed in the spectral algorithm provided it works, and its numerical performance is more critical.\footnote{We can justify the strategy with a simple thought experiment. Suppose $x_{1}^{(n)}$ is fixed at the true value $x_{1}^{*}$, and $x_{2}^{(n)}$ is not. Then, we should solve the equation $F_{2}(x_{2};x_{1}^{*})=0$ as an equation of $x_{2}$, and setting $\sigma_{2}^{(n)}=\frac{\left\Vert s_{2}^{(n)}\right\Vert _{2}}{\left\Vert y_{2}^{(n)}\right\Vert _{2}}$ is desirable. If we assume $\sigma_{1}^{(n)}=\sigma_{2}^{(n)}=\sigma^{(n)}$, we should set $\sigma^{(n)}=\frac{\left\Vert s^{(n)}\right\Vert _{2}}{\left\Vert y^{(n)}\right\Vert _{2}}=\frac{\sqrt{\left\Vert s_{1}^{(n)}\right\Vert _{2}^{2}+\left\Vert s_{2}^{(n)}\right\Vert _{2}^{2}}}{\sqrt{\left\Vert y_{1}^{(n)}\right\Vert _{2}^{2}+\left\Vert y_{2}^{(n)}\right\Vert _{2}^{2}}}$, which may not be equal to $\frac{\left\Vert s_{2}^{(n)}\right\Vert _{2}}{\left\Vert y_{2}^{(n)}\right\Vert _{2}}$.}

\subsection{Krylov iterative method\label{subsec:Krylov-iterative-method}}

Algorithm \ref{alg:Krylov} shows the basic structure of the Krylov iterative method. As the procedure shows, the algorithm does not require the direct computation of the matrix $A$ as long as the matrix-vector product $A\bm{v}$ is available for any vector $\bm{v}$.

\begin{algorithm}[H]
\caption{Basic structure of the Krylov method\label{alg:Krylov}}

\begin{itemize}
\item Goal: solve $Ad=b$ for $d\in\mathbb{R}^{m}$, where $A\in\mathbb{R}^{m\times m}$, $b\in\mathbb{R}^{m}$.
\item Input: $d_{0}$, $b$, $\left(A\ \text{or}\ g:\mathbb{R}^{m}\rightarrow\mathbb{R}^{m}\text{\ such that\ }g(\bm{v})=A\bm{v}\right)$ , and tolerance level $\epsilon$
\item Output: $d^{*}$
\end{itemize}
\begin{enumerate}
\item Compute $r_{0}\equiv Ad-b$. Let $w_{0}\equiv r_{0}$.
\item Iterate the following $(n=1,2,\cdots)$:
\begin{enumerate}
\item Compute $w_{n}\equiv Aw_{n-1}$ (which is equivalent to $A^{n}r_{0}$)
\item Compute 
\begin{eqnarray*}
\gamma_{n}^{*} & \equiv & \arg\min_{\gamma\equiv\left(\gamma_{n,i}\right)_{i=0,\cdots,n-1}}\left\Vert A\left(d_{0}+\sum_{i=0}^{n-1}\gamma_{n,i}A^{i}r_{0}\right)-b\right\Vert _{2}^{2}\\
 & = & \arg\min_{\gamma\equiv\left(\gamma_{n,i}\right)_{i=0,\cdots,n-1}}\left\Vert r_{0}+\sum_{i=0}^{n-1}\gamma_{n,i}w_{i+1}\right\Vert _{2}^{2}
\end{eqnarray*}
\item If $\left\Vert r_{0}+\sum_{i=0}^{n-1}\gamma_{n,i}w_{i+1}\right\Vert \leq\epsilon$, let $d^{*}=d_{0}+\sum_{i=0}^{n-1}\gamma_{n,i}^{*}A^{i}r_{0}=d_{0}+\sum_{i=0}^{n-1}\gamma_{n,i}^{*}w_{i+1}$ and exit the iteration.

Otherwise, go back to Step 2(a).
\end{enumerate}
\end{enumerate}
\end{algorithm}

\section{Further discussions on algorithms\label{sec:other_algorithms}}

This section discusses algorithms not described in the main part of the current paper. Unless explicitly mentioned, the following discussions mainly consider the single-agent model.

\subsection{Endogenous Value Function Iteration (EVFI) and Endogenous Policy Iteration (EPI)}

In general, states $s$ can be divided into two types of states: $x$ and $y$. $x$ denotes exogenous state variables whose state transition is not affected by the values of actions $a$ and states $y$. $y$ denotes endogenous state variables whose state transition depends on the values of actions $a$. We assume that the set of grid point $\widehat{\mathcal{S}}$ is the tensor product of $\widehat{\mathcal{X}}$ and $\widehat{\mathcal{Y}}$ (i.e. $\widehat{\mathcal{S}}=\widehat{\mathcal{X}}\times\widehat{\mathcal{Y}}$),\footnote{When we use the Smolyak method to construct grid points, as is done in the numerical experiments in Section \ref{subsec:investment-competition-conti-states}, the assumption does not hold, and it is not straightforward to apply the method.} where $\widehat{\mathcal{X}}$ and $\widehat{\mathcal{Y}}$ denotes the set of exogenous and endogenous grid points. Here, let $T_{a}$ be an operator such that $T_{a}V(\widehat{s})=r(a(\widehat{s}),\widehat{s})+\beta\int\overline{V}\left(s^{\prime}\right)p\left(s^{\prime}|\widehat{s},a(\widehat{s})\right)ds^{\prime}$ and let $T\equiv T_{a^{*}}$. In addition, let $\Lambda$ be an operator such that $\Lambda V(\widehat{s})\equiv V(\widehat{x},\widehat{y})-\frac{1}{\left|\widehat{\mathcal{Y}}\right|}\sum_{\widehat{y_{0}}\in\mathcal{\widehat{Y}}}\widetilde{V}\left(\widehat{x},\widehat{y_{0}}\right)$. \citet{bray2019markov} proposed to iterate the following until convergence: 
\begin{enumerate}
\item Compute $a^{*(n+1)}(\widehat{s})=\arg\max_{a(\widehat{s})}Q\left(a(\widehat{s}),\widehat{s};\widetilde{V}^{(n)}\right)$
\item Update $\widetilde{V}$ by 
\begin{eqnarray*}
\breve{V}^{(n+1)}(\widehat{s}) & = & \Lambda T_{a^{*(n+1)}}\breve{V}^{(n)}(\widehat{s})\\
 & = & Q\left(a^{*(n+1)}(\widehat{s}),\widehat{s};\breve{V}^{(n)}\right)-\frac{1}{\left|\widehat{\mathcal{Y}}\right|}\sum_{\widehat{y_{0}}\in\mathcal{\widehat{Y}}}Q\left(a^{*(n+1)}\left(\widehat{x},\widehat{y_{0}}\right),\widehat{x},\widehat{y_{0}};\breve{V}^{(n)}\right)
\end{eqnarray*}
\end{enumerate}
The method is known as Endogenous Value Function Iteration (we term the method as EVFI), which can be thought of as the generalization of the RVFI.\footnote{In the RVFI, $C(\widehat{x})$ is common for all $\widehat{x}$.} Note that $\breve{V}\equiv\left(V(\widehat{s})-C(\widehat{x})\right)_{\widehat{s}=\left(\widehat{x},\widehat{y}\right)\in\widehat{\mathcal{S}}}$, and $C(\widehat{x})$ satisfies\footnote{$C(\widehat{x})$ cannot be analytically represented as a function of $\breve{V}$, unlike the case of the RVFI. Hence, if we want to recover $V$, we should solve $T_{a}V(\widehat{s})=r(a(\widehat{s}),\widehat{s})+\beta\int\overline{V}\left(s^{\prime}\right)p\left(s^{\prime}|\widehat{s},a(\widehat{s})\right)ds^{\prime}$ for $V$ using $a=a^{*}$.}
\begin{eqnarray*}
 &  & \frac{1}{\left|\widehat{\mathcal{Y}}\right|}\sum_{\widehat{y_{0}}\in\mathcal{\widehat{Y}}}\left[r\left(a^{*}\left(\widehat{x},\widehat{y_{0}}\right),\widehat{x},\widehat{y_{0}}\right)+\beta\int\overline{\breve{V}}\left(s^{\prime}\right)p\left(s^{\prime}|\widehat{x},\widehat{y_{0}},a^{*}\left(\widehat{x},\widehat{y_{0}}\right)\right)ds^{\prime}\right]\\
 & = & C(\widehat{x})-\beta\left(\int C(x^{\prime})p\left(x^{\prime}|\widehat{x}\right)dx^{\prime}\right).
\end{eqnarray*}

Because 
\begin{eqnarray*}
a^{*}(\widehat{s}) & = & \arg\max_{a(\widehat{s})}\left[r\left(a\left(\widehat{s}\right),\widehat{s}\right)+\beta\int\left(\overline{V}\left(x^{\prime},y^{\prime}\right)\right)p\left(x^{\prime}|\widehat{x}\right)p\left(y^{\prime}|\widehat{x},\widehat{y},a\left(\widehat{x},\widehat{y}\right)\right)dx^{\prime}dy^{\prime}\right]\\
 & = & \arg\max_{a(\widehat{s})}\left[r\left(a\left(\widehat{s}\right),\widehat{s}\right)+\beta\int\left(\overline{V}\left(x^{\prime},y^{\prime}\right)\right)p\left(x^{\prime}|\widehat{x}\right)p\left(y^{\prime}|\widehat{x},\widehat{y},a\left(\widehat{x},\widehat{y}\right)\right)dx^{\prime}dy^{\prime}-\right.\\
 &  & \left.\beta\left(\int C(x^{\prime})p\left(x^{\prime}|\widehat{x}\right)dx^{\prime}\right)\cdot\left(\int p\left(y^{\prime}|\widehat{x},\widehat{y},a\left(\widehat{x},\widehat{y}\right)\right)dy^{\prime}\right)\right]\ \left(\because\int p\left(y^{\prime}|\widehat{x},\widehat{y},a\left(\widehat{x},\widehat{y}\right)\right)dy^{\prime}=1\right)\\
 & = & \arg\max_{a(\widehat{s})}\left[r\left(a\left(\widehat{s}\right),\widehat{s}\right)+\beta\int\left(\overline{V}\left(x^{\prime},y^{\prime}\right)-C\left(x^{\prime}\right)\right)p\left(x^{\prime}|\widehat{x}\right)p\left(y^{\prime}|\widehat{x},\widehat{y},a\left(\widehat{x},\widehat{y}\right)\right)dx^{\prime}dy^{\prime}\right]
\end{eqnarray*}
 and $\left(V\left(\widehat{s}\right)-C\left(\widehat{x}\right)\right)=r\left(a^{*}\left(\widehat{s}\right),\widehat{s}\right)+\beta\int\left(\overline{V}\left(s^{\prime}\right)-C\left(x^{\prime}\right)\right)p\left(s^{\prime}|\widehat{s},a^{*}\left(\widehat{s}\right)\right)ds^{\prime}-C(\widehat{x})+\beta\left(\int C(x^{\prime})p\left(x^{\prime}|\widehat{x}\right)dx^{\prime}\right)$ holds for any $C\left(\widehat{x}\right)\in\mathbb{R}\ \left(\widehat{x}\in\widehat{\mathcal{X}}\right)$, we can think of solving for $\breve{V}\equiv\left(V(\widehat{s})-C\left(\widehat{x}\right)\right)_{s\in\widehat{\mathcal{S}}}$ and $a^{*}\equiv\left(a^{*}(\widehat{s})\right)_{s\in\widehat{\mathcal{S}}}$ satisfying the following equations:

\begin{eqnarray*}
a^{*}(\widehat{s}) & = & \arg\max_{a(\widehat{s})}\left[r\left(a^{*}\left(\widehat{s}\right),\widehat{s}\right)+\beta\int\overline{\breve{V}}\left(s^{\prime}\right)p\left(s^{\prime}|\widehat{s},a^{*}\left(\widehat{s}\right)\right)ds^{\prime}\right]\\
\breve{V}\left(\widehat{s}\right) & = & \left[r\left(a^{*}\left(\widehat{s}\right),\widehat{s}\right)+\beta\int\overline{\breve{V}}\left(s^{\prime}\right)p\left(s^{\prime}|\widehat{s},a^{*}\left(\widehat{s}\right)\right)ds^{\prime}\right]\\
 &  & -\frac{1}{\left|\widehat{\mathcal{Y}}\right|}\sum_{\widehat{y_{0}}\in\mathcal{\widehat{Y}}}\left[r\left(a^{*}\left(\widehat{x},\widehat{y_{0}}\right),\widehat{x},\widehat{y_{0}}\right)+\beta\int\overline{\breve{V}}\left(s^{\prime}\right)p\left(s^{\prime}|\widehat{x},\widehat{y_{0}},a^{*}\left(\widehat{x},\widehat{y_{0}}\right)\right)ds^{\prime}\right]\\
 & = & T_{a^{*}}\breve{V}\left(\widehat{s}\right)-\frac{1}{\left|\widehat{\mathcal{Y}}\right|}\sum_{\widehat{y_{0}}\in\mathcal{\widehat{Y}}}T_{a^{*}}\breve{V}\left(\widehat{x},\widehat{y_{0}}\right)\\
 & = & \Lambda T_{a^{*}}\breve{V}\left(\widehat{s}\right)
\end{eqnarray*}

Here, $\left\{ C\left(\widehat{x}\right)\right\} _{\widehat{x}\in\mathcal{\widehat{X}}}$ satisfies 
\begin{eqnarray*}
 &  & \frac{1}{\left|\widehat{\mathcal{Y}}\right|}\sum_{\widehat{y_{0}}\in\mathcal{\widehat{Y}}}\left[r\left(a^{*}\left(\widehat{x},\widehat{y_{0}}\right),\widehat{x},\widehat{y_{0}}\right)+\beta\int\overline{\breve{V}}\left(s^{\prime}\right)p\left(s^{\prime}|\widehat{x},\widehat{y_{0}},a^{*}\left(\widehat{x},\widehat{y_{0}}\right)\right)ds^{\prime}\right]\\
 & = & C(\widehat{x})-\beta\left(\int C(x^{\prime})p\left(x^{\prime}|\widehat{x}\right)dx^{\prime}\right)\cdot
\end{eqnarray*}

\citet{bray2019markov} showed that the convergence speed of the EVFI is faster than the RVFI.\footnote{More specifically, $\left\Vert \left(\Lambda T_{a^{*}}\right)^{n}\breve{V}-\breve{V}^{*}\right\Vert $ is $O(\gamma^{n}\beta^{n})$ for for all $\gamma>\phi\left(\Lambda Q(a^{*})\right)$ for EVFI, where $\phi(\cdot)$ denotes the largest eigenvalue modulus, and $Q(a^{*})$ denotes the state transition probability matrix given the optimal policy function $a^{*}$. In contrast, $\left\Vert \left(\Delta T_{a^{*}}\right)^{n}\widetilde{V}-\widetilde{V}^{*}\right\Vert $ is $O(\gamma^{n}\beta^{n})$ for for all $\gamma>\sigma\left(Q(a^{*})\right)$ for RVFI, where $\sigma(\cdot)$ denotes the second largest eigenvalue modulus. Because $\phi\left(\Lambda Q(a^{*})\right)<\sigma\left(Q(a^{*})\right)$, EVFI converges faster than RVFI.}\footnote{\citet{chen2024identifying} applied the EVFI algorithm to solve a single-agent continuous action dynamic model, and showed that EVFI is roughly 100 times faster than VFI, and several times faster than RVFI.}

Note that we can also use the endogenous value function to the PI, as discussed in \citet{bray2019markov}. The discussions above imply that we can obtain the solution $a^{*}$ by the following iteration:
\begin{enumerate}
\item Compute $a^{*(n+1)}(\widehat{s})=\arg\max_{a(\widehat{s})}Q\left(a(\widehat{s}),\widehat{s};\breve{V}^{(n)}\right)$
\item Update $\breve{V}$ by solving the following equation:
\begin{eqnarray*}
 &  & \breve{V}^{(n+1)}(\widehat{s})\\
 & = & \Lambda T_{a^{*(n+1)}}\breve{V}^{(n)}(\widehat{s})\\
 & = & Q\left(a(\widehat{s}),\widehat{s};\breve{V}^{(n)}\right)-\frac{1}{\left|\widehat{\mathcal{Y}}\right|}\sum_{\widehat{y_{0}}\in\mathcal{\widehat{Y}}}Q\left(a\left(\widehat{x},\widehat{y_{0}}\right),\widehat{x},\widehat{y_{0}};\breve{V}^{(n)}\right)\\
 & = & \beta\left[\int\overline{\breve{V}^{(n+1)}}\left(s^{\prime}\right)p\left(s^{\prime}|\widehat{s},a^{*}\left(\widehat{s}\right)\right)ds^{\prime}-\frac{1}{\left|\widehat{\mathcal{Y}}\right|}\sum_{\widehat{y_{0}}\in\mathcal{\widehat{Y}}}\left[\int\overline{\breve{V}^{(n+1)}}\left(s^{\prime}\right)p\left(s^{\prime}|\widehat{x},\widehat{y_{0}},a^{*}\left(\widehat{x},\widehat{y_{0}}\right)\right)ds^{\prime}\right]\right]+\\
 &  & \left[r\left(a^{*(n+1)}\left(\widehat{s}\right),\widehat{s}\right)-\frac{1}{\left|\widehat{\mathcal{Y}}\right|}\sum_{\widehat{y_{0}}\in\mathcal{\widehat{Y}}}r\left(a^{*(n+1)}\left(\widehat{x},\widehat{y_{0}}\right),\widehat{x},\widehat{y_{0}}\right)\right]
\end{eqnarray*}
\end{enumerate}
We can also use the endogenous value function to other value function-based algorithms, such as the VF-PGI-Spectral.

\subsection{Optimistic Policy Iteration (OPI)}

In the policy evaluation step of the PI algorithm (Algorithm \ref{alg:PI_single_agent}), the mapping $\Phi_{V}$ is iteratively applied to $V^{(n)}$ until convergence. Optimistic Policy Iteration (OPI) is the algorithm where $V^{(n+1)}$ is computed by $V^{(n+1)}=\Phi_{V}^{m}\left(V^{(n)};a^{*(n+1)}\right)$ $(m\in\mathbb{N})$. As discussed in Section 2.3.3 of \citet{bertsekas2012dynamic}, OPI is convergent under regularity conditions. By choosing the appropriate $m$, we can typically achieve faster convergence than the PI. However, as discussed by \citet{sargent2024dynamic}, the choice is not easy: it depends on the structure of the model and the computational environment.

Note that we can also use the idea of OPI even when using Krylov-based methods in the PI. Namely, we can terminate the Krylov-based iterations before the convergence, by setting an arbitrary maximum number of iterations.\footnote{\citet{gargiani2024inexact} proposed a variant of the OPI where iterative methods (e.g., GMRES) are used in the policy evaluation step and the number of iterations in the step is determined based on the stopping criteria regulating the level of inexactness of the convergence. They show local and global contraction properties in single-agent models with finite state and finite actions. Though it is not clear whether the theoretical properties apply to the continuous state models, the approach seems promising.} The results of the numerical experiments show that it leads to slightly faster convergence than the original PI using Krylov-based methods. In the numerical experiments of the single-agent growth model, I choose $m=100$ for the PI and $m=5$ for the PI-Krylov. 

\subsection{Model-Adaptive approach for the policy evaluation step of the PI\label{subsec:Model-Adaptive-PI}}

\citet{chen2025model} proposed a model-adaptive approach for solving the linear equation of the policy evaluation step in dynamic discrete choice models The algorithm is applicable with slight modification in our setting.\footnote{Let $f(s^{\prime}|s)$ be the density of the state transition probability given actions. In the case of continuous states, \citet{chen2025model} proposes to use $\widehat{f}\left(\widehat{s}^{\prime}|\widehat{s}\right)\equiv\frac{f\left(\widehat{s}^{\prime}|\widehat{s}\right)}{\sum_{\widehat{s}^{\prime}\in\widehat{\mathcal{S}}}f\left(\widehat{s}^{\prime}|\widehat{s}\right)}$ as the discretized state transition probability matrix. However, in the single-agent growth model, $f(\widehat{k}^{\prime}|\widehat{k})$ can take 0, and the denominator of $\frac{f\left(\widehat{s}^{\prime}|\widehat{s}\right)}{\sum_{\widehat{s}^{\prime}\in\widehat{\mathcal{S}}}f\left(\widehat{s}^{\prime}|\widehat{s}\right)}$ can be zero, implying that the proposed specification is not directly applicable.} In our setting, the idea is to solve $AA^{\prime}y=r$ for $y$ by a conjugate gradient-based method,\footnote{Conjugate gradient method for solving linear equations is one kind of the Krylov-based methods. Note that the original conjugate gradient method is not directly applicable in general because the coefficient matrix should be symmetric.} and then recover $V=A^{\prime}y$. 

Though theoretical properties of the method are deeply investigated, one potential problem of the method, other than the necessity of computing the matrix $A$ in our setting, is that the method largely relies on solving the linear equation $AA^{\prime}y=r$, whose coefficient matrix's condition number can be very huge because $cond(A^{\prime}A)=\left(cond(A)\right)^{2}$.\footnote{$cond(A)$ denotes the condition number of the matrix $A$. See Section 2.6 of Kelley \citet{kelley1995iterative}.} In contrast, other major Krylov-based methods, such as the GMRES, attempt to directly solve $AV=r$ for $V$. Generally, large condition number of the coefficient matrix of a linear equation implies that numerical precision problems would be more severe (cf. Section 3.5 of \citealp{judd1998numerical}). In addition, as discussed in Section 2.3 of \citet{kelley1995iterative}, the convergence speed of the conjugate gradient-based methods generally gets slower as the condition number gets large. In fact, the numerical results in Appendix \ref{subsec:Additional-comparison-of-PI} show that the model-adaptive approach exhibited worse performance than the GMRES. Though the idea of conjugate gradient-based algorithms might be useful because their theoretical properties are established well, the original model-adaptive algorithm might not be necessarily as fast as the GMRES, at least in our setting.

\subsection{Refactored value function iteration}

\citet{ma2021dynamic} proposed a refactored value function iteration algorithm, which is much faster than the standard VFI when there exists an exogenous state that is IID or conditionally independent given other states. The idea is that we can reduce the dimension of variables to be solved by changing the algorithm when the conditions are met. Though the current study does not show numerical results because the models presented in Section \ref{subsec:PI-single-agent} do not satisfy the condition, the algorithm would be helpful when the conditions are met.

\subsection{Approach using cubic splines}

\citet{Goettler2011} and others\footnote{\citet{Goettler2011} solved a dynamic oligopoly model of durable goods firms introducing dynamic demand structure and firms' continuous investment decisions.} use cubic splines to approximate the shape of the action value functions $Q$ with respect to $a$. Because cubic spline functions are cubic functions, their derivatives are quadratic functions, which implies that the counterpart of $\frac{\partial Q}{\partial a}=0$ are quadratic equations and their analytical solutions exist. Hence, analysts need not solve nonlinear equations numerically. However, the introduction of cubic splines needs many grid points to evaluate the values of $Q$ for good approximation, thereby requiring many function evaluations.

\subsection{Projection method and deep learning-based methods}

The projection method (e.g., \citealp{judd1992projection}) parameterizes variables as functions of states, such as value functions and actions, and solves for the parameters that satisfy the nonlinear equations or minimize the residuals of the nonlinear equations. The method has recently been extended by introducing deep learning techniques (e.g., \citealp{azinovic2022deep}). As discussed in Section \ref{subsec:Spectral-algorithm}, in the VF-PGI-Spectral method, we consider Bellman equations and optimality conditions concerning actions as nonlinear equations to be solved, and the proposed method resembles projection-based methods. Moreover, actions are updated in the directions of gradients in the proposed algorithm, and deep learning techniques also rely on gradient descent-type algorithms. Note that VFI- and PI-based algorithms would be more intuitive than the projection-based algorithms, because we can provide an intuitive explanation of the iteration process until convergence, as discussed in Sections \ref{sec:Single-agent-dynamic-models} and \ref{sec:Multi-agent-dynamic-games}. 

\subsection{Reinforcement learning}

The concept of using the gradient of ``long-term reward'' concerning actions for updating actions is similar to the Deterministic Policy Gradient method (DPG; \citealp{silver2014deterministic}), which has been widely applied in the literature of reinforcement learning.\footnote{The DPG method utilizing deep learning techniques is known as DDPG (\citealp{lillicrap2015continuous}). DDPG was extended to the multi-agent deterministic policy gradient method (MADDPG) by \citet{lowe2017multi}.} However, the settings and ideas are slightly different, and I clarify the point here by considering a single-agent dynamic model.

Unlike the dynamic model examined in the current study, DPG is developed in model-free settings, where agents do not know the parametric forms of per-period profits and state transitions. DPG approximates the action value function $Q\left(a,s\right)$ by $Q\left(a,s\right)\approx\widetilde{Q}\left(a,s;\theta_{Q}\right)$ and action $a(s)$ by $a(s)\approx\widetilde{a}\left(s;\theta_{a}\right)$. The method iterates the joint updates of the parameters $\theta_{Q}$ and $\theta_{a}$ until convergence. Regarding $\theta_{a}$, the parameter is updated by $\theta_{a}^{(n+1)}=\theta_{a}^{(n)}+\lambda_{a}\nabla_{\theta_{a}}\widetilde{a}\left(s;\theta_{a}\right)\nabla_{a}\widetilde{Q}\left(s,a\right)|_{a=\widetilde{a}\left(s;\theta_{a}\right)}$, where $\lambda_{a}$ denotes the learning rate.\footnote{See \citet{silver2014deterministic} for details.} The updating equation implies that the values of $\theta_{a}$ are updated so that they step in the direction of the gradient of the action value function $Q$, and the idea is analogous to the VF-PGI-Spectral algorithm. Note that VF-PGI-Spectral is more intuitive because it directly updates actions rather than the parameters related to the actions.

\subsection{Other methods for single-agent models\label{subsec:Other-methods-sigle-agent}}

\subsubsection{Accelerated Value Iteration (AVFI)}

\citet{goyal2023first} proposed the Accelerated Value Iteration (which we term AVFI) method for solving single-agent dynamic optimization problems with finite states and finite actions. They show that the method is faster than the VFI with Anderson acceleration, and faster than PI-Exact in a relatively large state space setting. Note that they proposed a safeguarded version of the algorithm, which guarantees convergence in the finite-state and finite-action setting. However, the current study found that introducing the safeguard procedure makes the computation time longer, and the current study shows the performance of the algorithm without the safeguard procedure. 

\subsubsection{Gauss-Seidal VFI\protect\footnote{See Section 12.4 of \citet{judd1998numerical} for details.}}

To describe the Gauss-Seidal VFI, consider the case where the state space $\mathcal{S}$ is finite. Under the setting, in Gauss-Seidal VFI, $V$ is updated by $V^{(n+1)}(s)=\max_{a(s)}\frac{r\left(a\left(s\right),s\right)+\beta\sum_{s^{\prime}\in\mathcal{S}-\{s\}}V\left(s^{\prime}\right)p\left(s^{\prime}|s,a(s)\right)}{1-\beta p\left(s|s,a(s)\right)}$, where $p\left(s^{\prime}|s,a^{*}(s)\right)$ denotes the state transition probability. Though it has been known as a classical acceleration method of the VFI, the convergence is not necessarily very fast compared to recently proposed methods (e.g., Safeguarded AVFI, Anderson acceleration method) in finite-state finite-action models, as shown in \citet{goyal2023first}.

\subsubsection{Globally convergent Type-I Anderson acceleration}

\citet{zhang2020globally} proposed a Type-I Anderson acceleration method for fixed-point iterations. The method is applicable to VFI, and the iteration is guaranteed to be convergent under finite-state finite-action settings. Though the steps are more involved and it is unclear whether the global convergence results apply to the VFI for the continuous states model, the idea would be useful for guaranteeing convergence even in the continuous state setting. 

\subsubsection{Deflated Dynamics Value Iteration (DDVI)}

\citet{lee2024deflated} proposed the DDVI (Deflated Dynamics Value Iteration) algorithm, which can be thought of as the generalization of the RVFI. They show numerical results in finite-state finite-action settings that DDVI-AutoQR is roughly twice as fast as rank-1 DDVI, which corresponds to RVFI, and VFI with Anderson acceleration. Though DDVI-AutoQR requires more steps to implement than RVFI and the computation of the counterpart of $A$ discussed in the PI, the method might be promising. 

\subsubsection{Euler equation (EE) method}

The Euler Equation (EE) method attempts to directly solve for $a^{*}$, based on the Euler equation of the single-agent dynamic model. See Appendix \ref{subsec:Details-of-growth-model} for the algorithm in the single-agent growth model.

\subsubsection{Envelope Condition Method (ECM)}

In general, states $s$ can be divided into two types of state: $x$ and $y$. $x$ denotes exogenous state variables whose state transition is not affected by the values of actions $a$ and states $y$, and $y$ denotes endogenous state variables whose state transition depends on the values of actions $a$. Here, we assume that the state transitions of $y$ is deterministic as in the macroeconomics literature, and let $y^{\prime}(x,y,a)$ be the law of motion. 

Then, the original problem $V(s)=\max_{a(s)}Q(a(s),s;V)\equiv r(a^{*}(s),s)+\beta\int V\left(s^{\prime}\right)p\left(s^{\prime}|s,a^{*}(s)\right)ds^{\prime}$ can be reformulated as:

\begin{eqnarray*}
V(s) & = & \max_{y^{\prime}(s)}Q(a(s),s;V)\equiv\exists\widetilde{r}(y^{\prime}(s),x,y)+\beta\int V\left(x^{\prime},y^{\prime}(s)\right)p\left(x^{\prime}|x\right)dx^{\prime},
\end{eqnarray*}
where $r\left(a^{*}(s),s\right)=\widetilde{r}(y^{\prime}(s),x,y)$. Then, we can derive:

\begin{eqnarray*}
\frac{\partial V(s)}{\partial y} & = & \frac{\partial\widetilde{r}(y^{\prime}(s),x,y)}{\partial y}+\frac{\partial y^{\prime}(x,y)}{\partial y}\left[\frac{\partial\widetilde{r}(y^{\prime}(s),x,y)}{\partial y^{\prime}(s)}+\beta\int\frac{\partial V\left(x^{\prime},y^{\prime}(s)\right)}{\partial y^{\prime}(s)}p\left(x^{\prime}|x\right)dx^{\prime}\right]\\
 & = & \frac{\partial\widetilde{r}(y^{\prime}(s),x,y)}{\partial y}\ \ \ \left(\because\ \frac{\partial\widetilde{r}(y^{\prime}(s),x,y)}{\partial y^{\prime}(s)}+\beta\int\frac{\partial V\left(x^{\prime},y^{\prime}(s)\right)}{\partial y^{\prime}(s)}p\left(x^{\prime}|x\right)dx^{\prime}=0\right).
\end{eqnarray*}

In the VFI and PI algorithms, the mapping $\Phi_{a}$ such that $\Phi_{a}\left(V,a\right)\left(\widehat{s}\right)\equiv\left[a\left(s\right)\ \text{satisfying }\frac{\partial r\left(a\left(\widehat{s}\right),\widehat{s}\right)}{\partial a\left(\widehat{s}\right)}+\beta\frac{\partial\left[\int\overline{V}\left(s^{\prime}\right)p\left(s^{\prime}|\widehat{s},a\left(\widehat{s}\right)\right)ds^{\prime}\right]}{\partial a\left(\widehat{s}\right)}=0\right]$ is used. In some settings, we can easily derive the function $\widetilde{r}(y^{\prime}(s),x,y)$. In this case, we can use the following mapping $\Phi_{a,ECM}$ in the VFI and PI algorithms, as the alternative to the mapping $\Phi_{a}$: 

\[
\Phi_{a,ECM}\left(V,a\right)\left(\widehat{s}\right)\equiv\left[a\left(s\right)\ \text{satisfying }\frac{\partial V(s)}{\partial y}=\frac{\partial\widetilde{r}(y^{\prime}(s),x,y)}{\partial y}\ \text{and}\ y^{\prime}(s)=y^{\prime}(s,a(s))\right]
\]

Let VFI-ECM and PI-ECM be the algorithms using the mapping $\Phi_{a,ECM}$ in the VFI and PI algorithms. The advantage of using the mapping $\Phi_{a,ECM}$, rather than $\Phi_{a}$, is that we do not have to evaluate the integral concerning the realizations of future states. It leads to a smaller computational burden.\footnote{In addition, in some specific models (e.g., growth model with inelastic labor supply; cf. \citealp{coleman2021matlab}), the evaluation of $\Phi_{a,ECM}$ does not require nonlinear root findings, and it largely reduces the computational burden of the ECM.} The disadvantage is that the convergence property is less clear than the VFI and PI, as discussed in \citet{arellano2016envelope}. Appendix \ref{subsec:Details-of-growth-model} shows the algorithm in the single-agent optimal growth model. 

\subsubsection{Endogenous Gridpoint Method (EGM)}

Endogenous Gridpoint Method (EGM) is the algorithm where the grid points $\widehat{s}$ are endogenously constructed given the points $s^{\prime}$ in the next period, and apply the VFI-based algorithms. Appendix \ref{subsec:Details-of-growth-model} shows the algorithm in the single-agent optimal growth model. 

\subsubsection{ECM-DVF and EGM-DVF}

As discussed in \citet{maliar2013envelope} and \citet{maliar2014numerical}, we can think of algorithms where we solve for the derivatives of value functions concerning endogenous states, rather than value functions themselves (we term them as DVFI-ECM, DVFI-EGM). \citet{maliar2014numerical} numerically showed that the algorithms attain superior accuracy than the VFI. Section \ref{subsec:Details-of-growth-model} discusses the details of the algorithms in the context of the single-agent growth model with elastic labor supply. 

\subsection{Detailed algorithms for solving the single-agent growth model\label{subsec:Detailed-algorithms-growth}}

This subsection describes the derivations and details of the algorithms for solving the single-agent growth model with elastic labor supply.

\subsubsection*{Basic equations for the algorithms}

Concerning the implementations of the algorithms, we rely on the following equations:
\begin{itemize}
\item FOC concerning $l$:

\begin{eqnarray}
0 & = & u_{l}(c,l)+\beta zf_{l}(k,l)\cdot E_{z^{\prime}}\left[V_{k^{\prime}}\left(k^{\prime},z^{\prime}\right)\right]\nonumber \\
 & = & -B\left(1-l\right)^{-\mu}+\beta zAk^{\alpha}l^{-\alpha}\cdot E_{z^{\prime}}\left[V_{k^{\prime}}\left(k^{\prime},z^{\prime}\right)\right]\label{eq:FOC_l_growth}
\end{eqnarray}

\item FOC concerning $c$:

\begin{eqnarray}
0 & = & u_{c}(c,l)-\beta E_{z^{\prime}}\left[V_{k^{\prime}}\left(k^{\prime},z^{\prime}\right)\right]\nonumber \\
 & = & c^{-\gamma}-\beta E_{z^{\prime}}\left[V_{k^{\prime}}\left(k^{\prime},z^{\prime}\right)\right]\label{eq:FOC_c_growth}
\end{eqnarray}

\item Envelope condition:

\begin{eqnarray}
V_{k}(k,z) & = & u_{c}(c,l)\frac{\partial}{\partial k}\left[(1-\delta)k+zf(k,l)-c\right]\nonumber \\
 & = & c^{-\gamma}\cdot\left[1-\delta+\alpha zAk^{\alpha-1}l^{1-\alpha}\right]\label{eq:Envelope_cdn}
\end{eqnarray}

\item Euler equation, which can be derived from (\ref{eq:FOC_l_growth}) and (\ref{eq:Envelope_cdn}):
\end{itemize}
\begin{eqnarray*}
0 & = & -B\left(1-l\right)^{-\mu}+\beta zf_{l}(k,l)E\left[V_{k^{\prime}}(k^{\prime},z^{\prime})\right]\\
 & = & -B\left(1-l\right)^{-\mu}+\beta zAk^{\alpha}l^{-\alpha}E\left[\left(c^{\prime}\right)^{-\gamma}\cdot\left[1-\delta+\alpha z^{\prime}A\left(k^{\prime}\right)^{\alpha-1}\left(l^{\prime}\right)^{1-\alpha}\right]\right]
\end{eqnarray*}

Equations (\ref{eq:FOC_l_growth}) and (\ref{eq:FOC_c_growth}) imply:

\begin{eqnarray*}
c & = & \left(\frac{B(1-l)^{-\mu}}{Az(1-\alpha)k^{\alpha}l^{-\alpha}}\right)^{-\frac{1}{\gamma}}
\end{eqnarray*}

Furthermore, (\ref{eq:FOC_c_growth}) and (\ref{eq:Envelope_cdn}) imply:

\begin{eqnarray*}
V_{k}(k,z) & = & \beta\left[(1-\delta)+zf_{k}(k,l)\right]\cdot E_{z^{\prime}}\left[V_{k^{\prime}}(k^{\prime},z^{\prime})\right]
\end{eqnarray*}

\subsubsection*{Detailed steps of the algorithms}

\begin{algorithm}[H]
\begin{enumerate}
\item Set grid points $(k,z)$ and initial $V(k,z)$.
\item Iterate the following until convergence:
\begin{enumerate}
\item Solve for $l$ satisfying $-B\left(1-l\right)^{-\mu}+\beta zAk^{\alpha}l^{-\alpha}\cdot E_{z^{\prime}}\left[V_{k^{\prime}}\left(k^{\prime},z^{\prime}\right)\right]=0$.
\item Compute $c=\left(\frac{B(1-l)^{-\mu}}{(1-\alpha)Azk^{\alpha}l^{-\alpha}}\right)^{\frac{-1}{\gamma}}$ and $k^{\prime}=(1-\delta)k+zf(k,l)-c$
\item Update $V(k,z)$ by $V(k,z)\leftarrow u(c,l)+\beta E_{z^{\prime}}\left[\overline{V}(k^{\prime},z^{\prime})\right]$
\end{enumerate}
\end{enumerate}
\caption{VFI, PI for the single-agent growth model\label{alg:VFI-growth}}
\end{algorithm}

\begin{algorithm}[H]
\begin{enumerate}
\item Set grid points $(k,z)$ and initial $V(k,z),l(k,z),c(k,z)$.
\item Iterate the following until convergence:
\begin{enumerate}
\item Compute $\frac{\partial Q}{\partial l}(l,c;k,z;V)=-B\left(1-l\right)^{-\mu}+\beta zAk^{\alpha}l^{-\alpha}\cdot E_{z^{\prime}}\left[V_{k^{\prime}}\left(k^{\prime},z^{\prime}\right)\right]$ and update $l$ by $l(k,z)\leftarrow l(k,z)+\lambda\frac{\partial Q}{\partial l}(l,c;k,z;V)$
\item Compute $\frac{\partial Q}{\partial c}(l,c;k,z;V)=c^{-\gamma}-\beta E_{z^{\prime}}\left[V_{k^{\prime}}\left(k^{\prime},z^{\prime}\right)\right]$ and update $c$ by $c(k,z)\leftarrow c(k,z)+\lambda\frac{\partial Q}{\partial c}(l,c;k,z;V)$
\item Compute $k^{\prime}=(1-\delta)k+zf(k,l)-c$
\item Update $V(k,z)$ by $V(k,z)\leftarrow u(c,l)+\beta E_{z^{\prime}}\left[\overline{V}(k^{\prime},z^{\prime})\right]$
\end{enumerate}
\end{enumerate}
\caption{VF-PGI for the single-agent growth model\label{alg:VF-PGI-growth}}
\end{algorithm}

\begin{algorithm}[H]
\begin{enumerate}
\item Set grid points $(k,z)$ and initial $V(k,z),l(k,z)$.
\item Iterate the following until convergence:
\begin{enumerate}
\item Compute $\frac{\partial Q}{\partial l}(l,c;k,z;V)=-B\left(1-l\right)^{-\mu}+\beta zAk^{\alpha}l^{-\alpha}\cdot E_{z^{\prime}}\left[V_{k^{\prime}}\left(k^{\prime},z^{\prime}\right)\right]$ and update $l$ by $l(k,z)\leftarrow l(k,z)+\lambda\frac{\partial Q}{\partial l}(l,c;k,z;V)$
\item Compute $c=\left(\frac{B(1-l)^{-\mu}}{(1-\alpha)Azk^{\alpha}l^{-\alpha}}\right)^{\frac{-1}{\gamma}}$ and $k^{\prime}=(1-\delta)k+zf(k,l)-c$
\item Update $V(k,z)$ by $V(k,z)\leftarrow u(c,l)+\beta E_{z^{\prime}}\left[\overline{V}(k^{\prime},z^{\prime})\right]$
\end{enumerate}
\end{enumerate}
\caption{VF-PGI for the single-agent growth model\label{alg:VF-PGI-growth-1}}
\end{algorithm}

\begin{algorithm}[H]
\begin{enumerate}
\item Set grid points $(k,z)$. Set initial $V(k,z)$ for VFI-ECM, PI-ECM, or $V_{k}(k,z)$ for DVFI-ECM.
\item Iterate the following until convergence:
\begin{enumerate}
\item Solve for $l$ satisfying $V_{k}(k,z)=\frac{B(1-l)^{-\mu}}{(1-\alpha)zAk^{\alpha}l^{-\alpha}}\left[1-\delta+\alpha Azk^{\alpha-1}l^{1-\alpha}\right]$.
\item Compute $c=\left(\frac{B(1-l)^{-\mu}}{(1-\alpha)Azk^{\alpha}l^{-\alpha}}\right)^{\frac{-1}{\gamma}}$ and $k^{\prime}=(1-\delta)k+zf(k,l)-c$
\item Update $V(k,z)$ or $V_{k}(k,z)$: by $V(k,z)\leftarrow u(c,l)+\beta E_{z^{\prime}}\left[\overline{V}(k^{\prime},z^{\prime})\right]$
\begin{itemize}
\item VFI-ECM: $V(k,z)\leftarrow u(c,l)+\beta E_{z^{\prime}}\left[\overline{V}(k^{\prime},z^{\prime})\right]$
\item PI-ECM:

By applying an iterative method (e.g., Krylov), solve for $V(k,z)$ such that $V(k,z)=u(c,l)+\beta E_{z^{\prime}}\left[\overline{V}(k^{\prime},z^{\prime})\right]$
\item DVFI-ECM: $V_{k}\left(k(k^{\prime}),z\right)\leftarrow\beta\left[(1-\delta)+zf_{k}(k,l)\right]\cdot E_{z^{\prime}}\left[\overline{V_{k}}(k^{\prime},z^{\prime})\right]$.
\end{itemize}
\end{enumerate}
\end{enumerate}
\caption{VFI-ECM, PI-ECM, DVFI-ECM for the single-agent growth model\label{alg:ECM-growth}}
\end{algorithm}

\begin{algorithm}[H]
\begin{enumerate}
\item Set grid points $(k^{\prime},z)$ and initial $\theta_{V}$.
\item Iterate the following until convergence:
\begin{enumerate}
\item Compute $W\left(k^{\prime},z\right)\equiv E_{z^{\prime}}\left[V(k^{\prime},z^{\prime})|z\right]=E_{z^{\prime}}\left[\Psi\left(k^{\prime},z^{\prime}\right)\theta_{V}|z\right]$
\item Compute $c=\left[\beta W_{k^{\prime}}\left(k^{\prime},z\right)\right]^{-1/\gamma}$
\item Solve for $l$ satisfying $k^{\prime}=(1-\delta)\left(\frac{B(1-l)^{-\mu}}{\beta W_{k}(k^{\prime},z)z(1-\alpha)}\right)^{1/\alpha}l+\frac{B(1-l)^{-\mu}l}{\beta W_{k}(k^{\prime},z)(1-\alpha)}-c$.
\item Compute $c=\left(\frac{B(1-l)^{-\mu}}{(1-\alpha)Azk^{\alpha}l^{-\alpha}}\right)^{\frac{-1}{\gamma}}$ and $k=\frac{k^{\prime}+c-zf(k,l)}{1-\delta}$
\item Update $\theta_{V}$ by either of the following ways:
\begin{itemize}
\item VFI-EGM:
\begin{enumerate}
\item Compute $V\left(k(k^{\prime}),z\right)$ by $V\left(k(k^{\prime}),z\right)\leftarrow u(c,l)+\beta W(k^{\prime},z)$.
\item Update $\theta_{V}$ by $\theta_{V}=\arg\min_{\theta}\left\Vert \Psi\left(k,z\right)\theta_{V}-V\left(k(k^{\prime}),z\right)\right\Vert _{2}$.
\end{enumerate}
\item PI-EGM:

By applying an iterative method (e.g., Krylov), solve for $\theta_{V}$ such that:

\begin{eqnarray*}
V\left(k(k^{\prime}),z\right) & = & u(c,l)+\beta E_{z^{\prime}}\left[\Psi\left(k^{\prime},z^{\prime}\right)\theta_{V}|z\right]\\
\theta_{V} & = & \arg\min_{\theta}\left\Vert \Psi\left(k,z\right)\theta_{V}-V\left(k(k^{\prime}),z\right)\right\Vert _{2}
\end{eqnarray*}

\item DVFI-EGM:
\begin{enumerate}
\item Compute $V_{k}\left(k(k^{\prime}),z\right)$ by $V_{k}\left(k(k^{\prime}),z\right)\leftarrow\beta\left[(1-\delta)+zf_{k}(k,l)\right]\cdot E_{z^{\prime}}\left[\overline{V_{k}}(k^{\prime},z^{\prime})\right]$.
\item Update $\theta_{V}$ by $\theta_{V}=\arg\min_{\theta}\left\Vert \Psi_{k}\left(k,z\right)\theta_{V}-V_{k}\left(k(k^{\prime}),z\right)\right\Vert _{2}$.
\end{enumerate}
\end{itemize}
\end{enumerate}
\end{enumerate}
\caption{VFI-EGM, PI-EGM, DVFI-EGM for the single-agent growth model\label{alg:EGM-growth}}
\end{algorithm}

\begin{algorithm}[H]
\begin{enumerate}
\item Set grid points $(k,z)$ and initial $l(k,z)$.
\item Iterate the following until convergence:
\begin{enumerate}
\item Compute $c=\left(\frac{B(1-l)^{-\mu}}{(1-\alpha)Azk^{\alpha}l^{-\alpha}}\right)^{\frac{-1}{\gamma}}$ and $k^{\prime}=(1-\delta)k+zf(k,l)-c$
\item Compute $E_{z^{\prime}}\left[V_{k^{\prime}}\left(k^{\prime},z^{\prime}\right)\right]=E_{z^{\prime}}\left[\frac{B(1-l^{\prime}(k^{\prime},z^{\prime}))}{(1-\alpha)Az\left(k^{\prime}\right)^{\alpha}\left(l^{\prime}(k^{\prime},z^{\prime})\right)^{-\alpha}}\left[1-\delta+\alpha Az^{\prime}\left(k^{\prime}\right)^{\alpha-1}\left(l^{\prime}(k^{\prime},z^{\prime})\right)^{1-\alpha}\right]\right]$
\item Compute $\frac{\partial k^{\prime}}{\partial l}=(1-\alpha)Azk^{\alpha}l^{-\alpha}$
\item Update $l$ by solving $-B\left(1-l\right)^{-\mu}+\beta\frac{\partial k^{\prime}}{\partial l}E_{z^{\prime}}\left[V_{k^{\prime}}\left(k^{\prime},z^{\prime}\right)\right]=0$ for $l$:
\begin{itemize}
\item Numerical: Numerically solve for $l$
\item Analytical: $l\leftarrow1-\left(\frac{B}{\beta\frac{\partial k^{\prime}}{\partial l}E_{z^{\prime}}\left[V_{k^{\prime}}\left(k^{\prime},z^{\prime}\right)\right]}\right)^{\frac{1}{\mu}}$
\end{itemize}
\end{enumerate}
\end{enumerate}
\caption{EE for the single-agent growth model\label{alg:EE-growth}}

{\footnotesize{}Note. In the current setting, we can solve $-B\left(1-l\right)^{-\mu}+\beta\frac{\partial k^{\prime}}{\partial l}E_{z^{\prime}}\left[V_{k^{\prime}}\left(k^{\prime},z^{\prime}\right)\right]=0$ for $l$ analytically. However, in general, there is no guarantee that we can analytically obtain solutions, especially when employing complicated models. Hence, the current study experiments two algorithms: The one where the equation is analytically solved, and the one where the equation is numerically solved using the Newton's method.}{\footnotesize\par}
\end{algorithm}

\subsection{EE, ECM, and EGM in dynamic games\label{subsec:ECM,-EE,-EGM-dynamic-game}}

To ease the computation of single-agent dynamic optimization problems mainly applied in the macroeconomics literature, several methods (EE, ECM, EGM) have been proposed. These methods sometimes avoid nonlinear optimization problems in the iterations, leading to a lower computational burden. However, these methods do not seem to be directly applicable to multi-agent dynamic games. This section clarifies this point by considering the dynamic investment competition with continuous states considered in Section \ref{subsec:investment-competition-conti-states} as an example. To simplify equations, we consider the setting $c(i)=\theta_{1}i+\theta_{2}i^{2}$. Moreover, we assume that the exogenous states $z$ are constant over time, and we do not explicitly treat them as states.

\subsubsection{Euler Equation (EE) method (cf. \citealp{judd1998numerical})}

The Euler equation for the model is:\footnote{See Appendix \ref{subsec:Derivation-of-EE-EC} for the derivation. $k_{j}^{\prime\prime}$ denotes firm $j$'s capital two periods later than the current period.}

\begin{equation}
0=-c^{\prime}\left(k_{j}^{\prime}-(1-\delta)k_{j}\right)+\beta\frac{\partial\pi_{j}(k_{j}^{\prime},k_{-j}^{\prime})}{\partial k_{j}^{\prime}}+\beta(1-\delta)c^{\prime}\left(k_{j}^{\prime\prime}-(1-\delta)k_{j}^{\prime}\right)+\beta^{2}\frac{\partial k_{-j}^{\prime\prime}}{\partial k_{j}^{\prime}}\frac{\partial V_{j}(k_{j}^{\prime\prime},k_{-j}^{\prime\prime})}{\partial k_{-j}^{\prime\prime}}\label{eq:EE}
\end{equation}

If only one firm exists and $c(i)=\theta_{1}i+\theta_{2}i^{2}$, the last term disappears, and 

\[
0=-\theta_{1}-\theta_{2}i_{j}(k_{j})+\beta\frac{\partial\pi_{j}\left((1-\delta)k_{j}+i_{j}(k_{j})\right)}{\partial i_{j}}+\beta(1-\delta)\left[\theta_{1}+\theta_{2}i_{j}^{\prime}\right]
\]
holds.

Hence, we can consider the following algorithm: 
\begin{enumerate}
\item Take grid points $k_{j}$. Set initial values of $i_{j}^{(0)}(k_{j})$. Iterate the following until convergence $(n=0,1,\cdots)$:
\item Given $i_{j}^{(n)}(k_{j})$, analytically solve for $i_{j}^{(n+1)}(k_{j})$ satisfying $0=-\theta_{1}-\theta_{2}i_{j}^{(n+1)}(k_{j})+\beta\frac{\partial\pi_{j}\left(k_{j}^{\prime}(k_{j})\right)}{\partial k_{j}^{\prime}(k_{j})}+\beta(1-\delta)\left[\theta_{1}+\theta_{2}\overline{i_{j}^{(n)}}(k_{j}^{\prime}(k_{j}))\right]$,

where $k_{j}^{\prime}(k_{j})=(1-\delta)k_{j}+i_{j}^{(n)}(k_{j})$, and $\overline{i_{j}^{(n)}}$ denotes the interpolated value of $i_{j}$ using the values of $i_{j}^{(n)}(k)$.
\end{enumerate}
However, for dynamic games, the last term of equation (\ref{eq:EE}), $\beta^{2}\frac{\partial k_{-j}^{\prime\prime}}{\partial k_{j}^{\prime}}\frac{\partial V_{j}(k_{j}^{\prime\prime},k_{-j}^{\prime\prime})}{\partial k_{-j}^{\prime\prime}}$, remains. The term corresponds to the effect of strategic interactions under the Markov perfect equilibrium, and the term cannot be directly computed without introducing $V_{j}$. This implies the EE method is not directly applicable to dynamic games without explicitly introducing value functions.

\subsubsection{Envelope Condition Method (ECM; \citealp{maliar2013envelope}; \citealp{arellano2016envelope})}

In the dynamic investment competition model, envelope condition is:\footnote{See Appendix \ref{subsec:Derivation-of-EE-EC} for the derivation.}

\begin{equation}
\frac{\partial V_{j}\left(k_{j},k_{-j}\right)}{\partial k_{j}}=\frac{\partial\pi_{j}\left(k_{j},k_{-j}\right)}{\partial k_{j}}+(1-\delta)c^{\prime}\left(k_{j}^{\prime}-(1-\delta)k_{j}\right)+\beta\frac{\partial k_{-j}^{\prime}}{\partial k_{j}}\frac{\partial V_{j}\left(k_{j}^{\prime},k_{-j}^{\prime}\right)}{\partial k_{-j}^{\prime}}\label{eq:envelope_multi_agents}
\end{equation}

If firm $j$ alone exists, the last term disappears, and we have:

\[
\frac{\partial V_{j}(k_{j},k_{-j})}{\partial k_{j}}=\frac{\partial\pi_{j}(k_{j},k_{-j})}{\partial k_{j}}+(1-\delta)\left(\theta_{1}+2\theta_{2}i_{j}\right)
\]

It implies we can analytically solve for $i_{jt}$ given the value of $V_{j}(k_{j},k_{-j})$. Hence, we can consider the following algorithm to solve for $V$ when only firm $j$ exists:
\begin{enumerate}
\item Take grid points $k_{j}$. Set initial values of $V_{j}^{(0)}(k_{j})$. Iterate the following until convergence $(n=0,1,\cdots)$:
\item Given $V_{j}^{(n)}$, 
\begin{enumerate}
\item Analytically solve for $i_{j}^{(n+1)}(k_{j})$ satisfying $\frac{\partial V_{j}^{(n)}(k_{j})}{\partial k_{j}}=\frac{\partial\pi_{j}(k_{j})}{\partial k_{j}}+(1-\delta)\left(\theta_{1}+2\theta_{2}i_{j}^{(n+1)}\left(k_{j}\right)\right)$
\item Update $V_{j}(k_{j})$ by : $V_{j}^{(n+1)}(k_{j})=\pi_{j}(k_{j})-c\left(i_{j}^{(n+1)}(k_{j})\right)+\beta V_{j}^{(n)}\left((1-\delta)k_{j}+i_{j}^{(n+1)}(k_{j})\right)$
\end{enumerate}
\end{enumerate}
In contrast, in the model involving multiple firms, the term $\beta\frac{\partial k_{-j}^{\prime}}{\partial k_{j}}\frac{\partial V_{j}(k_{j}^{\prime},k_{-j}^{\prime})}{\partial k_{-j}^{\prime}}$ does not disappear, and we need knowledge of the values of the term $\frac{\partial k_{-j}^{\prime}}{\partial k_{j}}$ to analytically solve for $i_{j}^{(n+1)}(k_{j})$. However, the term cannot be directly recovered from $V$, and the idea of ECM is not directly applicable.

\subsubsection{Endogenous Gridpoint Method (EGM; \citealp{Carroll2006})}

The idea behind EGM is to solve the problem backward rather than forward. If we directly applied the EGM to the dynamic investment competition model, the algorithm would be as follows:
\begin{enumerate}
\item Take grid points $\widehat{k^{\prime}}\equiv\left(\widehat{k_{j}^{\prime}}\right)_{j\in\mathcal{J}}\in\widehat{\mathcal{S}_{E}}$. Set initial values of $\left(V_{j}^{(0)}\left(k\left(\widehat{k^{\prime}}\right)\right)\right)_{j\in\mathcal{J},\widehat{k^{\prime}}\in\widehat{\mathcal{S}_{E}}}$. Iterate the following until convergence $(n=0,1,\cdots,)$:
\item Given $V^{(n)}$, 
\begin{enumerate}
\item Analytically solve for $i_{j}^{(n+1)}\left(\widehat{k^{\prime}}\right)$ satisfying $-\theta_{1}-\theta_{2}i_{j}^{(n+1)}\left(\widehat{k^{\prime}}\right)+\beta\frac{\partial V_{j}\left(\widehat{k_{-j}^{\prime}},\widehat{k_{-j}^{\prime}}\right)}{\partial\widehat{k_{j}^{\prime}}}=0$
\item Compute endogenous grid point $k_{j}^{(n+1)}\left(\widehat{k^{\prime}}\right)=\frac{\widehat{k_{j}^{\prime}}-i_{j}^{(n+1)}\left(\widehat{k^{\prime}}\right)}{1-\delta}$
\item Update $V_{j}\left(k\left(\widehat{k^{\prime}}\right)\right)$ by $V_{j}^{(n+1)}\left(k\left(\widehat{k^{\prime}}\right)\right)=\pi_{j}\left(k\left(\widehat{k^{\prime}}\right)\right)-c\left(i_{j}^{(n+1)}\left(\widehat{k^{\prime}}\right)\right)+\beta V_{j}^{(n)}\left(\widehat{k^{\prime}}\right)$
\end{enumerate}
\end{enumerate}
Here, suppose that two equilibria $i^{(1)}\left(k^{*}\right)\equiv\left(i_{j}^{(1)}\left(k^{*}\right)\right)_{j\in\mathcal{J}}$ and $i^{(2)}\left(k^{*}\right)\equiv\left(i_{j}^{(2)}\left(k^{*}\right)\right)_{j\in\mathcal{J}}$exist at a state $k^{*}$. Let $k_{j}^{\prime(m=1,2)}=(1-\delta)i_{j}^{(m=1,2)}+k_{j}^{*}$ be the associated state of firm $j$ in the next period. We consider the case where we choose $k^{\prime(1)}\equiv\left(k_{j}^{\prime(1)}\right)_{j\in\mathcal{J}}$ and $k^{\prime(2)}\equiv\left(k_{j}^{\prime(2)}\right)_{j\in\mathcal{J}}$ as grid points. Then, if we use the first endogenous grid point, we evaluate $V_{j}(k^{*})$ as $\pi_{j}(k^{*})-c\left(i_{j}^{(1)}(k^{*})\right)+\beta V_{j}\left(k^{\prime(1)}\right)$. If we use the second endogenous grid point, we evaluate $V_{j}\left(k^{*}\right)$ as $\pi_{j}(k^{*})-c\left(i_{j}^{(2)}(k^{*})\right)+\beta V_{j}\left(k^{\prime(2)}\right)$. If we use both, we have two distinct values $V_{j}(k^{*})=\pi_{j}(k^{*})-c\left(i_{j}^{(1)}(k^{*})\right)+\beta V_{j}\left(k^{\prime(1)}\right)$ and $V_{j}(k^{*})=\pi_{j}(k^{*})-c\left(i_{j}^{(2)}(k^{*})\right)+\beta V_{j}\left(k^{\prime(2)}\right)$, which are hard to handle in the algorithm above. 

\section{Further discussions on the spectral algorithm\label{sec:Further-discussions-on-spectral}}

\subsubsection*{Line Search and Local/global convergence}

\citet{la2006spectral} proposed the DF-SANE algorithm, which introduces line search procedure to attain local and global convergence.\footnote{In the DF-SANE algorithm, the form of the step size $\sigma$ is not restricted to $\sigma_{2}^{(n)}=\frac{\left\Vert s_{2}^{(n)}\right\Vert _{2}}{\left\Vert y_{2}^{(n)}\right\Vert _{2}}$.}

In general, introducing variable-type-specific step sizes can be generalized to introducing vector step sizes $\sigma$. In the following, without loss of generality, let $\sigma F(x)$ represent $\left(\sigma^{(i)}F^{(i)}(x)\right)_{i=1,\cdots,n}$, where $\sigma$ is a $n$-dimensional vector. 

Here, we consider solving a nonlinear equation $F(x)=0$, where $F$ is continuously differentiable.

Algorithm \ref{alg:DF-SANE} shows the DF-SANE algorithm, allowing for the case where $\sigma$ is a vector.

\begin{algorithm}[H]
\caption{DF-SANE algorithm for solving the nonlinear equation $F(x)=0$\label{alg:DF-SANE}}

\begin{enumerate}
\item Set $x_{0}$, $\sigma_{0}\in[\sigma_{min}1_{n},\sigma_{max}1_{n}]$. Compute $f(x_{0})\equiv\left\Vert F(x_{0})\right\Vert ^{2}$.
\item Iterate the following until convergence $(k=0,1,2,\cdots)$:
\begin{enumerate}
\item Compute $\sigma_{k}\in[\sigma_{min}1_{n},\sigma_{max}1_{n}]$ (spectral coefficient) based on a predetermined formula. Let $d_{k}=\sigma_{k}F(x_{k})$ and $\overline{f_{k}}\equiv\max\left\{ f\left(x_{k}\right),\cdots,f\left(x_{\max\{0,k-k+1\}}\right)\right\} $.
\item Let$\alpha_{+}\leftarrow1_{n}$ and $\alpha_{-}\leftarrow1_{n}$. Iterate the following:
\begin{enumerate}
\item If $f\left(x_{k}+\alpha_{+}d_{k}\right)\leq\overline{f_{k}}+\eta_{k}-\gamma\alpha_{+}^{2}f(x_{k})$, let $x_{k+1}=x_{k}-\alpha_{+}\sigma_{k}F(x_{k})$, $\alpha_{k}^{+}\leftarrow\alpha_{+}$ and proceed to Step 2(c).
\item If $f\left(x_{k}-\alpha_{-}d_{k}\right)\leq\overline{f_{k}}+\eta_{k}-\gamma\alpha_{-}^{2}f(x_{k})$, let $x_{k+1}=x_{k}+\alpha_{-}\sigma_{k}F(x_{k})$, $\alpha_{k}^{-}\leftarrow\alpha_{-}$ and proceed to Step 2(c).
\item Choose $\alpha_{+,new}\in\left[\tau_{min}\alpha_{+},\tau_{max}\alpha_{+}\right]$ and $\alpha_{-,new}\in\left[\tau_{min}\alpha_{-},\tau_{max}\alpha_{-}\right]$.

Replace $\alpha_{+}\leftarrow\alpha_{+,new}$, $\alpha_{-}\leftarrow\alpha_{-,new}$ and go to Step 2(a).
\end{enumerate}
\item If $\left\Vert F(x_{k+1})\right\Vert \leq\epsilon$, exit the iteration.
\end{enumerate}
\end{enumerate}
\end{algorithm}

As the discussion in \citet{la2006spectral} and Appendix \ref{subsec:proof-Spectral} shows, by introducing the line search procedure, we can guarantee the followings:
\begin{enumerate}
\item There exist limit points $x_{*}$ of $\{x^{(n)}\}_{n\in\mathbb{N}}$ and $\sigma_{*}$ of $\{\sigma^{(n)}\}_{n\in\mathbb{N}}$ such that $\left\langle 2J(x)^{t}F(x),\sigma_{*}F(x_{*})\right\rangle =0$.
\item If some limit point of $\{x^{(n)}\}_{n\in\mathbb{N}}$ is a solution of $F(x)=0$, then every limit point is a solution.
\item If a limit point $x_{*}$ of $\{x^{(n)}\}_{n\in\mathbb{N}}$ is an isolated solution of $F(x)=0$,\footnote{$x_{*}$ is an isolated solution of $F(x)=0$ if there exists $\delta>0$ such that $F(x)\neq0$ whenever $0<\left\Vert x-x_{*}\right\Vert \leq\delta$.} then the whole sequence converges to $x_{*}$.
\item Let $x^{*}$ be the solution. Suppose that the there exists $\delta>0$ such that $\left\langle J(x)^{t}F(x),\sigma_{*}F(x)\right\rangle \neq0$ whenever $0<\left\Vert x-x_{*}\right\Vert \leq\delta$, for any limiting point of $\sigma_{*}\neq0$.\footnote{If $\sigma$ is a scalar, the condition reduces to the condition ``There exists $\delta>0$ such that $\left\langle J(x)^{t}F(x),F(x)\right\rangle \neq0$ whenever $0<\left\Vert x-x_{*}\right\Vert \leq\delta$''. \citet{la2006spectral} called the condition ``strong isolation of $x^{*}$''.} If the initial $x_{0}$ is close enough to $x_{*}$, $\{x^{(n)}\}_{n\in\mathbb{N}}$ converges to $x_{*}$.
\end{enumerate}
Proof of these statements are shown in Appendix \ref{subsec:proof-Spectral}.

\section{Extension of the VF-PGI: Box constraint\label{sec:Extension:-Box-constraint}}

So far, we have only considered settings in which first-order equality conditions of the optimization problems hold exactly. However, in some applications, the introduction of box constraints on the domain of actions, such as nonnegative constraints, may be suitable. For example, in the investment competition model discussed by \citet{Pakes_McGuire1994}, each firm's investment value should be nonnegative, and first-order equality conditions do not necessarily hold at boundary points. This section shows that we can easily incorporate box constraints into the proposed algorithm.

As discussed in Section \ref{subsec:Spectral-algorithm}, the proposed algorithm under no binding box constraints corresponds to solving a fixed point problem $x=\Phi(x)\equiv\left(\Phi_{V}(V,a),\Phi_{a}(V,a)\right)$, where $x\equiv(V,a)$. In contrast, in the setting where some of the box constraints are binding, we cannot directly derive fixed-point representations. For instance, in the case of the single-agent dynamic optimization problem, let $l$ and $u$ be the lower and upper bounds of the actions. Then, the following should hold:

\[
\frac{\partial Q(a^{*}\left(s\right),s;V)}{\partial a^{*}\left(s\right)}\begin{cases}
\leq0 & \text{if}\ a^{*}\left(s\right)=l\\
=0 & \text{if}\ a^{*}\left(s\right)\in(l,u)\\
\geq0 & \text{if}\ a^{*}\left(s\right)=u
\end{cases}
\]

Motivated by this, suppose that we want to solve the following general mathematical problem:

\begin{equation}
f_{i}(x)\begin{cases}
\leq0 & \text{if}\ x_{i}=l_{i}\\
=0 & \text{if}\ x_{i}\in(l_{i},u_{i})\\
\geq0 & \text{if}\ x_{i}=u_{i}
\end{cases},\label{eq:f_cdn}
\end{equation}
where $r:B\subset\mathbb{R}^{I}\rightarrow\mathbb{R}^{I}$ is a continuous function. Here, $B\equiv\Pi_{i\in\{1,\cdots,I\}}B_{i}$, $B_{i}\equiv[l_{i},u_{i}]$. 

Suppose that a function $\Phi:B\rightarrow\mathbb{R}^{I}$ satisfies:

\begin{equation}
f_{i}(x)=0\iff x_{i}=\Phi_{i}(x)\label{eq:Phi_cdn}
\end{equation}

Here, we define a function $\widehat{\Phi}:B\rightarrow B$ such that:

\[
\widehat{\Phi}_{i}(x)\equiv\begin{cases}
l_{i} & \text{if}\ f_{i}(x)\leq0\ \&\ x_{i}=l_{i}\\
u_{i} & \text{if}\ f_{i}(x)\geq0\ \&\ x_{i}=u_{i}\\
P_{B_{i}}\left(\Phi_{i}(x)\right) & \text{Otherwise}
\end{cases},
\]
where $P_{B}(x)$ denotes the projection of $x$ to a convex set $B$.

Then, the following statement holds:
\begin{prop}
\label{prop:Phi_hat_justification}$x^{*}$ is the solution to (\ref{eq:f_cdn}) $\iff$$x^{*}$ is the solution to $\widehat{\Phi}(x)=x$
\end{prop}
Regarding the dynamic optimization problems with continuous actions, we can construct an alternative mapping $\widehat{\Phi}$ from the fixed point mapping $\Phi$ under no binding constraints and the optimality conditions using $\frac{\partial Q}{\partial a}$. Then, the problem is equivalent to solving $x=\widehat{\Phi}(x)$, and we can solve the problem using fixed-point iterations $x^{(n+1)}=\widehat{\Phi}(x^{(n)})$. Note that we can also introduce the spectral algorithm.

Regarding the continuity of the function $\widehat{\Phi}$, the following statement applies.
\begin{prop}
\label{prop:continuity_Phi_hat}$\widehat{\Phi}$ is a continuous function, if $\Phi$ is continuous, $\lim_{x_{i}\rightarrow l_{i}+0,f_{i}(x)\leq0}\Phi_{i}(x)\leq l_{i}$, and $\lim_{x_{i}\rightarrow u_{i}-0,f_{i}(x)\geq0}\Phi_{i}(x)\geq u_{i}$.
\end{prop}
Then, we can easily derive the following statement for the mapping of VF-PGI, which we denote $\Phi^{VF-PGI}$:
\begin{cor}
Suppose that $\Phi^{VF-PGI}$ is a continuous function. Then, $\widehat{\Phi^{VF-PGI}}$, derived from $\Phi^{VF-PGI}$ by the box constraint on the domain of actions $a$, is continuous.
\end{cor}

\section{Details of the Numerical experiments\label{sec:Details-experiments}}

\subsection{Details of the numerical experiments in Section \ref{subsec:Single-agent-neoclassical}\label{subsec:Details-of-growth-model}}

The code for the experiment is based on the replication code of \citet{maliar2013envelope}. As in \citet{maliar2013envelope}, we use a rectangular, uniformly spaced grid of 10 $\times$ 10 points for capital $k$ and productivity $z$ within an ergodic range as a solution domain. Expectations concerning future exogenous variables are approximated by using a 3-node Gauss Hermite quadrature rule. We parameterize value functions with complete ordinary polynomials of degree 4. We assume the iteration converges when the unit-free norms of variables are less than 1E-8. For instance, we assume VFI converges when $\text{\ensuremath{\left\Vert \frac{V^{(n+1)}(s)}{V^{(n)}(s)}-1\right\Vert }}_{\infty}\leq$1E-8. Concerning the PI-based algorithms, we assume the iterations in the policy evaluation step converges when $\frac{\left\Vert V\left(\widehat{s}\right)-r\left(\widehat{s},a\left(\widehat{s}\right)\right)-\beta\int\overline{V}\left(s^{\prime}\right)p\left(s^{\prime}|\widehat{s},a\left(\widehat{s}\right)\right)ds^{\prime}\right\Vert _{\infty}}{||r\left(\widehat{s}\right)||_{\infty}}\leq$1E-9.

As in \citet{maliar2013envelope}, we use $z(1-\overline{l})$ as the initial labor function. Our initial guess on the consumption function is that a constant fraction $\pi_{c}$ of the period output goes to consumption and the rest goes to investment. As initial value functions, I use values consistent with the labor and consumption functions. 

When applying the VFI- and PI-based algorithms, we need to solve a nonlinear optimization problems for each state. To solve the problem, I apply Newton's method.

\subsection{Details of the numerical experiments in Section \ref{subsec:investment-competition-conti-states}}

I compute the expectations concerning future exogenous variables by using a 10-node Gauss Hermite quadrature rule. To ease the computation of the expectation, I apply the ``precomputation of integrals'' technique proposed by \citet{Judd2017}. The technique reduces the computational cost per iteration. 

Regarding basis functions, I use Chebyshev polynomials. As a solution domain, I choose a rectangular grid. The range of each firm's capital stock $k_{jt}$ is between $\left[1\slash1.5,1\times1.5\right]$. The range of the exogenous cost shock $\mu$ is between $\left[4\slash1.03,4\times1.03\right]$, and the range of the exogenous demand shock $\xi$ is between $\left[2\slash1.03,2\times1.03\right]$. To reduce the number of grid points without largely losing numerical accuracy, I apply the Smolyak method (\citealp{smolyak1963quadrature}; \citealp{judd2014smolyak}). The approximation level, which is denoted as $\mu$ in \citet{judd2014smolyak}, is set to 3. The number of grid points constructed based on the Smolyak method is 137($J=2$), 241($J=3$), 389($J=4$), 589($J=5$). 

As initial value functions, we use values assuming that exogenous states are constant over time and firms make constant investments so that the current capital stocks remain constant over time. As initial investment values, we use 0. The tolerance level for evaluating convergence is set to 1E-8. We assume the iteration converges when the unit-free norms of variables are less than 1E-8. Concerning the PI-based algorithms, we assume the iterations in the policy evaluation step converges when $\frac{\left\Vert V\left(\widehat{s}\right)-r\left(\widehat{s},a\left(\widehat{s}\right)\right)-\beta\int\overline{V}\left(s^{\prime}\right)p\left(s^{\prime}|\widehat{s},a\left(\widehat{s}\right)\right)ds^{\prime}\right\Vert _{\infty}}{||r\left(\widehat{s}\right)||_{\infty}}\leq$1E-9.

When applying the VFI{*} and PI{*} algorithms, we need to solve a nonlinear optimization problem for each firm and state. To solve the problem, I apply the Newton's method.

\section{Additional results\label{sec:Additional-results}}

\subsection{Single-agent neoclassical growth model with elastic supply\label{subsec:additional_results_growth_elastic}}

\subsubsection{Further results on the VF-PGI-Spectral algorithm\label{subsec:Further-results-VF-PGI-Spectral}}

Regarding the VF-PGI-Spectral, we must specify an exogenous tuning parameter $\lambda$. In the setting shown in the main part of the paper, we chose $\lambda=10^{-7}$.\footnote{Under $\alpha_{0}=1$, $a^{(n=1)}=a^{(n=0)}+\lambda\frac{\partial Q}{\partial a}\left(a^{(n=0)}\right)$ holds. If $\lambda$ is excessively large, $a^{(n=1)}$ can be too far from $a^{(n=0)}$ and the true solution, and the iteration may become unstable. In the current setting, $Q$ is of order $10^{5}$, and $\left|\frac{\partial Q}{\partial a}\right|$ can be very large. Hence, there is a need to choose small $\lambda$ like $10^{-7}$.} Table \ref{tab:growth_model_elastic_supply_lambda} shows how the results vary when $\lambda$ varies in the single-agent growth model.

The results suggest that taking a sufficiently small value of $\lambda$ is essential for the convergence of VF-PGI-Spectral. Note that the size of $\lambda$ has no significant effect on the convergence speed of the algorithm as long as it is sufficiently small, and choosing the adequate value of $\lambda$ is not critical as long as it leads to convergence.
\begin{center}
{\footnotesize{}}
\begin{table}[H]
{\footnotesize{}\caption{Effect of $\lambda$ in the VF-PGI-Spectral algorithm (Single-agent growth model with elastic labor supply)\label{tab:growth_model_elastic_supply_lambda}}
}{\footnotesize\par}
\begin{centering}
{\footnotesize{}}%
\begin{tabular}{ccccccccccc}
\hline 
{\footnotesize{}Method} & {\footnotesize{}Acceleration} & {\footnotesize{}$\log_{10}(\lambda)$} & {\footnotesize{}CPU (sec)} & {\footnotesize{}$L_{1}$} & {\footnotesize{}$L_{\infty}$} & {\footnotesize{}Conv.} & {\footnotesize{}Iter.} & {\footnotesize{}CPU / Iter.} & {\footnotesize{}Eval$\left(V\right)$} & {\footnotesize{}Eval$\left(\frac{\partial Q}{\partial a}\right)$}\tabularnewline
\hline 
\hline 
VF-PGI & {\footnotesize{}Spectral} & {\footnotesize{}-10} & {\footnotesize{}0.078} & {\footnotesize{}-5.425} & {\footnotesize{}-3.983} & {\footnotesize{}1} & {\footnotesize{}89} & {\footnotesize{}0.001} & {\footnotesize{}8900} & {\footnotesize{}8900}\tabularnewline
VF-PGI & {\footnotesize{}Spectral} & {\footnotesize{}-9} & {\footnotesize{}0.078} & {\footnotesize{}-5.425} & {\footnotesize{}-3.983} & {\footnotesize{}1} & {\footnotesize{}87} & {\footnotesize{}0.001} & {\footnotesize{}8700} & {\footnotesize{}8700}\tabularnewline
VF-PGI & {\footnotesize{}Spectral} & {\footnotesize{}-8} & {\footnotesize{}0.09} & {\footnotesize{}-5.425} & {\footnotesize{}-3.983} & {\footnotesize{}1} & {\footnotesize{}91} & {\footnotesize{}0.001} & {\footnotesize{}9100} & {\footnotesize{}9100}\tabularnewline
VF-PGI & {\footnotesize{}Spectral} & {\footnotesize{}-7} & {\footnotesize{}0.094} & {\footnotesize{}-5.425} & {\footnotesize{}-3.983} & {\footnotesize{}1} & {\footnotesize{}102} & {\footnotesize{}0.001} & {\footnotesize{}10200} & {\footnotesize{}10200}\tabularnewline
VF-PGI & {\footnotesize{}Spectral} & {\footnotesize{}-6} & {\footnotesize{}0.037} & {\footnotesize{}NaN} & {\footnotesize{}NaN} & {\footnotesize{}0} & {\footnotesize{}32} & {\footnotesize{}0.001} & {\footnotesize{}3200} & {\footnotesize{}3200}\tabularnewline
VF-PGI & {\footnotesize{}Spectral} & {\footnotesize{}-5} & {\footnotesize{}0.011} & {\footnotesize{}NaN} & {\footnotesize{}NaN} & {\footnotesize{}0} & {\footnotesize{}5} & {\footnotesize{}0.002} & {\footnotesize{}500} & {\footnotesize{}500}\tabularnewline
\hline 
\end{tabular}{\footnotesize\par}
\par\end{centering}
{\footnotesize{}Notes. $\alpha_{0}=1$.}{\footnotesize\par}

{\footnotesize{}$L_{1}$ and $L_{\infty}$ denotes the average and the maximum absolute values of Euler residuals in log 10 units on a stochastic simulation of 10,000 observations.}{\footnotesize\par}

{\footnotesize{}We let Conv.=0 if it diverges.}{\footnotesize\par}
\end{table}
{\footnotesize\par}
\par\end{center}

Table \ref{tab:VF-PGI-growth} shows the numerical results under the settings where we do not apply the spectral algorithm to the VF-PGI. The results imply that the iteration does not diverge when we set $\lambda\leq10^{-8}$. Nevertheless, the convergence is very slow, and the iterations terminated before reaching the convergence criteria. In contrast, when $\lambda\geq10^{-7}$, the iterations diverge. Hence, though VF-PGI-Spectral works well, VF-PGI-Spectral itself is not practical.
\begin{center}
{\footnotesize{}}
\begin{table}[H]
{\footnotesize{}\caption{Effect of the value of $\lambda$ in the VF-PGI algorithm (Single agent neoclassical growth model with elastic labor supply)\label{tab:VF-PGI-growth}}
}{\footnotesize\par}
\begin{centering}
{\footnotesize{}}%
\begin{tabular}{ccccccccccc}
\hline 
{\footnotesize{}Method} & {\footnotesize{}Acceleration} & {\footnotesize{}$\log_{10}(\lambda)$} & {\footnotesize{}CPU (sec)} & {\footnotesize{}$L_{1}$} & {\footnotesize{}$L_{\infty}$} & {\footnotesize{}Conv.} & {\footnotesize{}Iter.} & {\footnotesize{}CPU / Iter.} & {\footnotesize{}Eval$\left(V\right)$} & {\footnotesize{}Eval$\left(\frac{\partial Q}{\partial a}\right)$}\tabularnewline
\hline 
\hline 
VF-PGI & - & {\footnotesize{}-10} & {\footnotesize{}3.14} & {\footnotesize{}-3.266} & {\footnotesize{}-2.394} & {\footnotesize{}0} & {\footnotesize{}3000} & {\footnotesize{}0.001} & {\footnotesize{}300000} & {\footnotesize{}300000}\tabularnewline
VF-PGI & - & {\footnotesize{}-9} & {\footnotesize{}3.052} & {\footnotesize{}-3.852} & {\footnotesize{}-3.256} & {\footnotesize{}0} & {\footnotesize{}3000} & {\footnotesize{}0.001} & {\footnotesize{}300000} & {\footnotesize{}300000}\tabularnewline
VF-PGI & - & {\footnotesize{}-8} & {\footnotesize{}2.01} & {\footnotesize{}-5.425} & {\footnotesize{}-3.983} & {\footnotesize{}1} & {\footnotesize{}1963} & {\footnotesize{}0.001} & {\footnotesize{}196300} & {\footnotesize{}196300}\tabularnewline
VF-PGI & - & {\footnotesize{}-7} & {\footnotesize{}0.026} & {\footnotesize{}NaN} & {\footnotesize{}NaN} & {\footnotesize{}0} & {\footnotesize{}8} & {\footnotesize{}0.003} & {\footnotesize{}800} & {\footnotesize{}800}\tabularnewline
VF-PGI & - & {\footnotesize{}-6} & {\footnotesize{}0.008} & {\footnotesize{}NaN} & {\footnotesize{}NaN} & {\footnotesize{}0} & {\footnotesize{}4} & {\footnotesize{}0.002} & {\footnotesize{}400} & {\footnotesize{}400}\tabularnewline
VF-PGI & - & {\footnotesize{}-5} & {\footnotesize{}0.014} & {\footnotesize{}NaN} & {\footnotesize{}NaN} & {\footnotesize{}0} & {\footnotesize{}6} & {\footnotesize{}0.002} & {\footnotesize{}600} & {\footnotesize{}600}\tabularnewline
\hline 
\end{tabular}{\footnotesize\par}
\par\end{centering}
{\footnotesize{}Notes. $\alpha_{0}=1$.}{\footnotesize\par}

{\footnotesize{}$L_{1}$ and $L_{\infty}$ denotes the average and the maximum absolute values of Euler residuals in log 10 units on a stochastic simulation of 10,000 observations.}{\footnotesize\par}

{\footnotesize{}The maximum number of iterations is set to 3000.}{\footnotesize\par}

{\footnotesize{}We let Conv.=0 if it diverges.}{\footnotesize\par}
\end{table}
{\footnotesize\par}
\par\end{center}

Table \ref{tab:Effect-of-alpha0-VF-PGI-Spectral-growth} shows the results when we set $\lambda=1$. The results show that when we set $\alpha_{0}$ to be small, iterations converge. Note that the value of $\alpha_{0}$ does not largely affect the convergence speed as long as they converge. For relatively large values of $\alpha_{0}$, the iterations diverge.
\begin{center}
{\footnotesize{}}
\begin{table}[H]
{\footnotesize{}\caption{Effect of the value of $\alpha_{0}$ in the VF-PGI-Spectral algorithm (Single agent neoclassical growth model with elastic labor supply)\label{tab:Effect-of-alpha0-VF-PGI-Spectral-growth}}
}{\footnotesize\par}
\begin{centering}
{\footnotesize{}}%
\begin{tabular}{ccccccccccc}
\hline 
{\footnotesize{}Method} & {\footnotesize{}Acceleration} & {\footnotesize{}$\log_{10}(\alpha_{0})$} & {\footnotesize{}CPU (sec)} & {\footnotesize{}$L_{1}$} & {\footnotesize{}$L_{\infty}$} & {\footnotesize{}Conv.} & {\footnotesize{}Iter.} & {\footnotesize{}CPU / Iter.} & {\footnotesize{}Eval$\left(V\right)$} & {\footnotesize{}Eval$\left(\frac{\partial Q}{\partial a}\right)$}\tabularnewline
\hline 
\hline 
VF-PGI & {\footnotesize{}Spectral} & {\footnotesize{}-10} & {\footnotesize{}0.081} & {\footnotesize{}-5.425} & {\footnotesize{}-3.983} & {\footnotesize{}1} & {\footnotesize{}96} & {\footnotesize{}0.001} & {\footnotesize{}9600} & {\footnotesize{}9600}\tabularnewline
VF-PGI & {\footnotesize{}Spectral} & {\footnotesize{}-9} & {\footnotesize{}0.067} & {\footnotesize{}-5.425} & {\footnotesize{}-3.983} & {\footnotesize{}1} & {\footnotesize{}75} & {\footnotesize{}0.001} & {\footnotesize{}7500} & {\footnotesize{}7500}\tabularnewline
VF-PGI & {\footnotesize{}Spectral} & {\footnotesize{}-8} & {\footnotesize{}0.08} & {\footnotesize{}-5.425} & {\footnotesize{}-3.983} & {\footnotesize{}1} & {\footnotesize{}91} & {\footnotesize{}0.001} & {\footnotesize{}9100} & {\footnotesize{}9100}\tabularnewline
VF-PGI & {\footnotesize{}Spectral} & {\footnotesize{}-7} & {\footnotesize{}0.092} & {\footnotesize{}-5.425} & {\footnotesize{}-3.983} & {\footnotesize{}1} & {\footnotesize{}97} & {\footnotesize{}0.001} & {\footnotesize{}9700} & {\footnotesize{}9700}\tabularnewline
VF-PGI & {\footnotesize{}Spectral} & {\footnotesize{}-6} & {\footnotesize{}0.016} & {\footnotesize{}NaN} & {\footnotesize{}NaN} & {\footnotesize{}0} & {\footnotesize{}14} & {\footnotesize{}0.001} & {\footnotesize{}1400} & {\footnotesize{}1400}\tabularnewline
VF-PGI & {\footnotesize{}Spectral} & {\footnotesize{}-5} & {\footnotesize{}0.006} & {\footnotesize{}NaN} & {\footnotesize{}NaN} & {\footnotesize{}0} & {\footnotesize{}4} & {\footnotesize{}0.002} & {\footnotesize{}400} & {\footnotesize{}400}\tabularnewline
\hline 
\end{tabular}{\footnotesize\par}
\par\end{centering}
{\footnotesize{}Notes. $\lambda=1$.}{\footnotesize\par}

{\footnotesize{}$L_{1}$ and $L_{\infty}$ denotes the average and the maximum absolute values of Euler residuals in log 10 units on a stochastic simulation of 10,000 observations.}{\footnotesize\par}
\end{table}
{\footnotesize\par}
\par\end{center}

Table \ref{tab:same-alpha_comparison} shows the results under the alternative setting where we assume that $\alpha_{z}\ \left(z\in\left\{ V,a^{d=1,\cdots,D}\right\} \right)$ are common. The results show that the iteration becomes unstable when choosing a common value, and introducing variable-type-specific step sizes leads to better performance. 
\begin{center}
{\footnotesize{}}
\begin{table}[H]
{\footnotesize{}\caption{Effect of the choice of $\alpha_{z}\ \left(z\in\left\{ V,a^{d=1,\cdots,D}\right\} \right)$ in the VF-PGI-Spectral algorithm (Single agent neoclassical growth model with elastic labor supply)\label{tab:same-alpha_comparison}}
}{\footnotesize\par}
\begin{centering}
{\footnotesize{}}%
\begin{tabular}{ccccccccccc}
\hline 
{\footnotesize{}Method} & {\footnotesize{}Acceleration} & Same $\alpha_{z}$ & {\footnotesize{}CPU (sec)} & {\footnotesize{}$L_{1}$} & {\footnotesize{}$L_{\infty}$} & {\footnotesize{}Conv.} & {\footnotesize{}Iter.} & {\footnotesize{}CPU / Iter.} & {\footnotesize{}Eval$\left(V\right)$} & {\footnotesize{}Eval$\left(\frac{\partial Q}{\partial a}\right)$}\tabularnewline
\hline 
\hline 
VF-PGI & {\footnotesize{}Spectral} & {\footnotesize{}No} & {\footnotesize{}-5.425} & {\footnotesize{}-3.983} & {\footnotesize{}1} & {\footnotesize{}102} & {\footnotesize{}0.001} & {\footnotesize{}10200} & {\footnotesize{}10200} & {\footnotesize{}9600}\tabularnewline
VF-PGI & {\footnotesize{}Spectral} & Yes & {\footnotesize{}NaN} & {\footnotesize{}NaN} & {\footnotesize{}0} & {\footnotesize{}19} & {\footnotesize{}0.002} & {\footnotesize{}1900} & {\footnotesize{}1900} & {\footnotesize{}7500}\tabularnewline
\hline 
\end{tabular}{\footnotesize\par}
\par\end{centering}

{\scriptsize{}Notes. $\alpha_{0}=1,\lambda=10^{-7}$.}{\scriptsize\par}

{\scriptsize{}$L_{1}$ and $L_{\infty}$ denotes the average and the maximum absolute values of Euler residuals in log 10 units on a stochastic simulation of 10,000 observations.}{\scriptsize\par}

{\scriptsize{}``Root finding'' denotes whether a nonlinear root finding step exists in each algorithm.}{\scriptsize\par}

{\scriptsize{}Eval$\left(V\right)$ denotes the number of $\int\overline{V}^{(n)}\left(s^{\prime}\right)p\left(s^{\prime}|\widehat{s},a^{(n)}\left(\widehat{s}\right)\right)ds^{\prime}$ evaluations. Eval$\left(\frac{\partial Q}{\partial a}\right)$ denotes the number of $\frac{\partial\left[\int\overline{V}_{j}^{(n)}\left(\text{\ensuremath{s^{\prime}}}\right)p\left(s^{\prime}|\widehat{s},a^{(n*)}\left(\widehat{s}\right)\right)ds^{\prime}\right]}{\partial a^{(n*)}\left(\widehat{s}\right)}$ evaluations.}{\scriptsize\par}

\end{table}
{\footnotesize\par}
\par\end{center}

\subsubsection{Additional comparison of PI-based methods\label{subsec:Additional-comparison-of-PI}}

This section shows the numerical results of the single-agent growth model using PI where the matrix $A\equiv I_{|\widehat{\mathcal{S}}|}-\beta\widetilde{P}\widetilde{X}\left(X^{\prime}X\right)^{-1}X^{\prime}$ is explicitly computed. Because the matrix $X^{\prime}X$ can be ill-conditioned,\footnote{In fact, the current study found that the condition number of the matrix $X^{\prime}X$ sometimes exceeds $10^{10}$.} the current study uses QR factorization to more reliably compute $A$.\footnote{As discussed in \citet{nocedal1999numerical}, $(X^{\prime}X)^{-1}X^{\prime}=-PR^{-1}Q_{1}^{\prime}$ holds, where $XP=Q_{1}R$ is the QR factorization with column pivoting on the matrix $X$. Note that thee numerical accuracy got worse when directly computing $(X^{\prime}X)^{-1}X^{\prime}$ without relying on QR factorization or other methods. Note that mldivide function of MATLAB uses (truncated) QR factorization method with pivoting of columns to compute $\arg\min_{\theta}\left\Vert X\theta-V\right\Vert _{2}$.}

Other than the PI-Krylov, we additionally consider 3 algorithms: PI-Krylov{*}, PI-Exact, and PI-MA. Here, PI-Krylov{*} denotes the algorithm which directly compute the matrix $A\equiv I_{|\widehat{\mathcal{S}}|}-\beta\widetilde{P}\widetilde{X}\left(X^{\prime}X\right)^{-1}X^{\prime}$\footnote{Because $\widetilde{P}$ is a $|\widehat{\mathcal{S}}|\times\left(|\widehat{\mathcal{S}}|\times M\right)\text{-dimensional}$ matrix, it requires much memory when $|\widehat{\mathcal{S}}|$ is large. Hence, the current study alternatively compute $\widetilde{P}\widetilde{X}$ by $\sum_{m=1}^{M}w_{m}\widetilde{X_{m}}$.} to apply the Krylov method. 

PI-Exact is the algorithm which directly solves $AV=r$ for $V$ by a non-iterative method.\footnote{The current study uses the mldivide function.} Finally, PI-MA is the algorithm which corresponds to the one recently proposed by \citet{chen2025model} (see also Appendix \ref{subsec:Model-Adaptive-PI} for further discussions.).
\begin{center}
{\footnotesize{}}
\begin{table}[H]
{\footnotesize{}\caption{Speed and accuracy of algorithms (Single-agent neoclassical growth model; Comparison of PI-based algorithms)\label{tab:results_growth_elastic_labor-PI}}
}{\footnotesize\par}
\begin{centering}
{\scriptsize{}}%
\begin{tabular}{ccccccccccc}
\hline 
{\scriptsize{}$|\widehat{\mathcal{S}}|$} & {\scriptsize{}Method} & {\scriptsize{}Acceleration} & {\scriptsize{}CPU (sec)} & {\scriptsize{}$L_{1}$} & {\scriptsize{}$L_{\infty}$} & {\scriptsize{}Conv.} & {\scriptsize{}Iter.} & {\scriptsize{}Iter. (Policy eval)} & {\scriptsize{}\# of MVM} & {\scriptsize{}Eval$\left(\frac{\partial Q}{\partial a}\right)$}\tabularnewline
\hline 
\hline 
\multirow{4}{*}{{\scriptsize{}$10^{2}$}} & {\scriptsize{}PI-Krylov} & {\scriptsize{}-} & {\scriptsize{}0.077} & {\scriptsize{}-5.949} & {\scriptsize{}-4.836} & {\scriptsize{}1} & {\scriptsize{}5} & {\scriptsize{}50} & {\scriptsize{}50} & {\scriptsize{}3322}\tabularnewline
 & {\scriptsize{}PI-Krylov{*}} & {\scriptsize{}-} & {\scriptsize{}0.077} & {\scriptsize{}-5.949} & {\scriptsize{}-4.836} & {\scriptsize{}1} & {\scriptsize{}5} & {\scriptsize{}51} & {\scriptsize{}51} & {\scriptsize{}3322}\tabularnewline
 & {\scriptsize{}PI-Exact} & {\scriptsize{}-} & {\scriptsize{}0.064} & {\scriptsize{}-5.949} & {\scriptsize{}-4.836} & {\scriptsize{}1} & {\scriptsize{}5} & {\scriptsize{}NaN} & {\scriptsize{}NaN} & {\scriptsize{}3322}\tabularnewline
 & {\scriptsize{}PI-MA} & {\scriptsize{}-} & {\scriptsize{}0.071} & {\scriptsize{}-5.949} & {\scriptsize{}-4.836} & {\scriptsize{}1} & {\scriptsize{}5} & {\scriptsize{}164} & {\scriptsize{}492} & {\scriptsize{}3322}\tabularnewline
\multirow{4}{*}{{\scriptsize{}$10^{4}$}} & {\scriptsize{}PI-Krylov} & {\scriptsize{}-} & {\scriptsize{}5.56} & {\scriptsize{}-6.006} & {\scriptsize{}-4.801} & {\scriptsize{}1} & {\scriptsize{}5} & {\scriptsize{}50} & {\scriptsize{}50} & {\scriptsize{}328942}\tabularnewline
 & {\scriptsize{}PI-Krylov{*}} & {\scriptsize{}-} & {\scriptsize{}17.49} & {\scriptsize{}-6.006} & {\scriptsize{}-4.801} & {\scriptsize{}1} & {\scriptsize{}5} & {\scriptsize{}50} & {\scriptsize{}50} & {\scriptsize{}328944}\tabularnewline
 & {\scriptsize{}PI-Exact} & {\scriptsize{}-} & {\scriptsize{}37.149} & {\scriptsize{}-6.006} & {\scriptsize{}-4.801} & {\scriptsize{}1} & {\scriptsize{}5} & {\scriptsize{}NaN} & {\scriptsize{}NaN} & {\scriptsize{}328988}\tabularnewline
 & {\scriptsize{}PI-MA} & {\scriptsize{}-} & {\scriptsize{}31.637} & {\scriptsize{}-6.006} & {\scriptsize{}-4.801} & {\scriptsize{}1} & {\scriptsize{}5} & {\scriptsize{}184} & {\scriptsize{}552} & {\scriptsize{}328986}\tabularnewline
\hline 
\end{tabular}{\scriptsize\par}
\par\end{centering}

{\scriptsize{}Notes. $\alpha_{0}=1,\lambda=10^{-7}$.}{\scriptsize\par}

{\scriptsize{}$L_{1}$ and $L_{\infty}$ denotes the average and the maximum absolute values of Euler residuals in log 10 units on a stochastic simulation of 10,000 observations.}{\scriptsize\par}

{\scriptsize{}``Root finding'' denotes whether a nonlinear root finding step exists in each algorithm.}{\scriptsize\par}

{\scriptsize{}Eval$\left(V\right)$ denotes the number of $\int\overline{V}^{(n)}\left(s^{\prime}\right)p\left(s^{\prime}|\widehat{s},a^{(n)}\left(\widehat{s}\right)\right)ds^{\prime}$ evaluations. Eval$\left(\frac{\partial Q}{\partial a}\right)$ denotes the number of $\frac{\partial\left[\int\overline{V}_{j}^{(n)}\left(\text{\ensuremath{s^{\prime}}}\right)p\left(s^{\prime}|\widehat{s},a^{(n*)}\left(\widehat{s}\right)\right)ds^{\prime}\right]}{\partial a^{(n*)}\left(\widehat{s}\right)}$ evaluations.}{\scriptsize\par}

{\footnotesize{}``\# of MVM'' denotes the number of matrix-vector multiplications in the policy evaluation step. ``Iter. (Policy eval)'' denotes the mean number of iterations in the policy evaluation step.}{\footnotesize\par}
\end{table}
{\footnotesize\par}
\par\end{center}

Table \ref{tab:results_growth_elastic_labor-PI} shows the results. To validate how the results in a high-dimensional setting, the current study also experimented with the setting with 10,000 grid points (100 grid points for $k$ and 100 grid points for $z$). The results show that the PI-Krylov performs much better than the PI-Krylov and PI-MA in the high-dimensional setting ($|\widehat{S}|=10^{4}$).

Concerning non-iterative methods for solving linear equations (e.g., Methods using LU decomposition or QR factorization; corresponding to PA-Exact), it is known that the computational cost is of order $O(n^{3})$,\footnote{See \citet{trefethen2022numerical}.} where $n$ is the dimension of the equation. Hence, non-iterative methods are not so practical in high-dimensional settings. With regard to the PI-MA, the mean number of iterations in the policy evaluation step is over 3 times larger than that of the PI-Krylov{*}. In addition, the number of matrix-vector multiplication (MVM) in PI-MA is roughly 10 times larger than the PI-Krylov{*}, because PI-MA requires 3 MVM per iteration, but PI-Krylov using the GMRES algorithm requires only 1 MVM per iteration. Because MVM is the most costly part of these algorithms in large-dimensional settings, the difference leads to the larger computational costs of PI-MA.

\section{Proof\label{sec:Proof}}

\subsection{Proof of Proposition \ref{prop:Phi_hat_justification}}
\begin{proof}
(Proof of $\Rightarrow$)
\begin{itemize}
\item If $f_{i}(x^{*})\leq0$ and $x_{i}^{*}=l_{i}$, $\widehat{\Phi}_{i}(x^{*})-x_{i}^{*}=l_{i}-l_{i}=0$.
\item If $f_{i}(x^{*})\geq0$ and $x_{i}^{*}=u_{i}$, $\widehat{\Phi}_{i}(x^{*})-x_{i}^{*}=u_{i}-u_{i}=0$.
\item If $x_{i}^{*}\in(l_{i},u_{i})$, $f_{i}(x^{*})=0$ implies $x_{i}^{*}=\Phi_{i}(x^{*})$. Hence, $\widehat{\Phi}_{i}(x)-x_{i}^{*}=P_{B_{i}}\left(\Phi_{i}(x^{*})\right)-x_{i}^{*}=P_{B_{i}}\left(x_{i}^{*}\right)-x_{i}^{*}=0$.
\end{itemize}
Note that neither $\left(f_{i}(x^{*})>0\ \&\ x_{i}^{*}=l_{i}\right)$ nor $\left(f_{i}(x^{*})<0\ \&\ x_{i}^{*}=u_{i}\right)$ does not hold under (\ref{eq:f_cdn}).

Therefore, $\Rightarrow$ holds.

(Proof of $\Leftarrow$)
\begin{itemize}
\item Case of $x_{i}^{*}=l_{i}$: 

Suppose $f_{i}(x)>0$. Then, by the definition of $\widehat{\Phi}$, $l_{i}=x_{i}^{*}=P_{B_{i}}\left(\Phi_{i}(x^{*})\right)$ holds. Since $l_{i}\in B_{i}$, $l_{i}=P_{B_{i}}\left(\Phi_{i}(x^{*})\right)=\Phi_{i}(x^{*})$ holds. It implies $x_{i}^{*}=\Phi_{i}(x^{*})$, and $f_{i}(x^{*})=0$ holds by (\ref{eq:Phi_cdn}). This is a contradiction. Hence, $f_{i}(x)\leq0$ holds.
\item Case of $x_{i}^{*}=u_{i}$: We can show $f_{i}(x)\geq0$ as in the case of $x_{i}^{*}=l_{i}$.
\item Case of $x_{i}^{*}\in(l_{i},u_{i})$: Under $x_{i}^{*}\in(l_{i},u_{i})$, $x_{i}^{*}=\widehat{\Phi}_{i}(x^{*})=P_{B_{i}}\left(\Phi_{i}(x^{*})\right)=\Phi_{i}(x^{*})$ holds. Hence, by (\ref{eq:Phi_cdn}), $f_{i}(x^{*})=0$ holds.
\end{itemize}
\end{proof}

\subsection{Proof of Proposition \ref{prop:continuity_Phi_hat}}
\begin{proof}
It is sufficient to show the following.
\begin{itemize}
\item $\lim_{x_{i}\rightarrow l_{i}+0,f_{i}(x)\leq0}\widehat{\Phi}_{i}(x)=l_{i}$:

\begin{eqnarray*}
\lim_{x_{i}\rightarrow l_{i}+0,f_{i}(x)\leq0}\widehat{\Phi}_{i}(x) & = & \lim_{x_{i}\rightarrow l_{i}+0,f_{i}(x)\leq0}P_{B_{i}}\left(\Phi_{i}(x)\right)\\
 & = & P_{B_{i}}\left(\lim_{x_{i}\rightarrow l_{i}+0,f_{i}(x)\leq0}\Phi_{i}(x)\right)\\
 & = & P_{B_{i}}(l_{i})=l_{i}
\end{eqnarray*}

\item $\lim_{x_{-i}\rightarrow x_{-i}^{*}\ s.t.\ f_{i}(l_{i},x_{-i})>0,\ f_{i}(l_{i},x_{-i}^{*})=0}\widehat{\Phi}_{i}(l_{i},x_{-i})=l_{i}$:

Let $x_{-i}^{*}$ be such that $f_{i}(l_{i},x_{-i}^{*})=0$. Because $\Phi_{i}(l_{i},x_{-i})$ is continuous with respect to $x_{-i}$, for any $\epsilon>0$, there exists $\delta>0$ such that $|x_{-i}-x_{-i}^{*}|<\delta\Rightarrow\left|\Phi_{i}(l_{i},x_{-i})-\Phi_{i}(l_{i},x_{-i}^{*})\right|<\epsilon$. Then, for $x_{-i}$ such that $f_{i}(l_{i},x_{-i})>0$,

\begin{eqnarray*}
\left|\widehat{\Phi_{i}}(l_{i},x_{-i})-l_{i}\right| & = & \left|P_{B_{i}}\left(\Phi_{i}(l_{i},x_{-i})\right)-l_{i}\right|\\
 & \leq & \left|\Phi_{i}(l_{i},x_{-i})-l_{i}\right|\ \left(\because P_{B_{i}}\left(\Phi_{i}(l_{i},x_{-i})\right)\geq l_{i}\right)\\
 & = & \left|\Phi_{i}(l_{i},x_{-i})-\Phi_{i}(l_{i},x_{-i}^{*})\right|\\
 & \leq & \epsilon\ \text{ror}\ |x_{-i}-x_{-i}^{*}|<\delta
\end{eqnarray*}

\item $\lim_{x_{-i}\rightarrow x_{-i}^{*}\ s.t.\ f_{i}(u_{i},x_{-i})<0,\ f_{i}(u_{i},x_{-i}^{*})=0}\widehat{\Phi}_{i}(u_{i},x_{-i})=u_{i}$: Similar to the case above.
\end{itemize}
\end{proof}

\subsection{Derivation of the envelope condition and Euler equation in the dynamic investment competition model with continuous states\label{subsec:Derivation-of-EE-EC}}

In this section, we show equations (\ref{eq:envelope_multi_agents}) and (\ref{eq:EE}).

Let $k_{j}^{\prime}(k)$ be the optimal next period capital stock of firm $j$ given the current state $k$. By differentiating $V_{j}(k_{j},k_{-j})=\pi_{j}(k_{j},k_{-j})-c\left(k_{j}^{\prime}(k)-(1-\delta)k_{j}\right)+\beta V_{j}\left(k_{j}^{\prime}(k),k_{-j}^{\prime}(k)\right)$ with respect to $k_{j}$, we have:

\begin{eqnarray}
\frac{\partial V_{j}(k_{j},k_{-j})}{\partial k_{j}} & = & \frac{\partial\pi_{j}(k_{j},k_{-j})}{\partial k_{j}}+(1-\delta)c^{\prime}\left(k_{j}^{\prime}-(1-\delta)k_{j}\right)+\nonumber \\
 &  & \frac{\partial k_{j}^{\prime}}{\partial k_{j}}\left(-c^{\prime}\left(k_{j}^{\prime}-(1-\delta)k_{j}\right)+\beta\frac{\partial V_{j}(k_{j}^{\prime},k_{-j}^{\prime})}{\partial k_{j}^{\prime}}\right)+\beta\frac{\partial k_{-j}^{\prime}}{\partial k_{j}}\frac{\partial V(k_{j}^{\prime},k_{-j}^{\prime})}{\partial k_{-j}^{\prime}}\label{eq:dV_dk}
\end{eqnarray}

By firm $j$'s optimality condition regarding the choice of $k_{j}^{\prime}$,

\begin{equation}
-c^{\prime}\left(k_{j}^{\prime}-(1-\delta)k_{j}\right)+\beta\frac{\partial V(k_{j}^{\prime},k_{-j}^{\prime})}{\partial k_{j}^{\prime}}=0\label{eq:FOC_ecm_not_possible}
\end{equation}
 holds, and equations (\ref{eq:dV_dk}) and (\ref{eq:FOC_ecm_not_possible}) imply 

\[
\frac{\partial V_{j}\left(k_{j},k_{-j}\right)}{\partial k_{j}}=\frac{\partial\pi_{j}\left(k_{j},k_{-j}\right)}{\partial k_{j}}+(1-\delta)c^{\prime}\left(k_{j}^{\prime}-(1-\delta)k_{j}\right)+\beta\frac{\partial k_{-j}^{\prime}}{\partial k_{j}}\frac{\partial V_{j}\left(k_{j}^{\prime},k_{-j}^{\prime}\right)}{\partial k_{-j}^{\prime}},
\]
namely, equation (\ref{eq:envelope_multi_agents}).

Next, equations (\ref{eq:envelope_multi_agents}) and (\ref{eq:FOC_ecm_not_possible}) imply

\begin{eqnarray*}
0 & = & -c^{\prime}\left(k_{j}^{\prime}-(1-\delta)k_{j}\right)+\beta\frac{\partial\pi_{j}(k_{j}^{\prime},k_{-j}^{\prime})}{\partial k_{j}^{\prime}}+\beta(1-\delta)c^{\prime}\left(k_{j}^{\prime\prime}-(1-\delta)k_{j}^{\prime}\right)+\beta^{2}\frac{\partial k_{-j}^{\prime\prime}}{\partial k_{j}^{\prime}}\frac{\partial V_{j}(k_{j}^{\prime\prime},k_{-j}^{\prime\prime})}{\partial k_{-j}^{\prime\prime}},
\end{eqnarray*}
namely, equation (\ref{eq:EE}).

\subsection{Spectral algorithm\label{subsec:proof-Spectral}}

In the following, assume that $\left\{ x_{k}\right\} _{k\in\mathbb{N}}$ is generated by the DF-SANE{*} algorithm. In addition, let $g(x)\equiv2J(x)^{t}F(x)$, where $J(x)\equiv\frac{\partial F(x)}{\partial x}.$
\begin{thm}
(Global convergence; Counterpart of Theorem 1 in \citet{la2006spectral})

Assume that $\left\{ x_{k}\right\} _{k\in\mathbb{N}}$ is generated by the DF-SANE{*} algorithm. Then, every limit points $\left(x_{*},\sigma_{*}\right)$ of $\left\{ \left(x_{k},\sigma_{k}\right)\right\} _{k\in\mathbb{N}}$ satisfies $\left\langle g\left(x_{*}\right),\sigma_{*}F(x_{*})\right\rangle =0$.
\end{thm}
\begin{thm}
(Counterpart of Theorem 2 in \citet{la2006spectral})

Assume that $\left\{ x_{k}\right\} _{k\in\mathbb{N}}$ is generated by the DF-SANE{*} algorithm and that there exists a limit point $x_{*}$ of $\left\{ x_{k}\right\} _{k\in\mathbb{R}}$ such that $F(x_{*})=0$. Then, $\lim_{k\rightarrow\infty}F(x_{k})=0$.
\end{thm}
\begin{thm}
(Local convergence; Counterpart of Theorem 3 in \citet{la2006spectral})

Assume that $\left\{ x_{k}\right\} _{k\in\mathbb{N}}$ is generated by the DF-SANE{*} algorithm and that there exists a limit point $x_{*}$ of $\left\{ x_{k}\right\} _{k\in\mathbb{R}}$ such that $F(x_{*})=0$. Moreover, assume that there exists $\delta>0$ such that $F(x)\neq0$ whenever $0<\left\Vert x-x_{*}\right\Vert \leq\delta$. Then, $\lim_{k\rightarrow\infty}x_{k}=x_{*}$. Namely, if an isolated solution is a limit point of $\left\{ x_{k}\right\} _{k\in\mathbb{N}}$, then the whole sequence $x_{k}$ converges to $x_{*}$.
\end{thm}
\begin{thm}
(Local convergence; Counterpart of Theorem 4 in \citet{la2006spectral})

Let $x^{*}$ be the solution. Suppose that the there exists $\delta>0$ such that $\left\langle J(x)^{t}F(x),\sigma_{*}F(x)\right\rangle \neq0$ whenever $0<\left\Vert x-x_{*}\right\Vert \leq\delta$, for any limiting point of $\sigma_{*}\neq0$.
\end{thm}

\subsubsection*{Proof of Theorem 1}
\begin{proof}
Let $K$ be a set as defined in \citet{la2006spectral}. In the current study, let $K_{1}\subset K$ be an infinite sequence of indices such that $\lim_{k\in K_{1}}x_{k}=x^{*}$ and $\lim_{k\in K_{1}}\sigma_{k}=\sigma_{*}$. Note that $K_{1}$ defined in \citet{la2006spectral} does not impose the condition on $\sigma.$ Then, if $\{\alpha_{k}\}_{k\in K_{1}}$ does not tend to zero, $F(x_{*})=0$, namely, $\left\langle g\left(x_{*}\right),\sigma_{*}F(x_{*})\right\rangle $ holds, as shown in \citet{la2006spectral}.

Hence, we focus on the case where $\lim_{k\in K_{1}}\alpha_{k}=0$. In the setting, there exists $k_{0}\in K_{1}$ such that $\alpha_{k}<1$ for all $k\geq k_{0},k\in K_{1}$. By the discussion in the proof of Theorem 1 of \citet{la2006spectral}, the followings hold for all $k\in K_{1},k\geq k_{0}$:

\begin{eqnarray}
f\left(x_{k}-\alpha_{+}\sigma_{k}F\left(x_{k}\right)\right) & > & \overline{f_{k}}+\eta_{k}-\gamma\left(\alpha_{k}^{+}\right)^{2}f(x_{k})\label{eq:spectral_ineq_1}\\
f\left(x_{k}+\alpha_{-}\sigma_{k}F\left(x_{k}\right)\right) & > & \overline{f_{k}}+\eta_{k}-\gamma\left(\alpha_{k}^{-}\right)^{2}f(x_{k}).\label{eq:spectral_ineq_2}
\end{eqnarray}

By (\ref{eq:spectral_ineq_1}) and the discussion in \citet{la2006spectral},

\begin{eqnarray*}
\left\Vert F\left(x_{k}-\alpha_{k}^{+}\sigma_{k}F\left(x_{k}\right)\right)\right\Vert ^{2}-\left\Vert F\left(x_{k}\right)\right\Vert ^{2} & \geq & -\exists C^{+}\gamma\alpha_{k}^{+}
\end{eqnarray*}

Similarly, by (\ref{eq:spectral_ineq_2}),

\begin{eqnarray*}
\left\Vert F\left(x_{k}+\alpha_{k}^{-}\sigma_{k}F\left(x_{k}\right)\right)\right\Vert ^{2}-\left\Vert F\left(x_{k}\right)\right\Vert ^{2} & \geq & -\exists C^{-}\gamma\alpha_{k}^{-}
\end{eqnarray*}

They imply:

\begin{eqnarray*}
\frac{\left\Vert F\left(x_{k}-\alpha_{k}^{+}\sigma_{k}F\left(x_{k}\right)\right)\right\Vert ^{2}-\left\Vert F\left(x_{k}\right)\right\Vert ^{2}}{\alpha_{k}^{+}} & \geq & -C^{+}\gamma\alpha_{k}^{+}\\
\frac{\left\Vert F\left(x_{k}+\alpha_{k}^{-}\sigma_{k}F\left(x_{k}\right)\right)\right\Vert ^{2}-\left\Vert F\left(x_{k}\right)\right\Vert ^{2}}{\alpha_{k}^{-}} & \geq & -C^{-}\gamma\alpha_{k}^{-}
\end{eqnarray*}

By mean Value Theorem, there exist $\xi_{k}^{+},\xi_{k}^{-}\in[0,1]$ such that:

\begin{eqnarray*}
\left\langle g\left(x_{k}-\xi_{k}^{+}\alpha_{k}^{+}\sigma_{k}F\left(x_{k}\right)\right),-\sigma_{k}F(x_{k})\right\rangle  & \geq & -C^{+}\gamma\alpha_{k}^{+}\\
\left\langle g\left(x_{k}+\xi_{k}^{-}\alpha_{k}^{-}\sigma_{k}F\left(x_{k}\right)\right),\sigma_{k}F(x_{k})\right\rangle  & \geq & -C^{-}\gamma\alpha_{k}^{-}.
\end{eqnarray*}

Since $x_{k}\rightarrow x_{*}$, $\sigma_{k}\rightarrow\sigma_{*}$, $\alpha_{k}^{+}\rightarrow0$, $\alpha_{k}^{-}\rightarrow0$ and $\left\Vert \sigma_{k}F\left(x_{k}\right)\right\Vert $ is bounded, we obtain the following inequalities by taking limits of the two inequalities:

\begin{eqnarray*}
\left\langle g\left(x_{*}\right),-\sigma_{*}F(x_{*})\right\rangle  & \ge & 0\\
\left\langle g\left(x_{*}\right),\sigma_{*}F(x_{*})\right\rangle  & \geq & 0
\end{eqnarray*}

Hence, $\left\langle g\left(x_{*}\right),\sigma_{*}F(x_{*})\right\rangle =0$ holds. 
\end{proof}

\subsubsection*{Proof of Theorems 2 and 3}
\begin{proof}
The proof in \citet{la2006spectral} relies on $\left\Vert x_{k+1}-x_{k}\right\Vert \leq\sigma_{max}\left\Vert F(x_{k})\right\Vert $. 

Even when $\sigma_{k}$ is a $n$-dimensional vector, the following implies that $\left\Vert x_{k+1}-x_{k}\right\Vert \leq\sigma_{max}\left\Vert F(x_{k})\right\Vert $ also holds:

\begin{eqnarray*}
\left\Vert x_{k+1}-x_{k}\right\Vert _{2} & = & \sqrt{\sum_{i=1}^{n}\left(x_{k+1}^{(i)}-x_{k}^{(i)}\right)^{2}}=\sqrt{\sum_{i=1}^{n}\left(\sigma_{k}^{(i)}F^{(i)}\left(x_{k}\right)\right)^{2}}=\sqrt{\sum_{i=1}^{n}\left(\sigma_{k}^{(i)}\right)^{2}\left(F^{(i)}\left(x_{k}\right)\right)^{2}}\\
 & \leq & \sqrt{\sum_{i=1}^{n}\left(\sigma_{max}\right)^{2}\left(F^{(i)}\left(x_{k}\right)\right)^{2}}=\sigma_{max}\left\Vert F(x_{k})\right\Vert .
\end{eqnarray*}
\end{proof}

\subsubsection*{Proof of Theorem 4}

Let $\epsilon>0$ such that 

\begin{equation}
0<\left\Vert x-x_{*}\right\Vert \leq\epsilon\Rightarrow\left\langle g(x),\sigma^{*}F(x)\right\rangle \neq0\ \forall\sigma_{*}:\text{Limiting point of }\left\{ \sigma_{k}\right\} _{k}.\label{eq:strong_isolation_cdn}
\end{equation}

Then, by the same discussion in \citet{la2006spectral} using $\left\Vert x_{k+1}-x_{k}\right\Vert \leq\sigma_{max}\left\Vert F(x_{k})\right\Vert $, all the limiting points $\overline{x}$ of $\left\{ x_{k}\right\} _{k\in\mathbb{N}}$ are such that $\left\Vert \overline{x}-x_{*}\right\Vert <\epsilon$. By Theorem 1, $\overline{x}$ satisfies $\left\langle g(\overline{x}),\sigma_{*}F(\overline{x})\right\rangle =0$. Then, $\overline{x}=x_{*}$ should hold by (\ref{eq:strong_isolation_cdn}). Therefore, $\lim_{k\rightarrow\infty}x_{k}=x_{*}$.

\pagebreak{}

\bibliographystyle{apalike}
\bibliography{literature}

\end{document}